\documentclass[prd, 10pt,nofootinbib, floatfix, notitlepage,twocolumn,showpacs,aps]{revtex4-1}

\usepackage{amsmath,amsfonts,amssymb}
\usepackage{mathrsfs}
\usepackage{graphicx}
\usepackage[english]{babel} 
\usepackage{hyperref} 
\usepackage{verbatim}

\usepackage{color}
\usepackage{bbold}
\usepackage{overpic}
\newcommand{\be}{\begin{equation}}
\newcommand{\ee}{\end{equation}}
\newcommand{\bes}{\begin{equation*}}
\newcommand{\ees}{\end{equation*}}

  \usepackage[utf8]{inputenc} 
  
\newcommand{\bench}{$\mathcal{N}=3$ }
\newcommand{\smith}{$\mathcal{N}=1.6$}
\newcommand{\urena}{$\mathcal{N}=100$ }
\newcommand\mnras{MNRAS}

\begin{document}

\title{ How sound are our ultralight axion approximations?}

\author{Jonathan Cookmeyer$^{1,2}$}
\author{Daniel Grin$^1$}\email{dgrin@haverford.edu}
\author{Tristan L.~Smith$^{3}$}
\affiliation{$^1$Department of Physics and Astronomy, Haverford College, 370 Lancaster Avenue, Haverford, Pennsylvania 19041, USA}
\affiliation{$^2$Department of Physics, University of California, Berkeley, California 94720, USA}
\affiliation{$^3$Department of Physics and Astronomy, Swarthmore College, 500 College Avenue, Swarthmore, Pennsylvania 19081, USA}

\begin{abstract}
Ultralight axions (ULAs) are a promising dark-matter candidate. ULAs may have implications for small-scale challenges to the $\Lambda$CDM model, and arise in string scenarios. ULAs are already constrained by cosmic microwave background (CMB) experiments and large-scale structure surveys, and will be probed with much greater sensitivity by future efforts. It is challenging to compute observables in ULA scenarios with sufficient speed and accuracy for cosmological data analysis because the ULA field oscillates rapidly. In past work, an effective fluid approximation has been used to make these computations feasible. Here this approximation is tested against an exact solution of the ULA equations, comparing the induced error of CMB observables with the sensitivity of current and future experiments. In the most constrained mass range for a ULA dark-matter component ($10^{-27}~{\rm eV}\leq m_{\rm ax}\leq 10^{-25}~{\rm eV}$), the induced bias on the allowed ULA fraction of dark matter from \textit{Planck} data is less than $1\sigma$. In the cosmic-variance limit (including temperature and polarization data), the bias is $\lesssim 2\sigma$ for primary CMB anisotropies, with more severe biases (as high as $\sim 4\sigma$) resulting for less reliable versions of the effective fluid approximation. If all of the standard cosmological parameters are fixed by other measurements, the expected bias rises to $4-20\sigma$ (well beyond the validity of the Fisher approximation), though the required level of degeneracy breaking will not be achieved by any planned surveys. 

\end{abstract}

\date{\today}

\maketitle

\section{Introduction}

Measurements of cosmic microwave background (CMB) fluctuations \cite{Crites:2014prc,Louis:2016ahn,Akrami:2018vks,Aghanim:2018eyx}, the clustering/gravitational lensing of galaxies \cite{VanWaerbeke:2013eya,Alam:2016hwk,Abbott:2018xao,Abbott:2018wzc}, and the kinematics of cosmic acceleration (through Type Ia supernovae) \cite{Jones:2017udy,Abbott:2018wog} have ushered in the era of precision cosmology. Current data are consistent with the $\Lambda$CDM scenario, with cosmic density parameters of $\Omega_{b}h^{2}=(2.22\pm0.02)\times 10^{-3}$ for baryons, $\Omega_{c}h^{2}=0.120\pm 0.002$ for cold dark matter (CDM), and  $\Omega_{\Lambda}=0.685\pm0.007$ for dark energy.

The ``cold" in CDM refers to the fact that observations of cosmological large-scale structure require dark matter (DM) to be nonrelativistic when this structure forms. The standard model (SM) does not contain a DM candidate with this property and sufficiently weak couplings.

Many beyond the standard model candidates for DM have been proposed. The best motivated possibilities are weakly interacting massive particles (WIMPs) and axions \cite{Jungman:1995df,Feng:2010gw}. WIMPs (e.g.~neutralinos, gravitinos) arise in supersymmetric theories \cite{Jungman:1995df,Feng:2010gw} as well as some other scenarios, while axions provide a  solution to the strong $\mathcal{CP}$ (charge-parity) problem of quantum chromodynamics (QCD) \cite{Peccei:1977hh,Wilczek:1977pj,Weinberg:1977ma,Kim:1979if,Shifman:1979if,Zhitnitsky:1980tq,Dine:1981rt,Asztalos:2006kz,Kim:2008hd}. For particle masses $m_{\rm ax}\lesssim 10^{-2}~{\rm eV}$ (in units where $c=1$), axions would be produced nonthermally through oscillation of a scalar field, a distinct scenario from standard thermal production.

Direct detection experiments \cite{Aprile:2017iyp}, indirect detection efforts \cite{Fermi-LAT:2016uux,Hooper:2017gtd}, and Large-Hadron Collider searches for evidence of  supersymmetry \cite{Aad:2014mra} have all yielded increasingly stringent upper limits to WIMP properties \cite{Daylan:2014rsa}. The ample unexplored parameter space of QCD axion masses and couplings thus merits exploration, which is underway thanks to experimental efforts such as ADMX \cite{Du:2018uak}, IAXO \cite{Vogel:2013bta}, MADMAX \cite{TheMADMAXWorkingGroup:2016hpc}, CASPer \cite{Garcon:2017ixh}, and others \cite{Asztalos:2006kz}, as well as astrophysical tests using stellar cooling and other effects \cite{Raffelt:2006cw,Cadamuro:2012rm}. Most of these efforts probe values $m_{\rm ax}\gtrsim 10^{-14}~{\rm eV}$. 

Scalar fields of even lower masses are an interesting possibility \cite{Matos:1992qx,Frieman:1995pm,Matos:1999et,Amendola:2005ad}. Expectations for their standard-model couplings are model dependent, and so gravitational observables are a useful complement to detection efforts. These fields are referred to fuzzy dark matter (FDM) \cite{Hu:2000ke}, axion-like particles (ALPs), wave dark matter, or ultralight axions (ULAs). We use the latter nomenclature. ULAs could be astrophysically relevant on many scales, ranging from stellar-mass black holes to dwarf galaxies \cite{Grin:2019mub}.

ULAs would have unusual cosmological properties. ULAs maintain a constant energy density when $m_{\rm ax}/\hbar<H$ (where $H$ is the cosmic expansion rate) \cite{PhysRevD.28.1243,Frieman:1995pm}, and then redshift with the cosmic scale factor $a$ as $\rho_{\rm ax}\propto a^{-3}$ \cite{PhysRevD.28.1243}. If $m_{\rm ax}\lesssim 10^{-27}~{\rm eV}$, ULAs would only begin to dilute after matter-radiation equality, making them unsuitable as a dark-matter candidate. In this case, ULAs would contribute to the (early or late-time, depending on the mass) dark-energy (DE) density of the universe until they begin to dilute \cite{Coble:1996te,Hlozek:2014lca}. If $m_{\rm ax}\gtrsim 10^{-27}~{\rm eV}$, ULAs begin to dilute before equality, allowing them to cluster as dark matter \cite{Matos:1992qx,Frieman:1995pm,Matos:1999et,Hu:2000ke,Amendola:2005ad,Matos:2010zza,Hlozek:2014lca,Marsh:2015xka}. 

The existence of ULAs is predicted in string scenarios such as the ``axiverse" \cite{Arvanitaki:2009fg}, where they arise as Kaluza-Klein (KK) zero modes of antisymmetric tensors on compactified extra dimensions \cite{Witten:1984dg,Conlon:2006tq,Svrcek:2006yi}. In the axiverse, there is a broad mass spectrum of many axions, which may be important during different epochs (see Ref. \cite{Marsh:2015xka} and references therein). This motivates us to consider the observational consequences of a broad range of $m_{\rm ax}$ values.

The large de Broglie wavelengths of the ULA scalar field cause ULAs to exhibit suppressed gravitational clustering on galactic scales \cite{Hu:2000ke,Hwang:2009js,Hlozek:2014lca,Marsh:2015xka,Cembranos:2015oya,Cembranos:2018ulm,Poulin:2018dzj}. For masses around $m_{\rm ax}\sim 10^{-22}~{\rm eV}$, a large DM fraction in ULAs could address small-scale challenges to the $\Lambda$CDM scenario, such as cores in some galaxy density profiles, the paucity of Milky-way (MW) satellites, and the ``too big to fail" problem \cite{Marsh:2015xka}.  Other MW-scale dynamical probes also constrain ULAs \cite{Hui:2016ltb,Amorisco:2018dcn,Marsh:2018zyw}.

For masses $10^{-23}~{\rm eV}\lesssim m_{\rm ax}\lesssim 10^{-21}~{\rm eV}$, ULAs cause a suppression in neutral hydrogen (HI) density fluctuations at high redshift, suppressing the flux power spectrum of the Lyman-$\alpha$ (Ly$\alpha$) forest \cite{Armengaud:2017nkf,Kobayashi:2017jcf,Irsic:2017yje,Nori:2018pka,Leong:2018opi} in quasars. At lower masses still ($10^{-27}~{\rm eV}\lesssim m_{\rm ax}\lesssim 10^{-23}~{\rm eV})$, ULAs must be subdominant, but could still have a density that is $\sim~1\%-5\%$ of the DM density, comparable to the baryon and massive neutrino densities.

The dark sector may consist of numerous particle species (as the SM does) and in the axiverse scenario, the possibility of ULA dark matter that satisfies constraints in the range $m_{\rm ax}\sim 10^{-22}~{\rm eV}$ usually coincides with the existence of ULAs with $m_{\rm ax}\lesssim 10^{-23}~{\rm eV}$ \cite{Stott:2017hvl,Demirtas:2018akl}. 
In this lower mass range, limits of $\Omega_{\rm ax}h^{2}\leq 6\times 10^{-3}$ have been obtained using observations of CMB primary temperature/polarization anisotropies \cite{Hlozek:2014lca}.  Constraints of $\Omega_{\rm ax}h^{2}\leq 3 \times 10^{-3}$ have been obtained using CMB lensing potential reconstructions \cite{Hlozek:2017zzf}. 

ULAs could be a spectator field during inflation and source isocurvature perturbations, which are constrained by the CMB. The existence of ULAs could yield a new probe of the inflationary energy scale \cite{Marsh:2013taa,Marsh:2014qoa,Visinelli:2014twa,Hlozek:2017zzf,Visinelli:2017imh}, complementing constraints from B-mode polarization searches. All these probes (and future efforts) depend on reliable linear computations. Reliable simulations of nonlinear structure formation (relevant to MW scale observations) also require reliable linear initial conditions \cite{Schive:2014dra,Schive:2014hza,Zhang:2016uiy,Veltmaat:2016rxo,Martinez-Carrillo:2018ngl,Nori:2018hud}. 

Linear ULA perturbations are usually evolved using the effective fluid approximation (EFA). Stiff ordinary differential equations (ODEs) arise when ULAs are treated (exactly) as a classical scalar field (obeying the Klein-Gordon equation), due to disparate timescales ($m_{\rm ax }/\hbar \gg H_{0}$). The EFA is obtained by cycle-averaging out the fast timescale ($\sim m_{\rm ax}^{-1}$) to yield fluid equations with a time-dependent equation of state $w(a)$ (which interpolates between $w=-1$ at early times and $w=0$ when $m/\hbar\gg H$), and a scale-dependent sound speed $c_{s}^{2}$ \cite{Khlopov:1985jw,Nambu:1989kh,Hu:2000ke,Hwang:2009js,Marsh:2010wq,Suarez:2011yf,Park:2012ru,Hlozek:2014lca,Marsh:2015xka,Cembranos:2015oya,Suarez:2015fga,Fan:2016rda,Noh:2017sdj,Hlozek:2017zzf,Cembranos:2018ulm,Poulin:2018dzj}. 

This method is essentially the Wentzel, Kramers, Brillouin (WKB) approximation, and is implemented in the \textsc{AxionCAMB} \cite{Hlozek:2014lca,Hlozek:2017zzf} code used in a number of works. An alternative cycle-averaging formulation is proposed (and implemented) in Ref. \cite{Urena-Lopez:2015gur}, and further applied in Ref. \cite{Cedeno:2017sou}. We find that it is equivalent to the EFA (see Sec. \ref{sec:URvar}), an important conclusion of our work.

The next generation of CMB experiments, e.g.~CMB Stage-4 (CMB-S4)~\cite{Abazajian:2019eic} could yield cosmic-variance limited measurements of CMB polarization out to $\ell < 5000$ \cite{Abazajian:2016yjj}, with significant signal to noise coming from lensing in the nonlinear regime \cite{Nguyen:2017zqu}. Analysis of the Ly-$\alpha$ forest and nonlinear observables requires precise computations of the matter power spectrum in ULA models. 

It is timely to ask if the EFA is accurate enough: Can we trust observables predicted using the EFA? Is it sound to use a scale-dependent sound speed? How large is the bias induced by the EFA in ULA abundance measurements? In this work, we solve the exact ODEs for ULA DM and compare with results from the EFA in the range $m_{\rm ax}\gtrsim 10^{-27}~{\rm eV}$. 

Mode evolution near the era of recombination is significantly altered for some of the CMB-scale $k$ values. At the most constrained ULA masses ($m_{\rm ax}\sim 10^{-27}~{\rm eV})$, we find that CMB anisotropy power spectra between the two computational approaches vary significantly, for sufficiently high ULA mass fraction $r_{\rm ax}=\Omega_{\rm ax}/\Omega_{\rm DM}$, where $\Omega_{\rm DM}=\Omega_{\rm ax}+\Omega_{c}$ is the total DM density parameter. 

We examined three different EFA implementations, distinguished by the time at which exact equations are matched to EFA equations. For the fiducial switch at $m_{\rm ax}/\hbar=3H$, we evaluate the resulting bias in CMB-based determinations of $\Omega_{\rm ax}/\Omega_{\rm DM}$ for a cosmic-variance limited experiment (including cosmological parameter degeneracies). We find this bias to be $2 \sigma$ for $m_{\rm ax}= 3.2 \times 10^{-25}~{\rm eV}$ and $0.2 \sigma $ for $m_{\rm ax}= 10^{-24}~{\rm eV}$, with intermediate results elsewhere in the mass range $10^{-27}~{\rm eV}\lesssim m_{\rm ax}\lesssim 10^{-24}~{\rm eV}$. At \textit{Planck} noise levels, this implementation exhibits bias $\lesssim 1\sigma$ (if parameter degeneracies are included), validating the CMB-only constraints of Ref. \cite{Hlozek:2014lca}. The other two EFA implementations have larger biases (with $m_{\rm ax}$ dependence) by a factor of $\sim 2$. In the idealized case that external data perfectly break all CMB parameter degeneracies, larger biases of $4~-~20\sigma$ could result for the fiducial switch.

The organization of this paper is as follows. In Sec.~\ref{sec:ULAphys} we review the dynamics of cosmological ULAs, introduce the EFA, show the equivalence of an alternate cycle-averaging procedure to the EFA, and summarize the set of EFA implementations to be compared. In Sec. \ref{sec:boltz} we explain the details of our Boltzmann code. In Sec.~\ref{Sec:mode/CMB} we assess the impact of the EFA on perturbation evolution and CMB observables. In Sec.~\ref{Sec:zstatbias}, we estimate parameter bias induced by the use of the EFA. We conclude in
Sec.~\ref{sec:disc-conc}. A short derivation of the EFA is given in Appendix \ref{app:WKB}, while the numerical equivalence between the EFA and the alternate cycle-averaging approach is demonstrated in Appendix \ref{app:URnum}. The Z statistic (used to bound errors in CMB predictions) is discussed in Appendix \ref{app:Zstat_deriv}. Additional numerical results are shown in Appendix \ref{sec:exfigs_cmb}.

\section{ULA physics}
\label{sec:ULAphys}
We briefly summarize the relevant axion physics. For a more in-depth review, see Refs.~\cite{Marsh:2015xka,Hlozek:2014lca}. Depending on details of the production mechanism \& cosmological bounds, the QCD axion introduced to solve the strong $\mathcal{CP}$ problem must have a mass $m_{\rm ax} \gtrsim 10^{-12}$ eV \cite{Marsh:2015xka}. ULAs arise as KK zero modes from the compactified extra dimensions predicted by string theory \cite{Arvanitaki:2009fg}. The masses of these particles can be extremely small ($m_{\rm ax}\lesssim 10^{-18}~{\rm eV}$), motivating the term ``ultralight axions".

\subsection{ULA equations of motion}
\label{Sec:EOMs}
The equations of motion (EOMs) for ULAs are those of a scalar field in a perturbed Friedmann-Robertson-Walker metric, the homogeneous and perturbed Klein-Gordon (KG) equations. Using conformal time and synchronous gauge, these
equations take the form \cite{Hu:2004xd} ($\hbar = c=1$)
\begin{align}
\phi_0'' &= -2 \frac{a'}{a} \phi_0' - a^2 \tilde{m}^2 \phi_0,\label{eq:kg}\\
\phi_1 '' &= -2 \frac{a'}{a} \phi_1' - (k^2 + a^2 \tilde{m}^2) \phi_1 - \frac{h_L' \phi_0'}{2},\label{eq:kga}
\end{align}
where the full field $\phi(\vec{k},t)=\phi_{0}(t)+\phi_{1}(\vec{k},t)$ is expanded in terms of a background field $\phi(t)$ and a perturbation $\phi_{1}(\vec{k},t)$. Here $a$ is the usual cosmological scale factor, $k$ is the comoving wave number of a mode, and $h_L$ is one of the degrees of freedom of the metric perturbation \cite{Hu:2004xd}. The derivatives in these equations are with respect to conformal time, and denoted with the $'$ symbol. The axion mass (usually given in eV) is converted to cosmological units of ${\rm Mpc}^{-1}$ using the conversion $\tilde{m}=m_{\rm ax}/\hbar $, where $\hbar$ is given in ${\rm eV}{\rm s}$.

The stress-energy tensor components associated with the homogeneous field are
\begin{equation} \label{eq:axquants}
\begin{aligned}
\rho_{\rm ax} &= \frac12(\phi')^2a^{-2}+\frac{\tilde{m}^2}2 \phi^2, \\
p_{\rm ax} &=  \frac12(\phi_0')^2a^{-2}-\frac{\tilde{m}^2}2 \phi_0^2,\end{aligned}\end{equation} while those corresponding to perturbations are \begin{equation}\begin{aligned}
\delta \rho_{\rm ax} &= \phi_0' \phi_1' a^{-2} + \tilde{m}^2 \phi_0 \phi_1,\\
\delta p_{\rm ax} &= \phi_0' \phi_1' a^{-2} - \tilde{m}^2 \phi_0 \phi_1,\\
u_{\rm ax} &= (1+w_{\rm ax})v_{\rm ax} = k \frac{\phi_0' \phi_1}{\rho_{\rm ax} a^2}.
\end{aligned}
\end{equation}
The fluid variables are the homogeneous ULA density $\rho_{\rm ax}$, pressure $P_{\rm ax}$, density perturbation $\delta \rho_{\rm ax}$, and pressure perturbation $\delta p_{\rm ax}$. The equation of state parameter is
$w_{\rm ax} = p_{\rm ax}/\rho_{\rm ax}$ and $v_{\rm ax}$ is the scalar associated with the ULA velocity. Additionally, we have the usual Friedmann equation
\begin{equation}
    H^{2}=\frac{8 \pi G}{3}\sum_{i}\rho_{i},
\end{equation}where the sum is over all relevant particle species. Computationally, it is often helpful to use the conformal Hubble parameter $\mathcal{H}$, defined by $a^{\prime}=a\mathcal{H}$, and related to the standard Hubble parameter via $\mathcal{H}=aH$.

We have taken the potential to be harmonic $V = \frac12 m_{\rm ax}^2 \phi^2$, a reasonable approximation near the minimum of the periodic instanton-generated $V\propto [1-\cos{(\phi/f_{\rm ax})}]$ potentials typical of ULAs. As noted in Refs. \cite{Hlozek:2014lca,Hlozek:2017zzf}, most of the posterior parameter space consistent with CMB observations has $\phi\ll f_{\rm ax}$, and so this is a sensible approximation. At higher masses, anharmonic corrections could have important implications for predictions at Lyman-$\alpha$ forest scales or nonlinear structure formation \cite{Desjacques:2017fmf}. We will explore anharmonic potentials in future work.

As Eqs.~(\ref{eq:kg})-(\ref{eq:kga}) are just the homogeneous and perturbed Klein-Gordon equations in an expanding universe, we expect oscillations with frequency $\tilde{m}$. Evolving these equations of motion up until the present day from the early universe to the present day is thus numerically expensive if $\tilde{m}/H \gg 1$, due to the stiff differential equations which result. Cosmological parameter constraints and tests of novel models typically require repeated calls to a Boltzmann code, as a likelihood surface is explored using Monte Carlo Markov chain techniques (MCMC) or related methods. For constraints to be obtained, the Boltzmann code should have an execution time of $\lesssim 10~{\rm s}$. For ULA models, achieving this has required the use of the EFA, which essentially averages over the fast timescale of the oscillations.

\subsection{Effective fluid approximation}
\label{sec:efa}
\begin{figure}[h]
    \centering
     \includegraphics[scale=0.4]{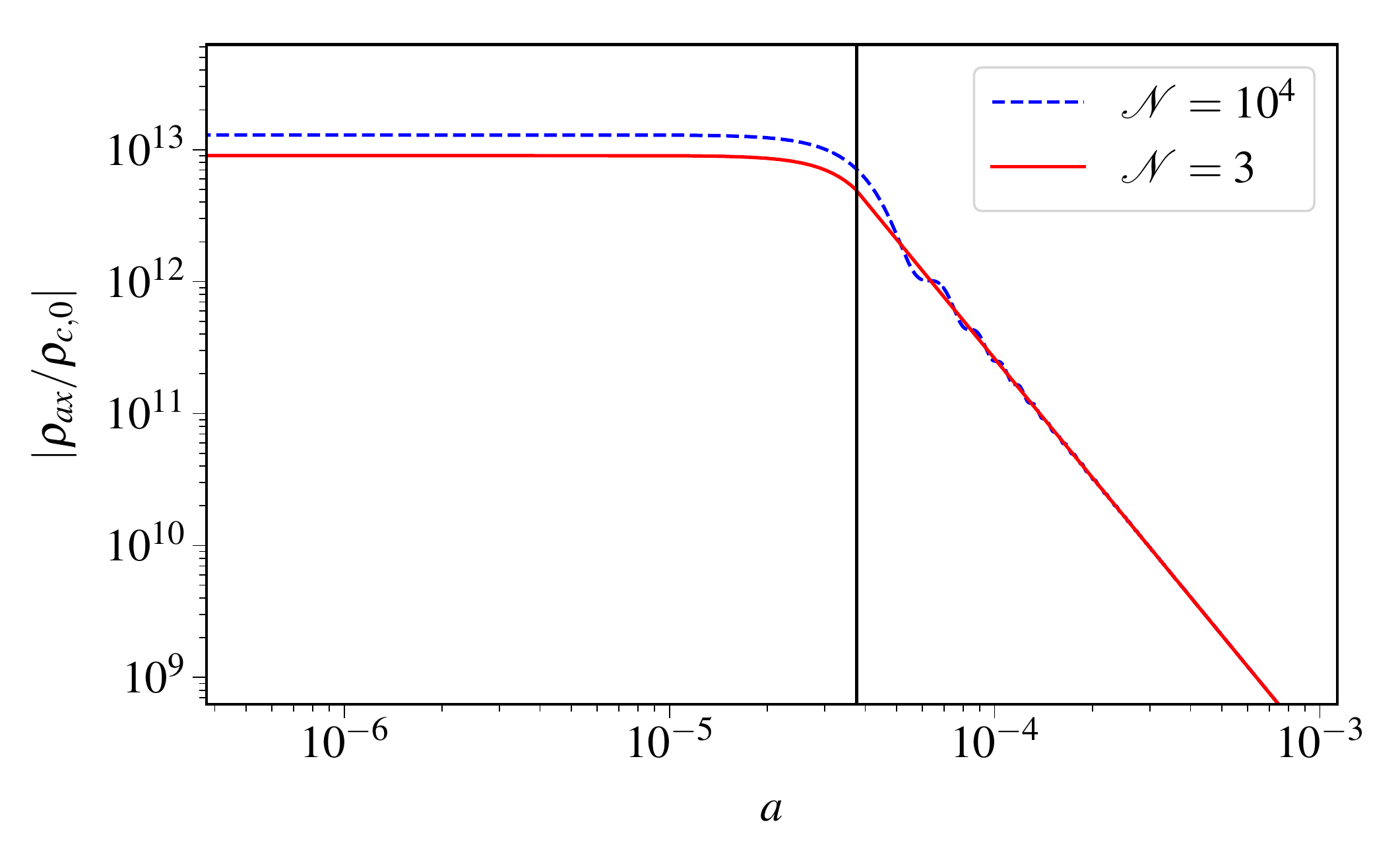}
    \caption{(Color online). Density evolution for ``exact" and effective fluid approximation. Here $\rho_{c,0}$ is the critical density today, and the ULA parameters are $m_{\rm ax}=3.16\times 10^{-25}$ eV and $\Omega_{\rm ax} = \Omega_\text{DM}$ (with initial $\phi_{0}$ values chosen accordingly). The black line indicates when $\tilde{m}/H = 3$, which is the beginning of the use of the effective fluid approximation. }
    \label{fig:axback}
\end{figure}

In Appendix~\ref{app:WKB}, we briefly review a derivation of the cycle-averaged ULA background \cite{PhysRevD.28.1243}, which is:
\begin{equation}
    \rho_{\rm ax} \approx \begin{cases} \rho_{\rm ax}(a=0) & \text{ if $H/\tilde{m}\gg1$}, \\
    \rho_{\rm ax}(a=a_\text{osc})\left(\frac{a_{\rm osc}}{a}\right)^3 & \text{ if $H/\tilde{m} \ll 1$} \end{cases},
\end{equation}
where $H(a_\text{osc})\approx \tilde{m} $ (see Fig.~\ref{fig:axback}).
If $H/\tilde{m} \gg 1$, then the field is overdamped and does not oscillate.

To state the EFA for perturbations, it is helpful to rewrite the ULA's exact equations of motion in terms of synchronous gauge fluid variables to obtain \cite{Hlozek:2014lca}
\begin{align}\label{eq:fluidax}
    \rho_{\rm ax}' &= -3 \frac{a'}{a} (1+w_{\rm ax}) \rho_{\rm ax}, \\
    \delta_{\rm ax}' &= - ku_{\rm ax} - (1+w_{\rm ax}) \frac{h_L}{2} -3 \frac{a'}{a}(c_s^2-w_{\rm ax})\delta_{\rm ax}, \label{eq:ncont}\\
    u_{\rm ax}' &= \frac{a'}{a} (3 w_{\rm ax} - 1) u_{\rm ax} + k c_s^2 \delta_{\rm ax}, 
    \label{eq:euler}\\
    c_{s}^{2}\delta_{{\rm ax}}&=\delta_{{\rm ax}}+3\frac{a^{\prime}}{a}\left(1-c_{\rm ad}^{2}\right)\frac{u_{{\rm ax}}}{k},\label{eq:ncs}
\end{align}
with $w_{\rm ax} = p_{\rm ax} / \rho_{\rm ax}$, $c_\text{ad}^2 =  p^{\prime}/\rho^{\prime}$, $\delta_{\rm ax} = \delta \rho_{\rm ax} / \rho_{\rm ax}$, and $c_s^2 = \delta p_{\rm ax} / \delta \rho_{\rm ax}$. Note that these equations with fluid variables $(\rho_{\rm ax}, w_{\rm ax}, \delta_{\rm ax}, u_{\rm ax})$ are still \textit{exact}, and equivalent to Eqs.~(\ref{eq:kg})-(\ref{eq:kga}) with field variables $(\phi_0, \phi_0', \phi_1, \phi_1'$). This formulation is the generalized dark matter (GDM) formulation of Ref. \cite{Hu:1998kj}.

The oscillations at frequency $\tilde{m}$ can be removed from Eq.~\eqref{eq:fluidax} by using the fact that the ULA fluid equations are well described by an effective fluid with cycle-averaged $\langle w_{\rm ax}\rangle$ and $\langle c_s^2\rangle $ which are not oscillating with the ULA field. We have already seen that when $H/\tilde{m}\ll 1$, $\rho_{\rm ax}' \approx -3 a' \rho_{\rm ax} / a$ implying that the cycle-averaged $\left \langle w_{\rm ax}\right \rangle = 0$. 

Applying the ansatz $\phi_1 = \phi_+ \cos (\tilde{m} t) + \phi_- \sin( \tilde{m}t) $, the authors of Ref.~\cite{Hwang:2009js} find the cycle-averaged sound speed to be
\begin{equation}\label{eq:cycavsound}
\left \langle c_s^2\right \rangle \equiv \frac{\left \langle \delta p_{\rm ax}\right \rangle}{\left \langle \delta \rho_{\rm ax}\right \rangle} =\frac{k^2/(4\tilde{m}^2 a^2)}{1+k^2/(4\tilde{m}^2 a^2)},
\end{equation} where $\langle\delta p_{\rm ax}\rangle$ and $\langle\delta \rho_{\rm ax}\rangle$ are the cycle-averaged ULA pressure and density fluctuations in the ULA's cycle-averaged rest frame.

Using more general assumptions about equipartition between kinetic and potential terms (and without assumptions about the functional form of scalar-field oscillations), the same expression for the cycle average of $c_{s}^{2}$ is derived in Ref.~\cite{Poulin:2018dzj}, and generalized to potentials of the form $V\propto(1-\cos{\phi/f_{a}})^{n}$ near their minimum, where $V\propto \phi^{2n}$ and the field rapidly oscillates. Scalar fields described by such potentials are one way of resolving the tension between CMB data and more local measurements of $H_{0}$, by way of an era of early dark-energy (EDE) dominance \cite{Poulin:2018cxd,Agrawal:2019lmo,Lin:2019qug,Smith:2019ihp}. 

For $n\geq 2$, the hierarchy of timescales between scalar-field oscillation and the Hubble parameter is far less extreme, and the perturbed Klein-Gordon equation may be solved exactly in the course of an MCMC simulation \cite{Poulin:2018dzj,Agrawal:2019lmo}. The relative impact of anharmonic terms and a fluid approximation on observables and constraints in EDE models is the subject of ongoing discussion \cite{Poulin:2018cxd,Agrawal:2019lmo,Lin:2019qug,Smith:2019ihp}.  

Substituting $\langle w_{\rm ax} \rangle$ and $\langle c_s^2 \rangle$ into Eqs.~(\ref{eq:ncont})-(\ref{eq:euler}), we arrive at a set of effective (cycle averaged) equations of motion for the ULA fluid variables that are valid when $H/\tilde{m} \ll 1$. These define the EFA:\footnote{The expressions in Refs.~\cite{Hlozek:2014lca,Poulin:2018dzj} have additional terms in both perturbed fluid equations due to the transformation from the comoving gauge where $\langle c_{s}^{2}\rangle$ is derived to the synchronous gauge usually used in CMB calculations. It was verified there (and confirmed here) that these terms are negligible compared to those shown and do not affect any of the conclusions of this paper.}
\begin{align}\label{eq:fluidaxlate}
    \rho_{\rm ax}' &= -3 \frac{a'}{a} \rho_{\rm ax}, \\
    \delta_{\rm ax}' &= - ku_{\rm ax} - \frac{h_L}{2} -3 \frac{a'}{a}\left \langle c_s^2\right \rangle \delta_{\rm ax}, \label{eq:ncontlate}\\
    u_{\rm ax}' &= -\frac{a'}{a} u_{\rm ax} + k \left \langle c_s^2\right \rangle  \delta_{\rm ax}.
    \label{eq:eulerlate}\end{align}
The term proportional to $\langle c_{s}^{2}\rangle$ in Eq.~(\ref{eq:eulerlate}) is the linear-theory expansion of the quantum pressure term discussed in the fuzzy dark matter and ULA literature. 

The EFA is only valid once the field (or perturbation modes) begins to coherently oscillate, and so the exact field equations must be solved at early times, with a switch implemented from exact to cycle-averaged equations when $\tilde{m}/H\equiv\mathcal{N}$, where $\mathcal{N}$ is a constant that defines a specific EFA implementation. The ULA fluid variables $\delta_{\rm ax}$, $u_{\rm ax}$, and $\rho_{\rm ax}$ are matched to ensure continuity at the switch. 

\subsection{Alternative cycle-averaging procedure}
\label{sec:URvar}
In Ref.~\cite{Urena-Lopez:2015gur}, a different set of variables is used for mode evolution. Dimensionless angular coordinates $\theta$, $\vartheta$ are defined in lieu of $\phi_{0}$ and $\phi_{1}$, along with their difference $\tilde \theta = \theta-\vartheta$. Coherent oscillation (at times for which $\tilde{m}\gtrsim H$) is explicitly built into the formalism through terms like $\sin(\theta)$ and $\sin(\vartheta)$. At early times, the resulting equations are equivalent to the exact field equations [Eqs.~(\ref{eq:kg}) and (\ref{eq:kga})], and thus to our early-time GDM equations. At late times, these equations are stiff and an exact integration is still numerically intractable for use in fast cosmological parameter-space exploration. Some cycle-averaging procedure is thus still needed there. 

To that end, an alternative cycle-averaging procedure is introduced in Ref. \cite{Urena-Lopez:2015gur}. Oscillatory functions $f(x)$ ($x=\theta$, $\vartheta$, or $\tilde \theta$, depending on the equation) are replaced with 
\begin{equation}
    f_*(x) = \frac{1}{2}\left[1 - \tanh(x^2-x_*^2)\right] f(x),
\end{equation}
for $x\leq x_{*}$  (where $x_*=100$). When $x>x_{*}$, $f(x)$ is numerically set to zero. For example, the replacement \begin{widetext}
\begin{equation}
\cos(x)\to \left\{\begin{array}{ll}
\frac{1}{2}\left[1 - \tanh(x^2-x_*^2)\right] \cos(x)&\mbox{if $x\leq x_{*}$},\\
0&\mbox{if $x>x_{*}$}\end{array},\right\}\end{equation} \end{widetext} is made. 

Put simply, a smoothed switch is used to separate the fast ($m^{-1}$) and slow ($H$) timescales for ULA evolution. The chosen smoothing width is very narrow ($\delta t/t\simeq 5 \times 10^{-5}$), and thus qualitatively is likely to still exhibit whatever undesirable transient behavior occurs in the EFA discussed above. 

Since $\vartheta \approx \theta \approx 2 m_{\rm ax} t$, at times $t\gg 50/m_{\rm ax}$, the equations evolved in Ref.~\cite{Urena-Lopez:2015gur} are effectively
\begin{align}
 \frac{d\rho_{\rm ax}}{d \ln a} &= -3 \rho_{\rm ax}, \\
\frac{d \tilde \theta}{d \ln a} &= - \frac{k^2}{k_J^2} + e^{-\alpha} \frac{d h_L}{d \ln a} \cos \frac{\tilde \theta}{2},\label{eq:ulodd} \\
\frac{d \alpha}{d \ln a} &= \frac12 e^{-\alpha}\frac{d h_L}{d \ln a} \sin \frac{\tilde \theta}{2} ,
\label{eq:ulode}
\end{align}
where [noting that $(1+w_{\rm ax}) v_{\rm ax} = k u_{\rm ax}$]
\begin{equation}
\begin{aligned}
    \delta_{\rm ax} &= -e^\alpha \sin \frac{\tilde \theta}{2}; \\
    k u_{\rm ax} &= -\frac{k^2}{2 a m_{\rm ax}} e^\alpha \cos \frac{\tilde \theta}{2}. 
\end{aligned}\label{eq:uldef}
\end{equation}
Note the natural appearance of the ULA Jeans wave number $k_J$.

Differentiating Eq.~\eqref{eq:uldef} using Eqs.~(\ref{eq:ulodd}) and \eqref{eq:ulode}, we can represent the equations of motion for the variables $\alpha$ and $\tilde{\theta}$ in terms of more standard fluid variables, yielding
\begin{align}
    \rho_{\rm ax}' &= - 3 \frac{a'}{a} \rho_{\rm ax} \label{eq:fluidaxul},\\
    \delta_{\rm ax}' &= - k u_{\rm ax} -\frac{1}{2} h_L'\,\label{eq:ncontul}\\
    k u_{\rm ax}' &= -\frac{a'}{a} k u_{\rm ax} +\frac{k^4}{4 \tilde{m}^2 a^2} \delta_{\rm ax}.\label{eq:eulerul}
\end{align}
Equation~(\ref{eq:fluidaxul}) is just Eq~(\ref{eq:fluidaxlate}) in the $\langle w\rangle=0$ limit valid at late times. Equation~(\ref{eq:ncontul}) is Eq.~(\ref{eq:ncontlate}) if the term $\propto \langle c_{s}^{2} \rangle $ on the right-hand side of Eq.~(\ref{eq:ncontlate}) is dropped. Finally, Eq.~(\ref{eq:eulerul}) is Eq.~(\ref{eq:eulerlate}) in the nonrelativistic limit of Eq.~\eqref{eq:cycavsound}
that $\langle c_s^2 \rangle \simeq  k^2/(4 \tilde{m}^2 a^2)$. We verify in Appendix \ref{app:URnum} that these differences are numerically irrelevant for CMB observables, and so the method of Ref. \cite{Urena-Lopez:2015gur} is equivalent to the EFA.

\subsection{EFA implementation summary}
\label{sec:efa_details}
The choice \bench was used to obtain constraints on ULAs in Refs. \cite{Hlozek:2014lca,Hlozek:2017zzf}. In this work, we investigate the accuracy of the $\mathcal{N}=3$ EFA implementation as well as that of two other approaches. In one, the switch is imposed when $\phi_0(a) = 7\phi_0(a=0)/8$ (as in Ref. \cite{Poulin:2018dzj}), which is also when $\mathcal N\simeq 1.6$. In the other EFA implementation (see Sec. \ref{sec:URvar} and Ref. \cite{Urena-Lopez:2015gur}), the switch is imposed when $2\tilde{m}t=100$. This is equivalent (if the transition occurs during radiation domination) to $\mathcal{N}=100$, and so we use that notation for the remainder of this work.

For the ``exact" case, we solve the full KG field evolution [Eqs.~\eqref{eq:kg} and \eqref{eq:kga}] until $\tilde{m}/H = 10^4$ and then switch to an (extremely) late-time EFA. This is done as the logarithmic conformal time step needed to resolve rapid oscillations becomes prohibitively small for efficient computation at late times. For all but the largest mass considered ($m_{\rm ax} = 10^{-24}$ eV), the switch to the EFA occurs after recombination. The predicted CMB anisotropies thus accurately reflect the impact of exact scalar-field dynamics, though we expand on this issue further in Sec. \ref{Sec:zstatbias}.  

We summarize the various implementations in Table~\ref{tab:approx}. For the $m_{\rm ax}$ range under consideration here, we note that computing a full set of CMB power spectra for the ``exact" case is an order of magnitude more computationally expensive than the $\mathcal{N}=3$, $\mathcal{N}=1.6$ or $\mathcal{N}=100$ implementations.

\begin{table}[h]
\centering
\begin{tabular}{ c| c |c |c }
Name & $\tilde{m}/H\approx$ & Actual condition & Ref.\\ 
\hline
\hline
``Exact" or $\mathcal{N}=10^{4}$ & $10^4$ &$\tilde{m}/H = 10^4$  & \\  
 \bench & 3 & $\tilde{m}/H = 3$ & \cite{Hlozek:2014lca}\\
 \smith & 1.6 & $\phi_0(a)=7 \phi_0(a=0)/8$ & \cite{Poulin:2018dzj} \\
 \urena & 100 & $2 \tilde{m} t = 100$ & \cite{Urena-Lopez:2015gur}
\end{tabular}
\caption{The second column indicates the approximate value of $\tilde{m}/H$ (or other relevant criterion) when the switch to cycle-averaged equations is implemented.}
\label{tab:approx}
\end{table}

\section{Boltzmann code}\label{sec:boltz}
In order to investigate the effect of the EFA, we developed a \textsc{Python} Boltzmann code to calculate individual mode growth, CMB anisotropy power spectra, and the matter power spectrum in the presence of ULAs. We use our own Boltzmann code (as opposed to modifying existing codes such as \texttt{CLASS}~\cite{Blas:2011rf} or \textsc{CAMB}~\cite{lewis:2000}). We do this to clearly isolate the effect of different treatments of ULA perturbations (treated with sufficient time resolution) without requiring extensive modification to these more complete Boltzmann codes to resolve such time cales. Some preliminary comparisons were made in Refs. \cite{Hlozek:2014lca, Poulin:2018dzj}, but without assessing implications for parameter inference.

In our code, the other components of the universe (dark matter, baryons, radiation, neutrinos, dark energy) are evolved using the equations from Ref.~\cite{Ma:1995ey}, and we compute CMB power spectra using the line-of-sight method \cite{Seljak:1996is}. We keep seven photon modes and $12$ neutrino modes, and we truncate the hierarchy of equations as in Ref.~\cite{Ma:1995ey}. We use the initial conditions for a universe with ULAs as derived in Refs. \cite{Hlozek:2014lca,Hlozek:2017zzf} (using methods also applied in Ref. \cite{Doran:2003xq,Shaw:2009nf,Chluba:2013dna}). 

We ported a public C++ implementation of the \textsc{RECFAST} code \cite{Seager:1999km,Seager:1999bc,Chluba:2010ca,seager:2011} into \textsc{Python}, in order to compute cosmic recombination histories with reasonable precision. It is known that a variety of additional physical effects (beyond those included in the multi-evel atom computation underlying \textsc{RECFAST}) affects recombination at the $0.1\%$ level (e.g.~higher-$n$ two-photon transitions, resonant scattering in the Lyman-$\alpha$, deviations from statistical equilibrium between angular momentum substates of high-$n$ Rydberg states). These and other recombination effects are discussed in Refs. \cite{Chluba:2007qk,Chluba:2007yp,Hirata:2008ny,Chluba:2008vn,Grin:2009ik,Chluba:2009gv,Hirata:2009qy,AliHaimoud:2010ab,AliHaimoud:2010ym,AliHaimoud:2010dx,2010PhRvD..81h3004K,2011MNRAS.417.2417K} (and references therein).

We implemented the numerical methods used in Ref.~\cite{Callin:2006qx}, and have neglected neutrino mass. Current CMB data are consistent with this choice \cite{Gil-Botella:2015qaa} (though neutrino experiments indicate a nonzero neutrino mass whose absolute scale could be detected by future CMB experiments \cite{Abazajian:2016yjj}), allowing us to avoid the complications of following perturbation evolution  in a species that transitions from relativistic to nonrelativistic on cosmological timescales \cite{Lewis:2002nc,Shaw:2009nf,Blas:2011rf,Lesgourgues:2011rh}. We neglect smoothing of primary CMB anisotropies by weak gravitational lensing. We include homogeneous reionization as in Ref. \cite{Lewis:2013hha}. 

ULAs are added self-consistently as described in Sec.~\ref{Sec:EOMs}. We use the \textsc{SciPy} \cite{2019arXiv190710121V} ODE solver with \textsc{vode} \cite{vode} integrator using a \textsc{bdf} (stiff) method. Rather than stepping forward using derivatives computed at present-day function values, this implicit technique solves nonlinear equations for future function values, using expressions which approximate the relevant derivatives in terms of future and present function values. Implicit techniques are well suited to stiff differential equations -- the system under study here has two vastly disparate timescales [$\hbar/(m_{\rm ax}c^{2})$ and $c/H$], making it well suited to the chosen numerical method. We set a large ($10^{6}$) maximum number of steps. The most important parameters are the relative and absolute error tolerance parameters, which are set to values of $10^{-14}$ and $10^{-9}$, respectively.

When plotting spectra, we set all the parameters except the ULA mass $m_{\rm ax}$ and ULA dark matter fraction $r_{\rm ax}=\Omega_{\rm ax}/\Omega_\text{DM}$ to their best fit values from the 2015 \textit{Planck} data release  \cite{Ade:2015xua}.\footnote{While this work was nearing completion, the \textit{Planck} 2018 results were released \cite{Aghanim:2018eyx}. Because of the very small shifts in central values of parameters between the 2015 and 2018 \cite{Aghanim:2018eyx} data releases, and very weak dependence of bias results on fiducial parameter values over the allowed range, our conclusions should be unaffected by the difference between the two iterations of Planck results.} In Sec. \ref{Sec:zstatbias}, we explore variations in all cosmological parameters when assessing the impact of the EFA on ULA parameter inferences. 

Note that cold dark matter makes up a fraction of the energy density $\Omega_\text{CDM} = \Omega_\text{DM} - \Omega_{\rm ax}$.  We consider only ULA masses $m_{\rm ax}\gtrsim 10^{-27}$ eV. This mass restriction guarantees that ULAs begin to dilute as matter ($\rho_{\rm ax}\propto a^{-3}$) before matter-radiation equality (``equality" henceforth) \cite{Hlozek:2014lca}, keeping us in the ``DM-like" part of ULA parameter space. We will explore the impact of the EFA in the ``DE-like" part of the mass range ($m_{\rm ax}\lesssim 10^{-27}~{\rm eV}$) in future work. As initial conditions for the homogeneous axion field $\phi$, we use $\phi_0'=0$ and $\phi_0=C$ \cite{Marsh:2015xka}. 

We use a root-finding algorithm interfaced with our homogeneous scalar-field evolution module to find the value of $C$ which reproduces the desired value of $\Omega_{\rm ax}$. The initial guess is chosen to be $\phi_0=\sqrt{\Omega_{\rm ax} \rho_{\rm c}} /(m_{\rm ax}a_{\rm s}^{3/2})$, where $a_{s}$ is defined as the scale factor when $H/m_{\rm ax} = 3$, with $H$ computed using $\Lambda$CDM to obtain an initial guess but self-consistently including ULAs thenceforth. The algorithm assumes the existence of a root in the range $\phi_0/5 \to 5\phi_0$, which we have confirmed is sufficient for all values of the parameters explored in this work. For the perturbations, adiabatic initial conditions give $\phi_1=0$ \cite{Marsh:2015xka,Hlozek:2014lca}, and, because of large Hubble drag initially, $\phi_1'=0$ \cite{Poulin:2018dzj}. 

Our CMB and matter power spectrum in the absence of ULAs were compared against \textsc{CAMB}~\cite{lewis:2000} with quantitative agreement at the $5\%$ level, except for $\ell<10$, where agreement is at the $\sim 10\%$ level for the EE angular power spectrum $C_{\ell}^{\rm EE}$ and $\lesssim 5\%$ for the TT power spectrum $C_{\ell}^{\rm TT}$.  Although there is a difference between the code used in this work and \textsc{CAMB}, it is unimportant because our goal here is a comparison between the ``exact" treatment and the EFA with all other assumptions and computational tools held fixed for self-consistency. Our use of a simplified, relatively accurate recombination history (and simplifying approximations for neutrinos, as described above) should be sufficient for this goal.

\section{Mode growth and CMB}\label{Sec:mode/CMB}
We now explore the impact of the various EFA implementations on cosmological observables. Naively, one might expect that later switches (by following the exact EOMs for longer) would always perform better, but this is not the case, as we explain below.
\subsection{Mode evolution}
\begin{figure*}[ht]
    \centering
    \includegraphics[scale=0.4]{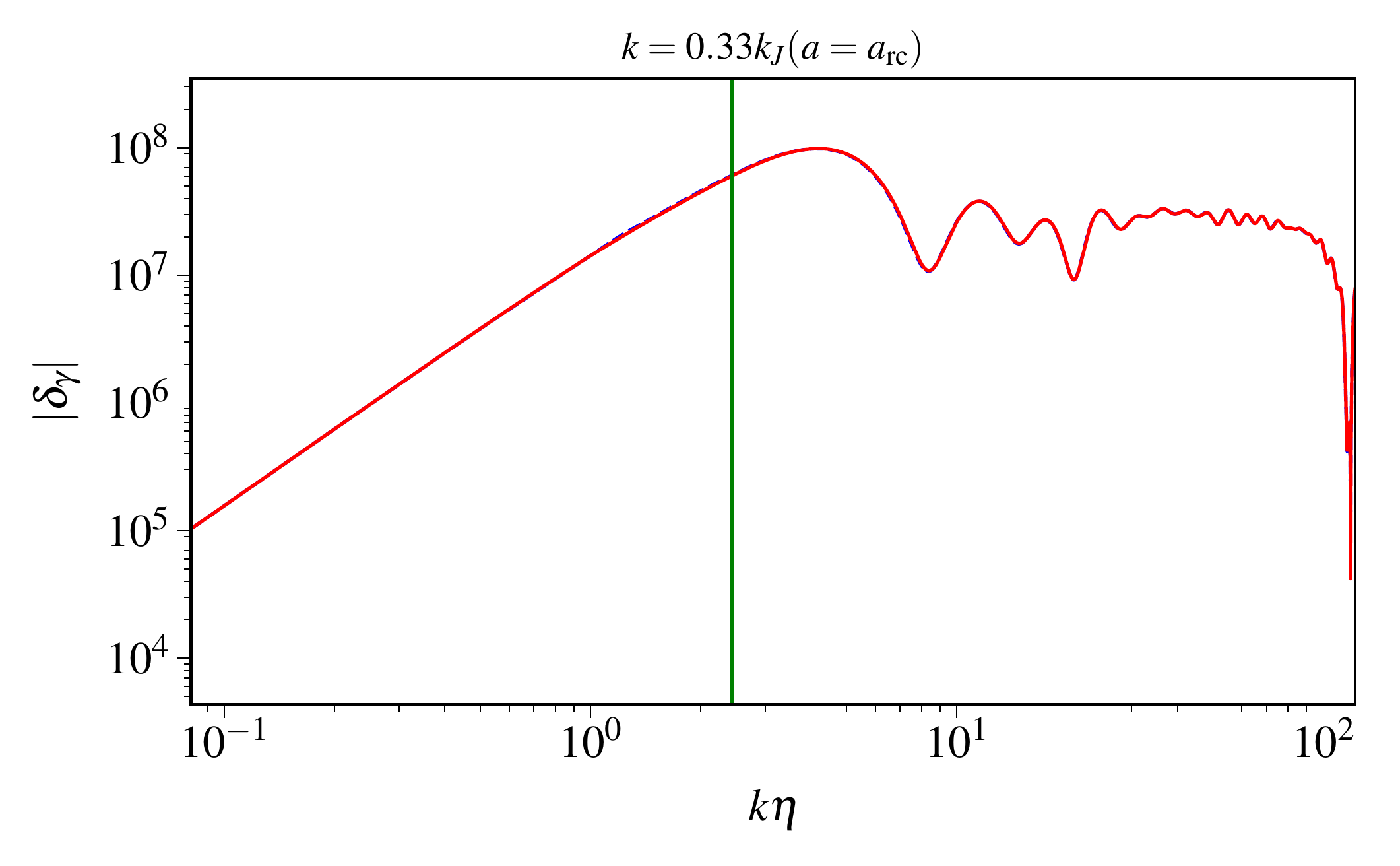}
    \includegraphics[scale=0.4]{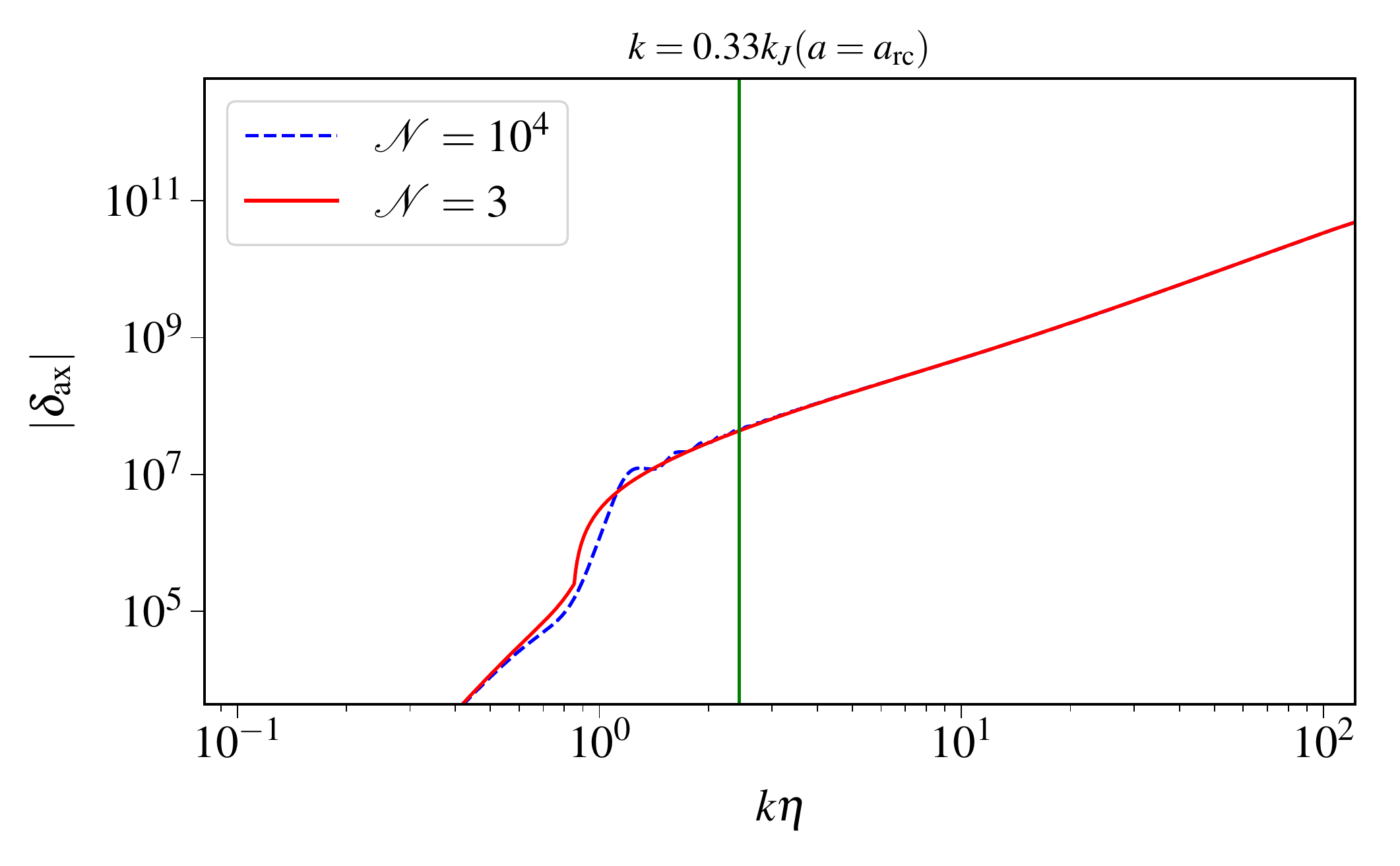}
    
    \includegraphics[scale=0.4]{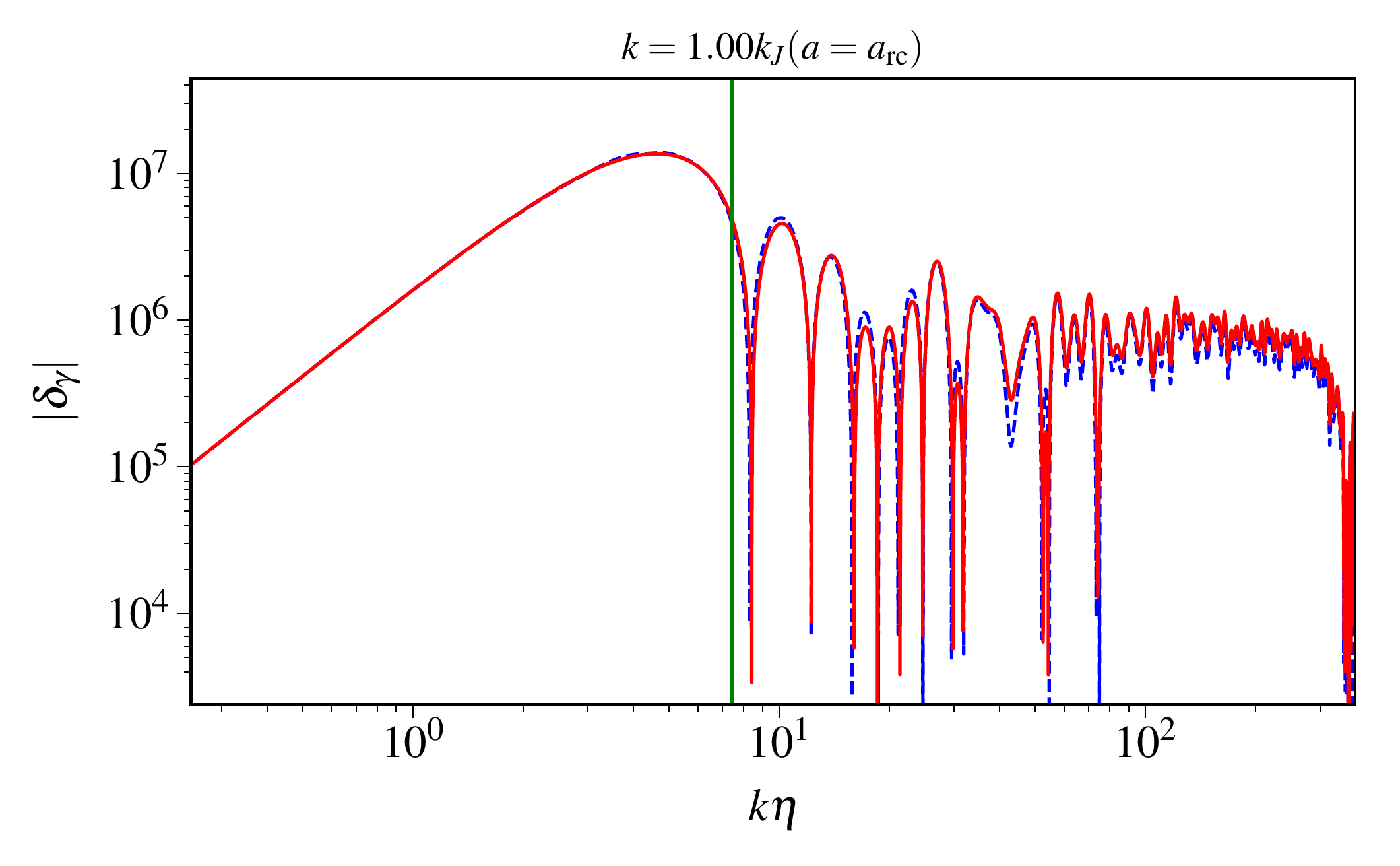}
    \includegraphics[scale=0.4]{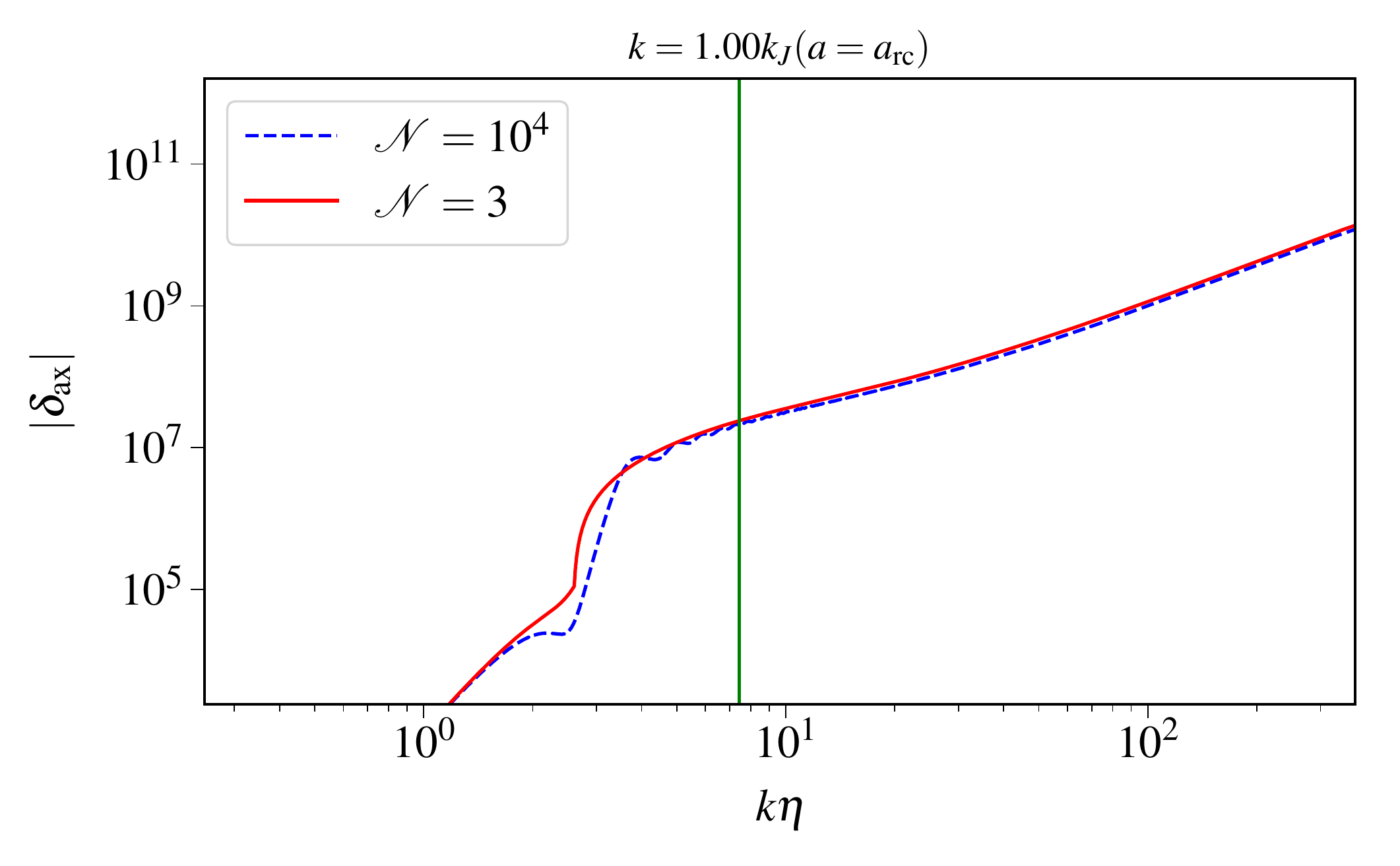}
    
    \includegraphics[scale=0.4]{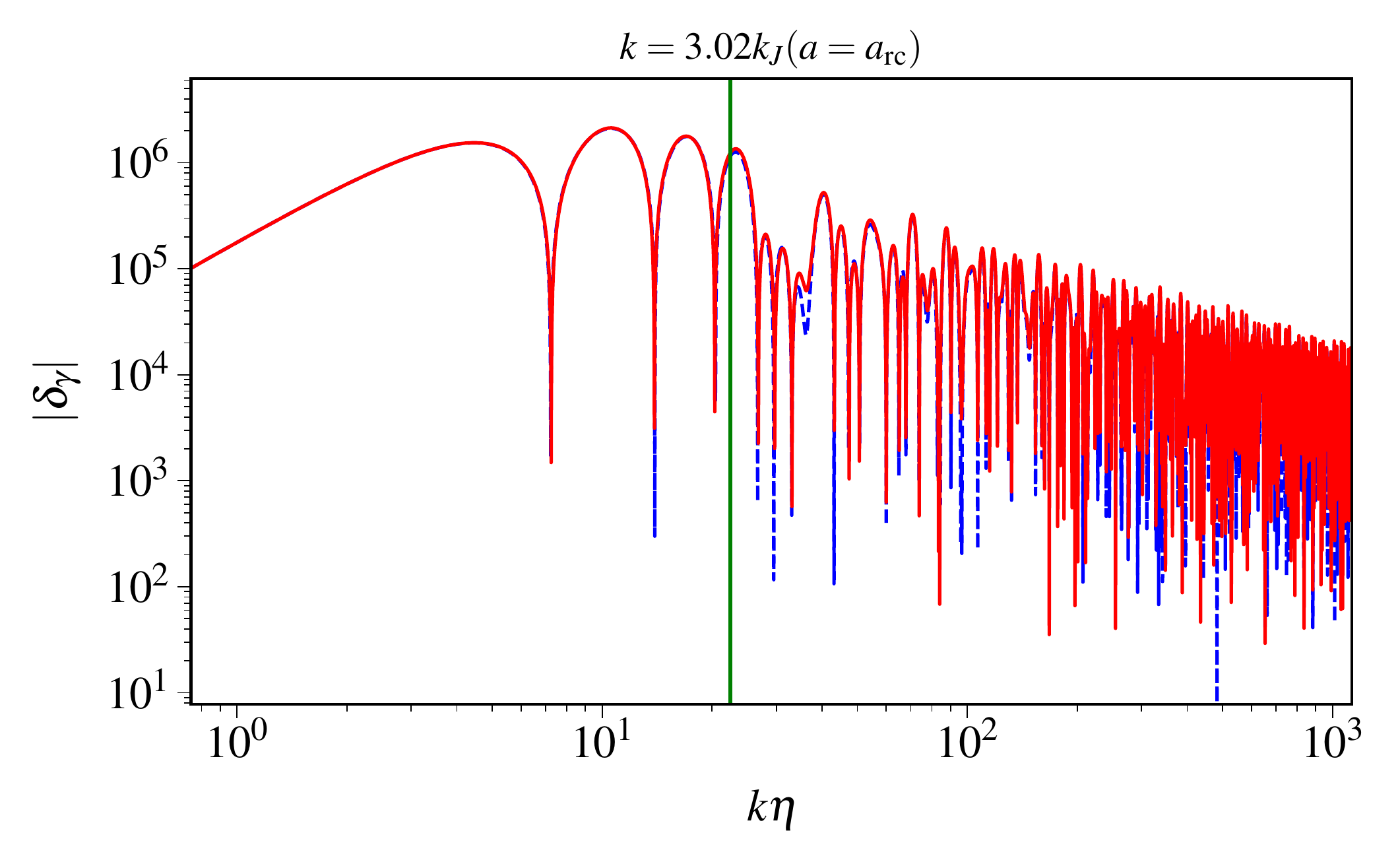}
    \includegraphics[scale=0.4]{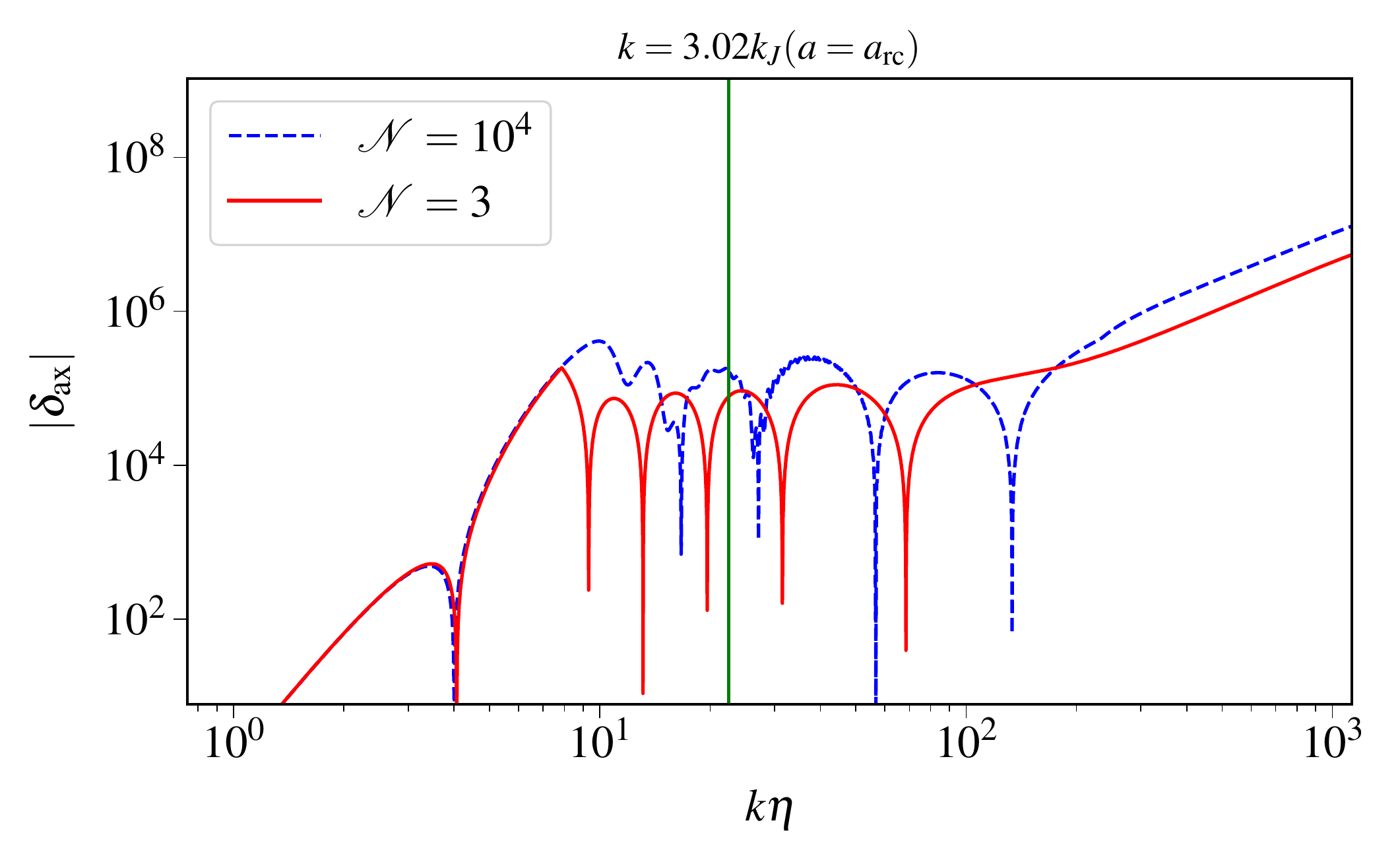}
    \caption{(Color online). Perturbation mode evolution for three different wave numbers in ULA models, using the \bench approximation and exact Klein-Gordon equation. The ULA parameters are $m_{\rm ax}=10^{-27}$ eV, and $\Omega_{\rm ax}/\Omega_{\rm DM}=1.0$. The absolute value of the photon overdensity $\delta_\gamma$ and ULA overdensity $\delta_{\rm ax}$ are plotted against $k\eta$, where $\eta$ is conformal time. The green vertical line indicates the time of recombination, with $a_\text{rc}$ being the scale factor at recombination. For $k \ll k_J$, the approximation does very well for ULA mode evolution ($\delta_{\rm ax}$), but as $k_J$ is approached and exceeded, the difference between the two cases at recombination and asymptotically becomes significant and then grows with $k$. The $y$-axis scale is arbitrary.}
    \label{fig:axmodes}
\end{figure*}
\begin{figure*}[ht]
\centering
    \includegraphics[scale=0.4]{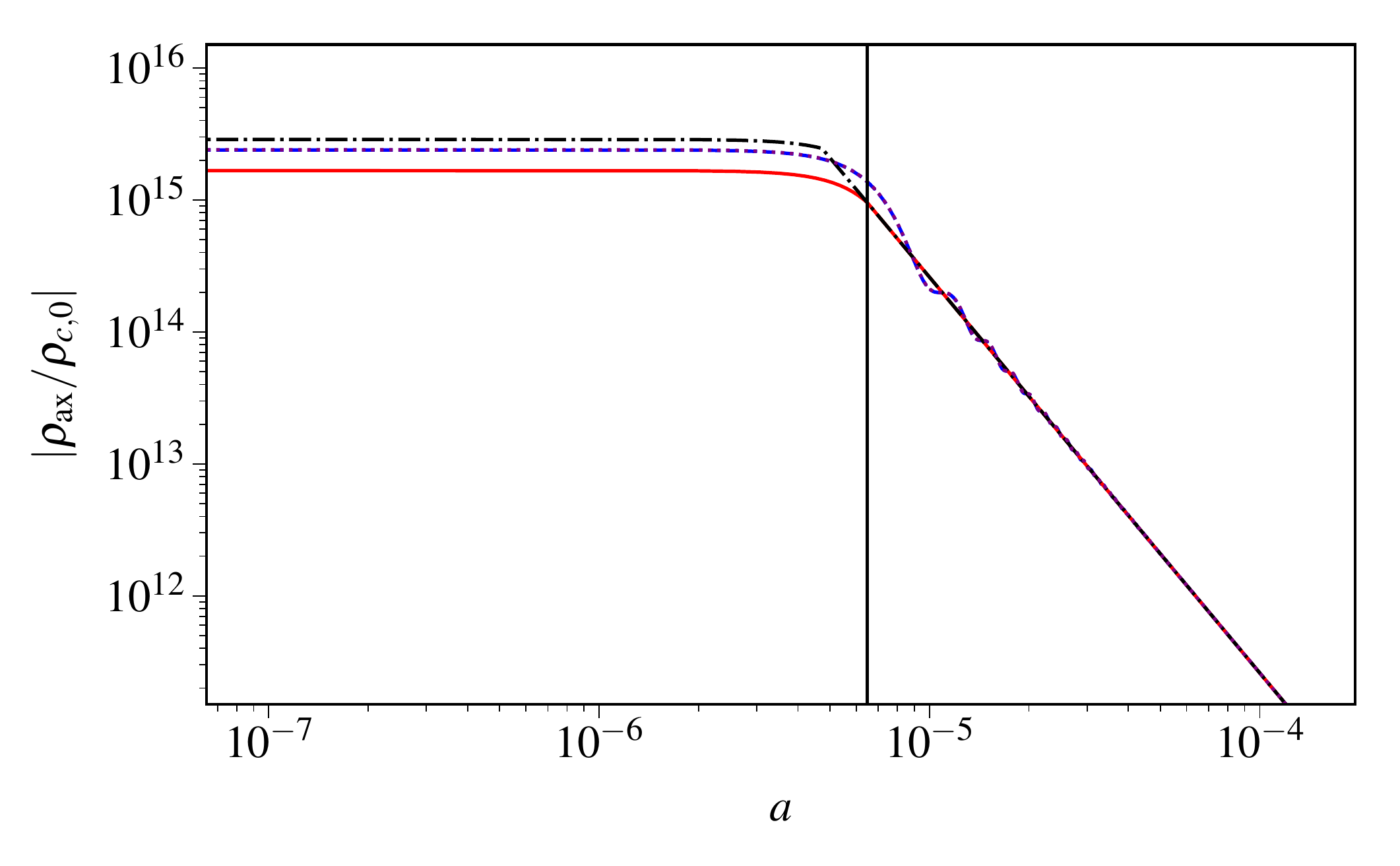}
    \includegraphics[scale=0.4]{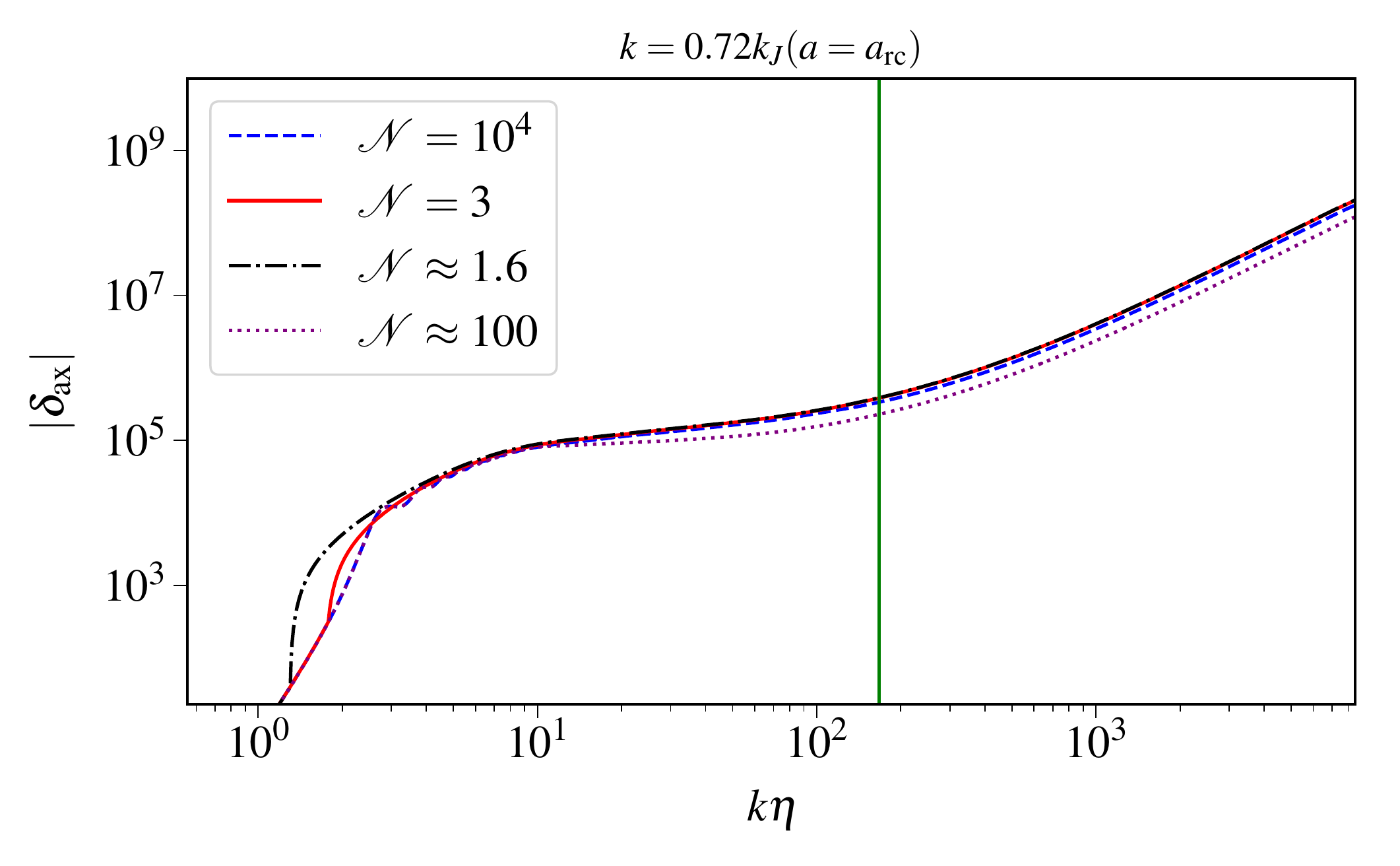}  
    \includegraphics[scale=0.4]{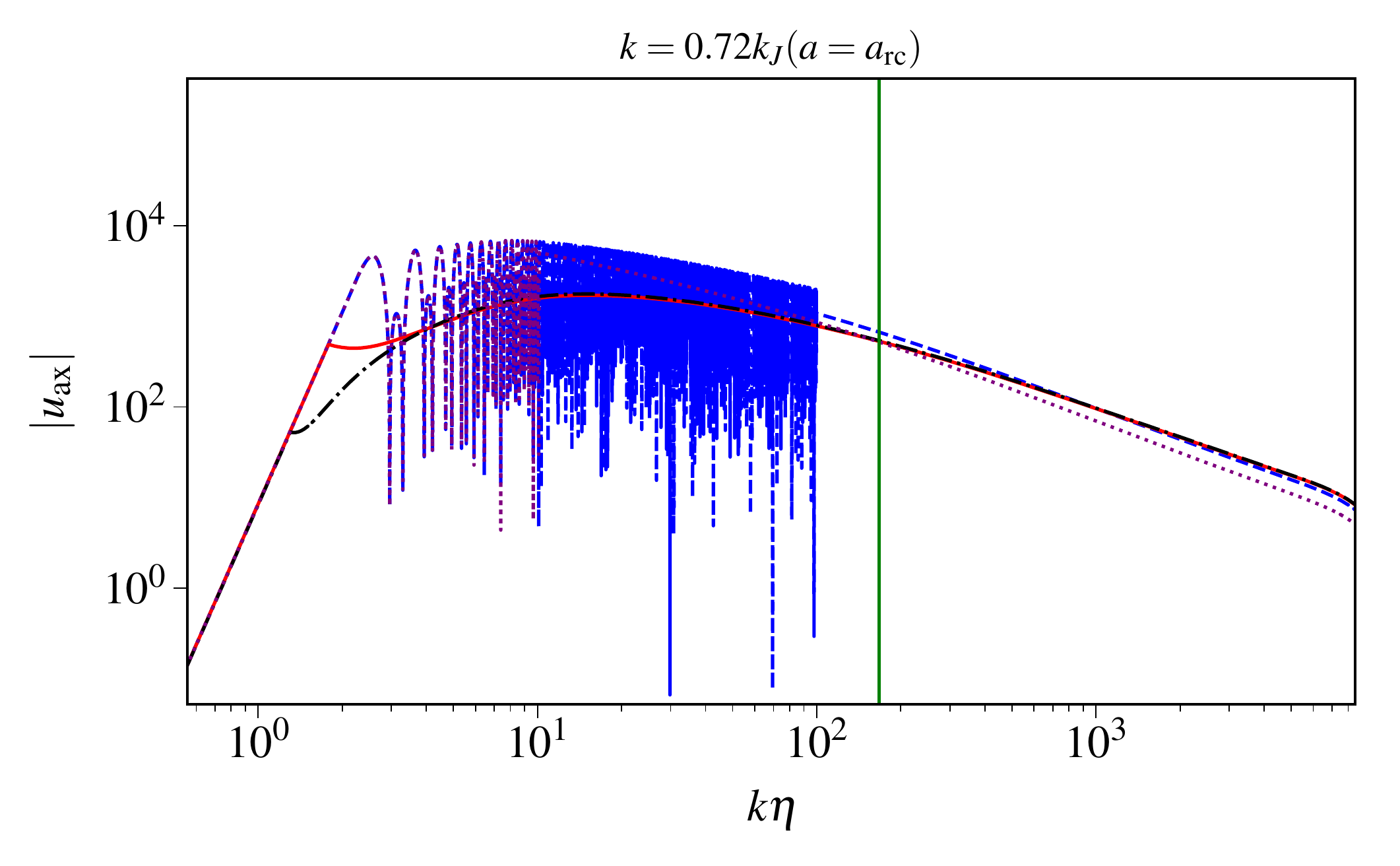}
    \includegraphics[scale=0.4]{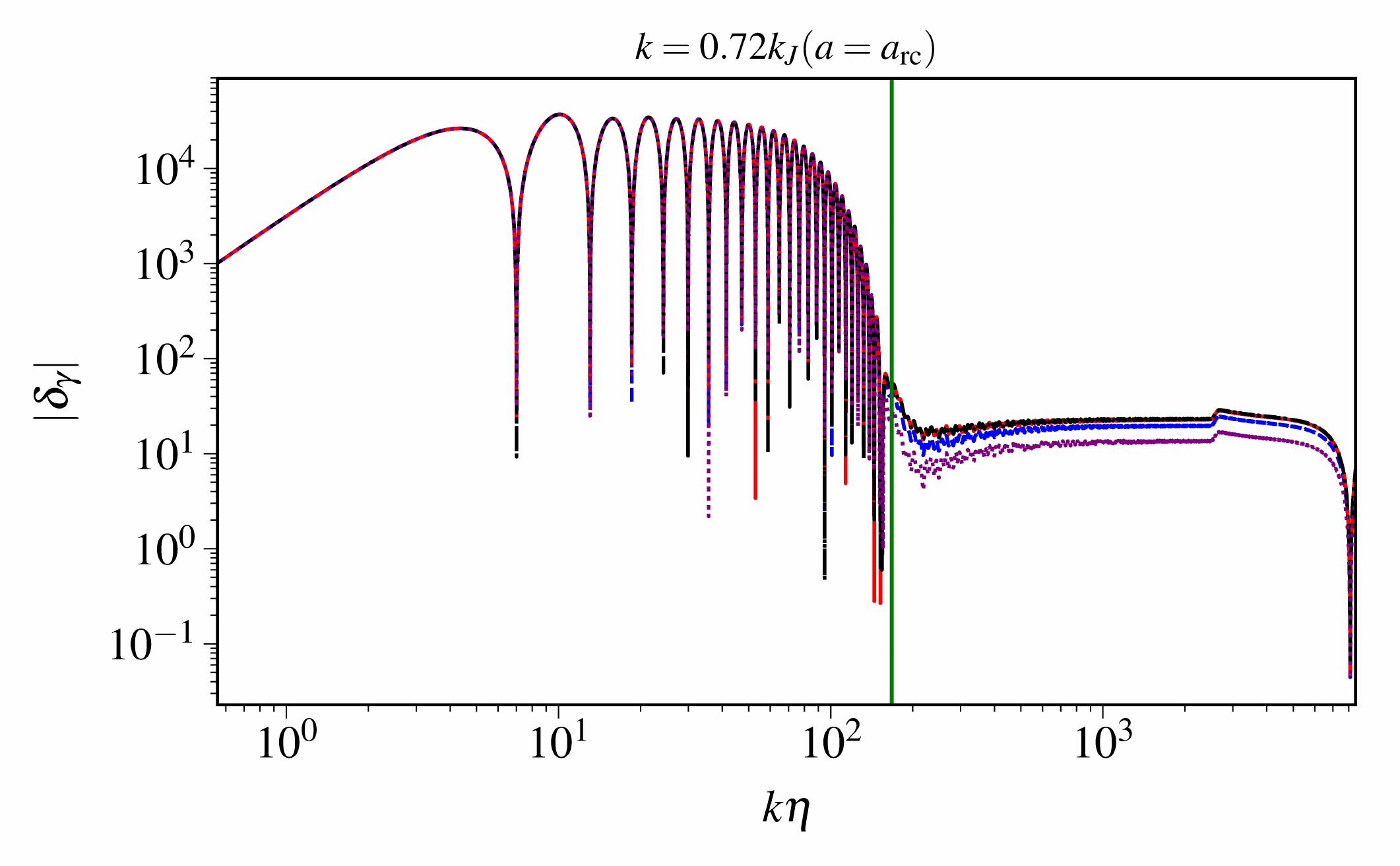}
      \caption{(Color online). Evolution of homogeneous ULA density $\rho_{\rm ax}$, as well as perturbation mode evolution of ULA overdensity $\delta_{\rm ax}$, ULA momentum $u_{\rm ax}$,  and photon overdensity $\delta_{\gamma}$. A single mode is shown with $k\approx 3k_{\rm J}(a=a_{\rm rc})/4$, where $k_{\rm J}(a)$ is the ULA Jeans scale at recombination. We compare results obtained using the exact Klein-Gordon equation with the \bench and \urena EFA implementations. The ULA parameters are $m_{\rm ax}=10^{-24}$ eV and $\Omega_{\rm ax}/\Omega_{\rm DM}=1.0$. The absolute value of $\delta_\gamma$, ULA overdensity $\delta_{\rm ax}$, and $u_{\rm ax}$ are plotted against $k\eta$, where $\eta$ is conformal time. The green line indicates the time of recombination, with $a_\text{rc}$ being the scale factor at recombination. The $y$-axis scale is arbitrary.}
    \label{fig:axulmodes}
\end{figure*}
\label{sec:moev}
We first study the effect of the EFA on the growth of individual modes. To understand general features, we show individual mode growth for different values of $k$ in Fig.~\ref{fig:axmodes}. We show the evolution of $\delta_{\rm a}$ given our obvious interest in its dynamics and of $\delta_{\gamma}$, as it is the fluid variable most important for observed properties of the CMB. A useful reference scale is given by the comoving ULA Jeans wave number
\begin{equation}
    k_J(a) = a \sqrt{\tilde{m} H}.\label{eq:jdef}
\end{equation}
Modes with $k>k_J$ have suppressed growth relative to $\Lambda$CDM, as the perturbation lies within the ULAs wavelength \cite{Nambu:1989kh}. We expect that $k_J(a=a_\text{rc})$ (where $a_\text{rc}$ is the scale factor at recombination) will be the important scale for computations of CMB anisotropies. We find $a_\text{rc}$ using the peak of the visibility function $g= \kappa'e^{\kappa}$ with $\kappa = a n_e \sigma_T$ with $n_e$ being the density of electrons and $\sigma_T$ the Thomson cross section.

For $k\lesssim k_J(a=a_\text{rc})$, the approximation and the ``exact" treatment agree asymptotically, so there is only a small change in the photon overdensity, $\delta_\gamma$. Once $k \gtrsim k_J$, there are noticeable differences between the two approaches, as seen both in the ULA overdensity $\delta_{\rm ax}$ and in $\delta_\gamma$, which could have an observable impact on CMB anisotropy measurements. Although we show mode evolution for $m_{\rm ax} = 10^{-27}$ eV, the qualitative features hold for other values.

We compare all three EFA implementations with the ``exact" calculation in Fig. \ref{fig:axulmodes}, for the value $m_{\rm ax}=10^{-24}~{\rm eV}$. We see that the \urena implementation captures early-time homogeneous density evolution more accurately than the \bench implementation, as well as early-time perturbation evolution (at least for the mode shown). This is unsurprising given that this implementation follows the exact scalar-field EOMs for a much larger number of oscillation cycles. The \bench implementation captures the later-time evolution of both ULA and photon variables more accurately, as there is more time for numerical transients (introduced by a discontinuous swap in the second derivatives of various fluid quantities) to dissipate. Given the differences between perturbation evolution in different implementations, a more detailed comparison of CMB power spectra is necessary.

\subsection{CMB anisotropies}
We now compute CMB anisotropies for our full grid of models using the standard line-of-sight formalism as implemented in Sec.~\ref{sec:boltz}.
The ordering of the accuracy of different EFA implementations depends on the mass. The WKB approximation underlying the EFA is itself an expansion in the small (time-dependent) parameter $\epsilon_{\rm WKB}\equiv H/m_{\rm ax}\sim 1/\mathcal{N}$, which might be taken to imply that the \urena implementation should always outperform the \bench implementation, but this is not the case. Why?

All of these implementations excite spurious transients near the time of transition from the exact KG to EFA equations. The higher-$\mathcal{N}$ cases also have later switches, leaving less time for transient behavior to dissipate before recombination. There is thus a delicate balance at play (between suppression of transients and formal convergence of the WKB approximation) in determining which version of the EFA is best. This balance depends on when the switch occurs compared to the evolution of the observed CMB fluctuation modes, and is thus sensitive to both $\mathcal{N}$ and $m_{\rm ax}$.

\begin{figure*}
    \includegraphics[scale=0.40]{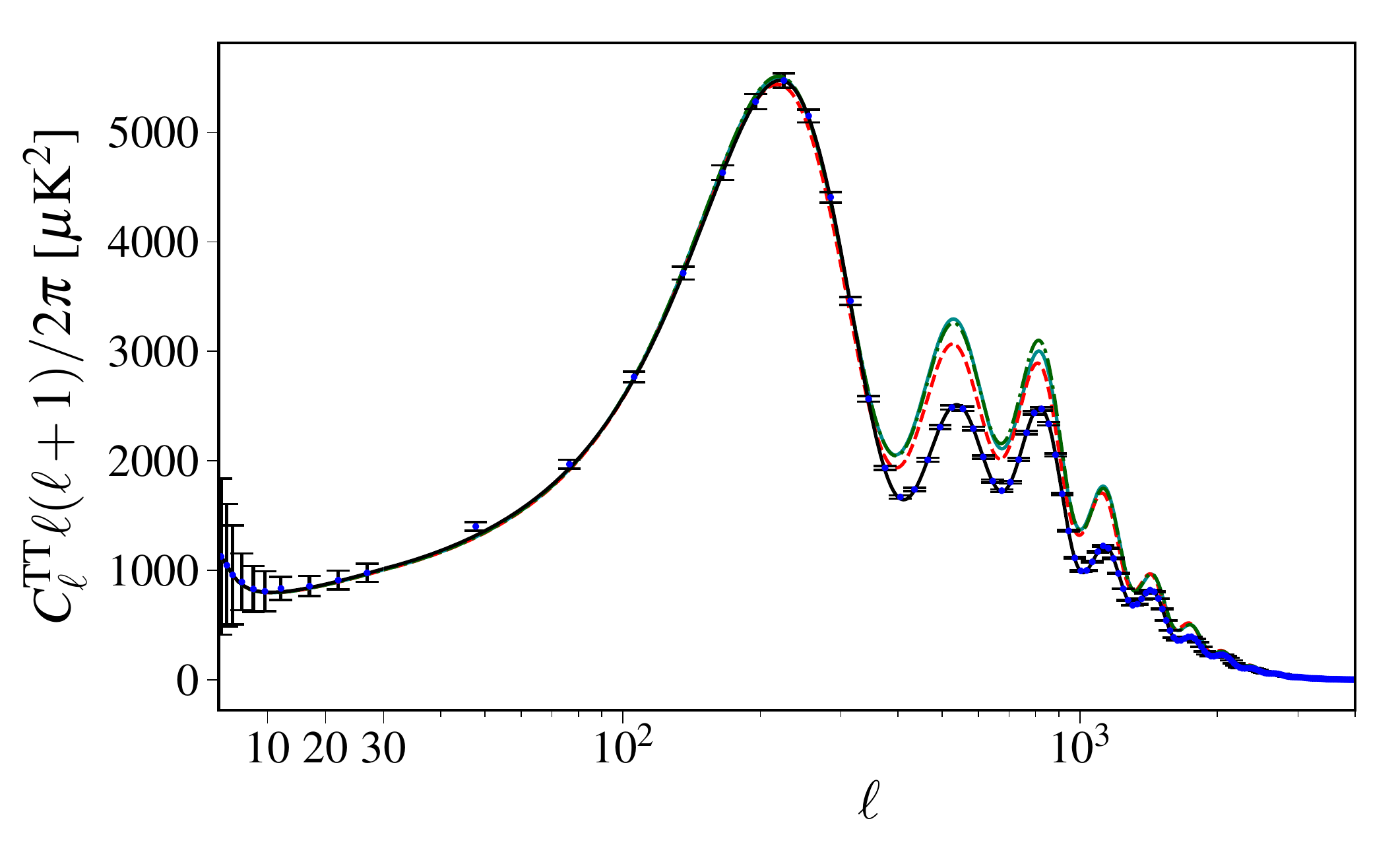}
    \includegraphics[scale=0.4025]{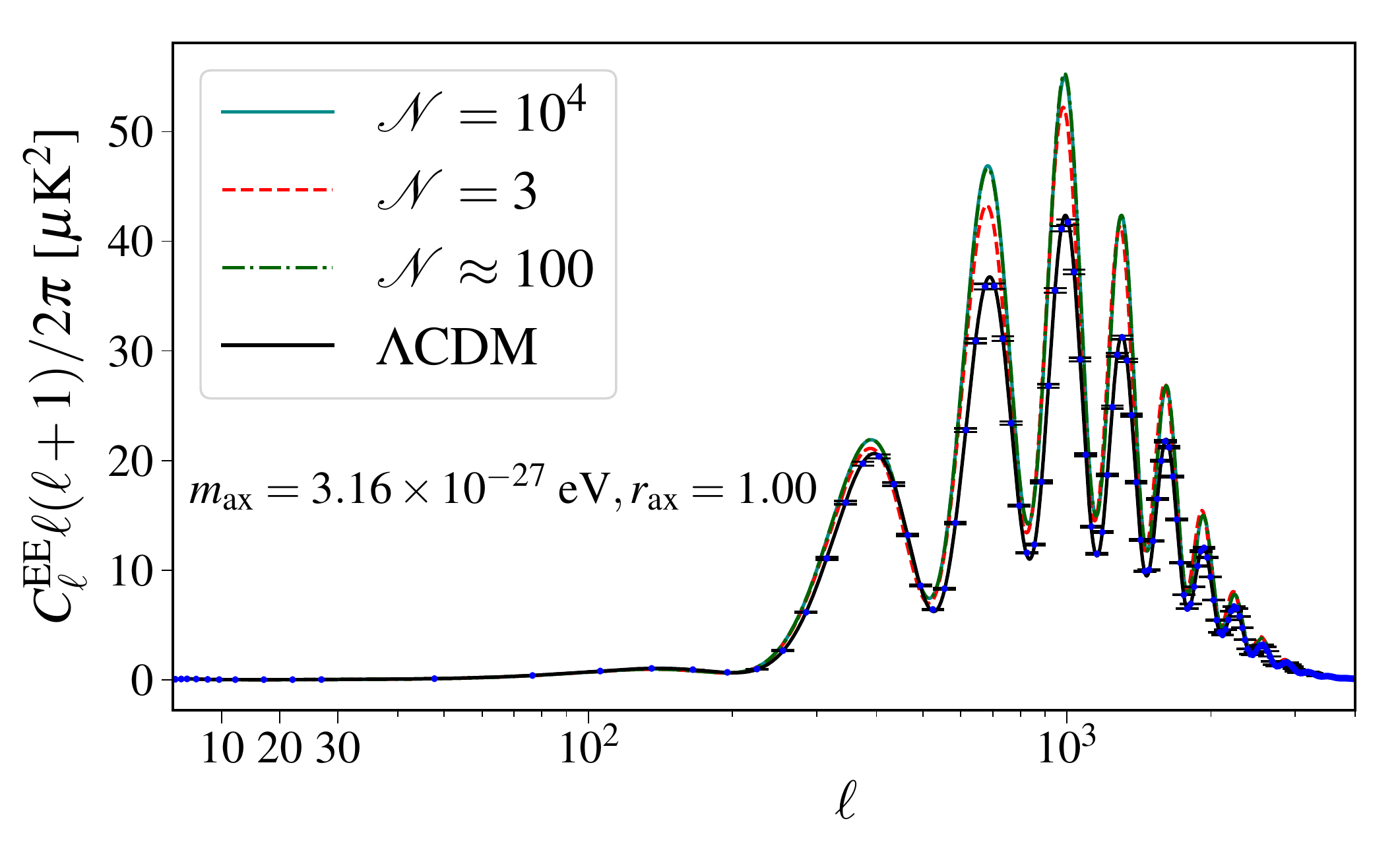}\\
        \includegraphics[scale=0.3825]{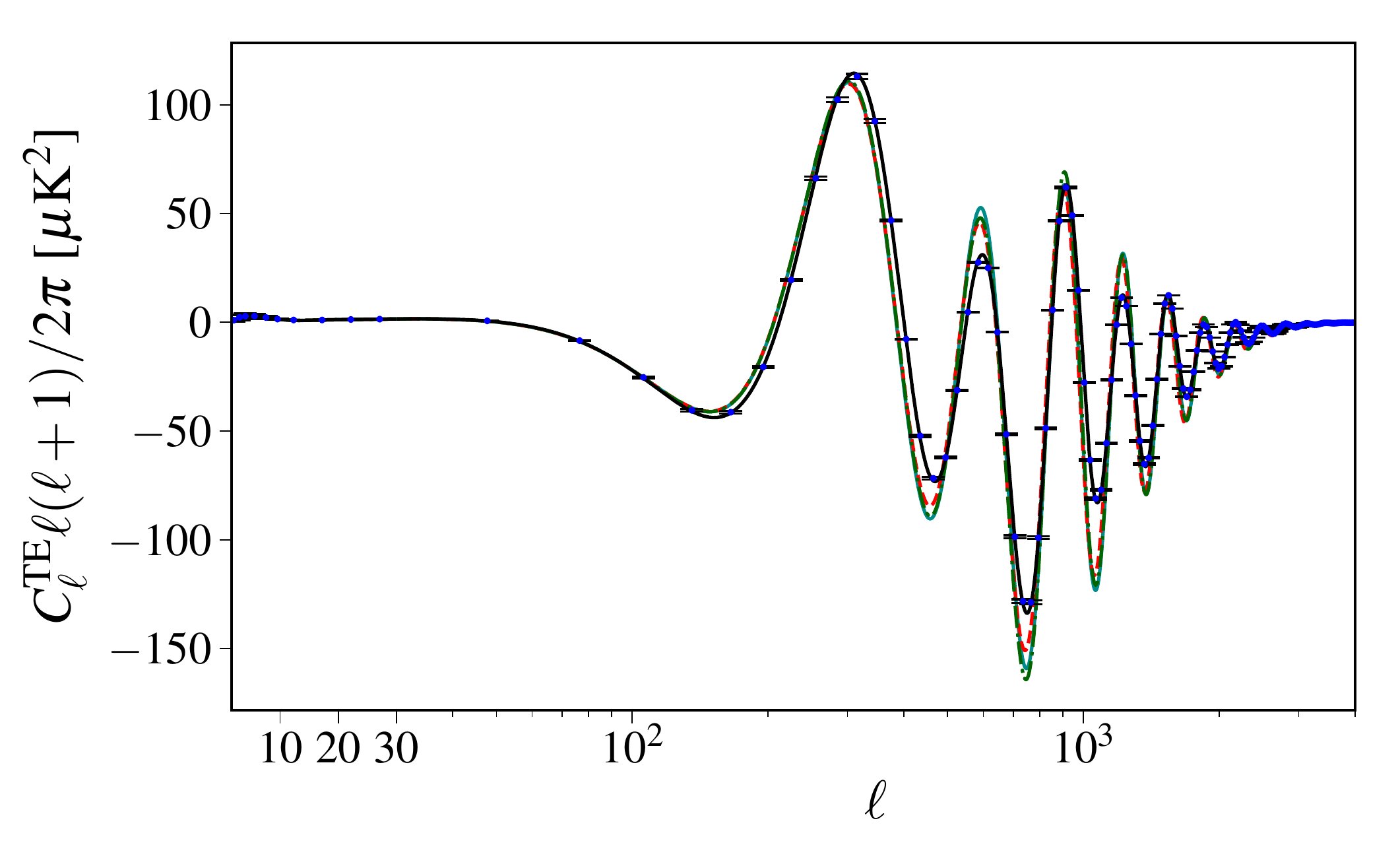}
          \includegraphics[scale=0.3825]{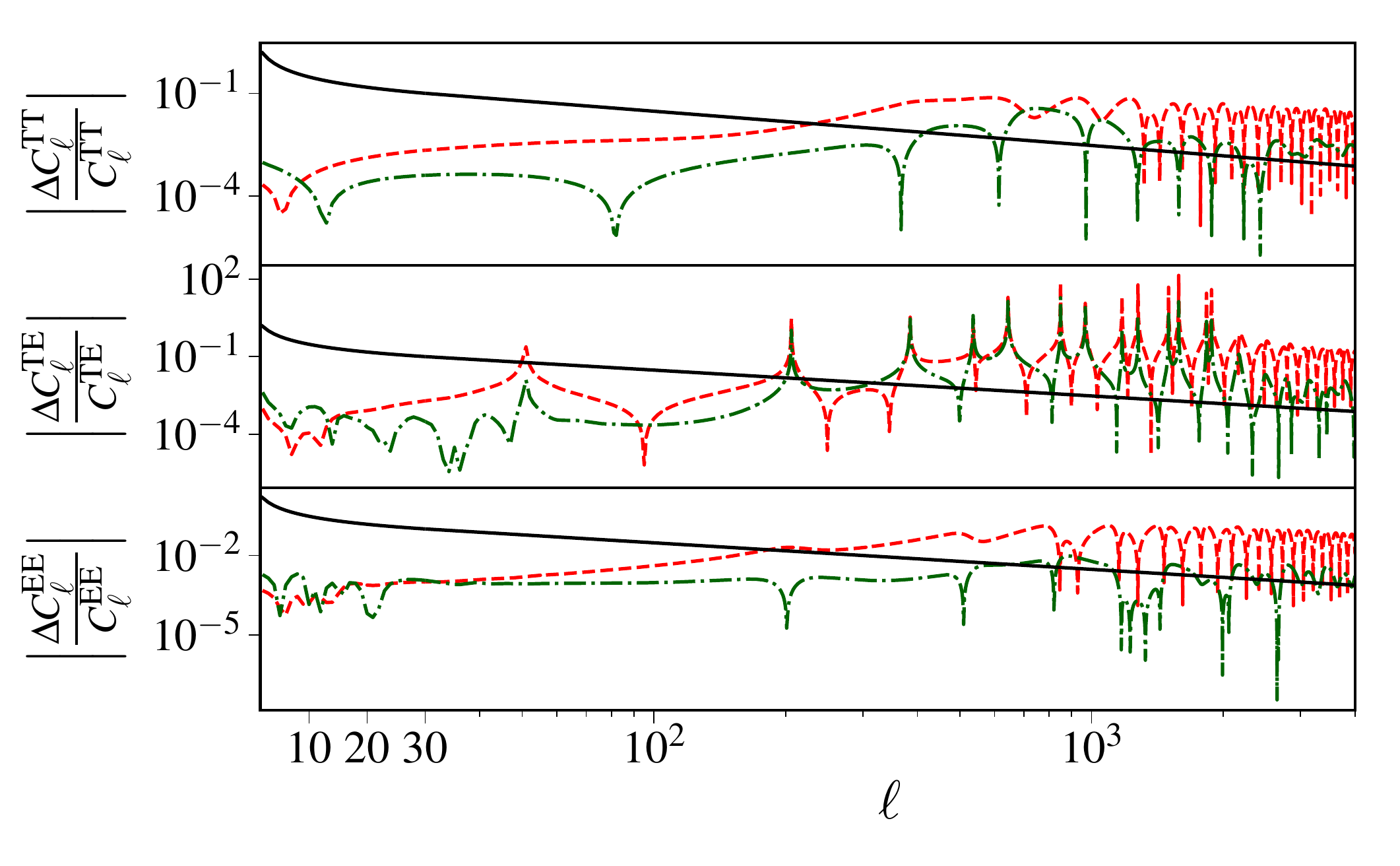}
    \caption{(Color online). We compare CMB power spectra with no axions ($\Lambda$CDM) to those with ULAs of the specified mass and fraction $r_{\rm ax}=\Omega_{\rm ax}/\Omega_{\rm DM}$ where the EFA is used when $m_{\rm ax}/\hbar > \mathcal{N}H$ (in units where $c=1$). The ``exact" implementation sets $\mathcal{N}=10^4$. The values of $r_{\rm ax}$ and $m_{\rm ax}$ shown here are ruled out by the data, but shown here for purposes of illustration. \textit{Top panels and bottom left panel}: Temperature/polarization auto and cross anisotropy power spectra as labeled, with Fisher-level binned $\Lambda$CDM error bars obtained as described in the text.
 \textit{Bottom right panel:} Here we show the relative error between the ``exact" solution and two EFA implementations ($\mathcal{N}=3$ and $\mathcal{N}\approx 100$). The black curve ($3/\ell$) is a rough precision threshold beyond which parameter biases may be significant \cite{Seljak:2003th}. If this curve is exceeded at many  $\ell$ values by the actual EFA relative errors, an explicit computation of bias is needed to assess the full implications of these errors for cosmological parameter inference and ULA constraints.}
    \label{fig:CMBlow}
\end{figure*}

For the $\mathcal{N}=3$ and $\mathcal{N}=100$ implementations, we show the primary TT, EE, and TE ($C_{\ell}^{\rm TE}$) power spectra for $m_{\rm ax}=3.16\times 10^{-27}~{\rm eV}$ and $r_{\rm ax} = \Omega_{\rm ax}/\Omega_\text{DM}=1$ in Fig.~\ref{fig:CMBlow}. This value of $r_{\rm ax}$ is not allowed by the constraints \cite{Hlozek:2014lca,Hlozek:2016lzm}, but is chosen to visually accentuate differences between approaches. We also show relative errors between different EFA implementations. In addition to the individual $C_\ell$, we plot fiducial model points with error bars corresponding to cosmic variance. Following Ref. \cite{Ade:2015xua}, the modes are binned into bins of sizes $\{1,2,5,30\}$ for $\ell$ in the range $\{[2,4],[5,10],[11,30],[31,4000]\}$, respectively. The value plotted is the weighted average of $C_\ell$ with cosmic variance error at the weighted average $\ell$ with weight proportional to $\ell(\ell+1)$, as in the Planck Collaboration pipeline \cite{Aghanim:2015xee}. For a variety of other $m_{\rm ax}$ values, we explore the relative accuracy of different EFA implementations in Appendix \ref{sec:exfigs_cmb} using Fig. \ref{fig:CMBlow_extra0}.

For $m_{\rm ax}= 3.16\times 10^{-27}~{\rm eV}$, the \bench implementation is less accurate than the \urena implementation for TT and EE spectra, and comparable for TE. For $m_{\rm ax}= 3.16\times 10^{-26}~{\rm eV}$, the \bench and \urena implementations are of comparable accuracy (compared with the exact KG equation) at most $\ell$ values. For $m_{\rm ax} = 3.16\times 10^{-25}~{\rm eV}$, the \bench implementation is more accurate than the \urena implementation. 

Similar comparisons can be made for the \smith~implementation, whose performance relative to other choices is a function of $m_{\rm ax}$. At higher $m_{\rm ax}$ values, the residuals are in between those of the \bench and \urena implementations, while at lower $m_{\rm ax}$ values, the residuals are worse than any of the other implementations. In the interests of simplifying the discussion, we have omitted these results from figures showing power-spectra comparisons but include it in our final computations of bias in ULA density constraints.
Depending on the implementation and $\ell$ value, the fractional error $\Delta C_{\ell}/C_{\ell}$ can be several orders of magnitude in excess of the approximate \textit{Planck} noise level ($\Delta C_{\ell}/C_{\ell}\sim 10^{-3}$) when $r_{\rm ax}=1$. Forthcoming experiments (e.g.~CMB-S4 \cite{Abazajian:2019eic}) will measure nearly $N_{\rm modes}\sim 10^{7}$ perturbation modes. Roughly speaking \cite{Shimon:2012xd}, the fractional effect of any systematic error in the computation of CMB anisotropies grows as $\sqrt{N_\text{modes}}\Delta C_{\ell}/C_{\ell}$, where $N_{\rm modes}\propto \ell^{2} $ is the total number of multipole values with $\ell'\lesssim \ell$. Small biases thus require that $\Delta C_{\ell}/C_{\ell}\lesssim 1/\ell$, with detailed numerical considerations yielding the useful rule of thumb that bias-free parameter inference requires $\Delta C_\ell / C_\ell \lesssim 3/\ell$ \cite{Seljak:2003th,Shimon:2012xd,shimon:2012vs}. We see in Figs.~\ref{fig:CMBlow} and \ref{fig:CMBlow_extra0} that this condition is violated by several orders of magnitude at high $\ell$ $\gtrsim 10^{2}$ when $r_{\rm ax}=1$. 

As $\Delta C_{\ell}/C_{\ell}\propto r_{\rm ax}$ and $r_{\rm ax}\sim \mathcal{O}(0.1)$, it stands to reason that the fractional error between all three EFA implementations and the ``exact" calculation exceeds the needed accuracy for unbiased parameter inference, depending on the detailed full structure of the likelihood function (e.g. parameter degeneracies). To determine if the errors induced by the EFA significantly affect cosmological constraints to (or measurements of) ULA DM, we must compare the scale-dependent error with the information content of the CMB, as represented by the Fisher matrix \cite{Tegmark:1996bz,Bond:1997wr,Eisenstein:1998hr}. We now estimate the EFA-induced parameter bias.

\label{sec:cmbaniso}

\section{$Z$ statistic and bias}\label{Sec:zstatbias}
We wish to estimate the systematic error in CMB measurements of (or limits on) the axion relic density $\Omega_{\rm ax}$ from the CMB that results from the use of the EFA. Although a full analysis would require mock datasets and MCMC analysis, reasonable estimates may be obtained using standard Fisher-analysis techniques \cite{Tegmark:1996bz,Bond:1997wr,Eisenstein:1998hr}.

The elements of the Fisher matrix for the CMB anisotropy power spectrum are given by \cite{Tegmark:1996bz,Kamionkowski:1996ks,Shimon:2012xd}:
\begin{eqnarray}
    F_{ij}&=& \sum_{A,A' \in \{\text{TT,TE,EE}\}}\sum_l f_{\text{sky}} \frac{\partial C_l^A}{\partial \lambda_i}\frac{\partial C_{l}^{A'}}{\partial \lambda_j}(\Xi_{\ell}^{-1})_{AA'},\nonumber\\
    \Xi_{\ell,AA'}&=&\left \langle \left(\hat{C}_{\ell}^{A}-C_{\ell}^{A}  \right)\left(\hat{C}_{\ell}^{A'}-C_{\ell}^{A'}  \right) \right \rangle ,
\end{eqnarray}
where $\lambda = (h,\Omega_Bh^2,\Omega_{\rm DM}h^2,z_{re},n_s,A_s,\Omega_{\rm ax})$ is a choice of $\Lambda$CDM parameters, along with the ULA density of $\Omega_{\rm ax}$ today, and $f_{\rm sky}$ is the sky fraction covered by the CMB experiment of interest. Here the data covariance matrix is $ \Xi_{\ell,AA'}$. The brackets $\left \langle \right \rangle$ denote an ensemble average and $\hat{C}_\ell^{A} = \sum_{m=-\ell}^\ell a_\ell^{X*} a_\ell^Y/(2\ell + 1)$ is the usual optimal estimator of angular power spectra using the multipole moments as data, where the observable $A=\left\{X,Y\right\}$ consists of the pair $X,Y\in\left\{{\rm T,~E}\right\}$.  We neglect $B$-mode (curl) anisotropies here, as our analysis neglects weak gravitational lensing of the CMB and primordial tensor modes. 

The ULA mass $m_{\rm ax}$ is varied and the full Fisher matrix $\mathcal{F}$ with elements $F_{ij}$ is recomputed at each value, to see how biases and parameter errors depend on ULA mass. The derivative $\partial C_\ell^A/\partial \lambda_i$ quantifies the response of the observables to the $\Lambda$CDM and ULA parameters of interest; typically this derivative must be obtained numerically using a Boltzmann code, though for $A_{s}$, the derivative may be obtained analytically ($\partial C_{\ell}^{A}/\partial A_{s}=C_{\ell}^{A}/A_{s}$). If we compute the Fisher matrix for a theoretical scenario and some experiment, we may forecast the error on the best fit parameters via $\sigma_{\lambda_i}=\sqrt{(\mathcal{F}^{-1})_{ii}}$, assuming that the parameter likelihood (given the data) distribution is Gaussian. 

The elements of the data covariance matrix are \cite{Kamionkowski:1996ks,Knox:1995dq}
\begin{equation}
\label{eq:data_covar}
\begin{aligned}
\Xi_{\ell,{AA}} &= \frac{2}{2\ell+1}\left(C^{A}_\ell + N^{A}_\ell\right)^2;\qquad A \in \{\text{TT,EE}\}\\
\Xi_{\ell,\text{TTEE}} &= \frac{2}{2\ell+1} (C^\text{TE}_\ell)^2\\
\Xi_{\ell,\text{TETE}} &= \frac{1}{2\ell+1}\left[ (C^\text{TE}_\ell)^2 + (C^\text{TT}_\ell + N^\text{TT}_\ell)(C^\text{EE}_\ell + N^\text{EE}_\ell)\right]\\
\Xi_{\ell,\text{TETT}} &= \frac{2}{2\ell+1} C^\text{TE}_\ell (C^\text{TT}_\ell + N^\text{TT}_\ell)\\
\Xi_{\ell,\text{TEEE}} &= \frac{2}{2\ell+1} C^\text{TE}_\ell (C^\text{EE}_\ell+N^\text{EE}_\ell) \\
N^{A}_\ell &= \delta_{AA}^2 e^{\ell(\ell+1) \theta_\text{FWHM} / (8 \ln 2)},
\end{aligned}
\end{equation}where $N^{A}_{\ell}$ is the noise power spectrum for the observable $AA$. The remaining elements follow trivially since $\Xi_{\ell,A'A} = \Xi_{\ell,AA'}$. The above covariance matrix takes into account both the cosmic variance, and the noise of the detector \cite{Kamionkowski:1996ks}. We use the usual approximations of Ref. \cite{Knox:1995dq}, with an overall amplitude noise amplitude $\delta_{A}^2$ in $(\mu K)^2$. The instrument beam is assumed to be Gaussian with a full-width half-max angular size of $\theta_\text{FWHM}$ in radians. 

The errors $\Delta C_l^{\rm XY}$ in the theoretical calculation of $C_{\ell}^{XY}$ drive the peak of the likelihood to a different set of parameter values than the true model parameters, resulting in parameter biases. Under a Gaussian likelihood approximation, we follow Refs. \cite{Knox:1998fp,Santos:2003jb,Zahn:2005fn,Taburet:2008vg,Shimon:2012xd,shimon:2012vs} to compute this bias, which is \cite{Taburet:2008vg} \begin{equation}\delta_{i} \simeq -\sum_j (\mathcal{F}^{-1})_{ij}V_{j},\label{eq:prebias}\end{equation} where 
\begin{equation}
\label{eq:bias}
    V_i = \sum_{A,A' \in \{\text{TT,TE,EE}\}} \sum_{\ell} f_\text{sky} \Delta C_l^A \frac{\partial C_l^{A'}}{\partial \lambda_i} (\Xi_{\ell}^{-1})_{AA'}
\end{equation}
and  $\Delta C_{\ell}^{A}=C_{\ell}^{A,{\rm cut}}-C_{\ell}^{A,{\rm exact} }$ is the shift in computed theoretical power spectra between the ``exact" treatment of ULAs and the EFA.\footnote{We follow Refs. \cite{Knox:1998fp,Santos:2003jb,Zahn:2005fn, Shimon:2012xd,shimon:2012vs} and drop higher-order terms of order $\Delta C_{\ell}$ in the Fisher matrix itself, as these lead to corrections of the form $\mathcal{O}(C_{\ell}^{2})$ to the bias. These terms are subdominant unless there is a ULA detection of high significance.} This expression takes into full account degeneracies between different cosmological parameters. If the bias in $\Omega_{\rm ax}$ resulting from use of the EFA can be absorbed by adjusting other cosmological parameters, Eq.~(\ref{eq:bias}) will indicate a negligible shift.

In the future, if other measurements (e.g.~large-scale structure surveys or the CMB weak lensing trispectrum) are combined with CMB power spectrum measurements (dominated by primary anisotropies) to break parameter degeneracies, a larger bias may result. The bias and error in $\Omega_{\rm ax}$ with all other cosmological parameters held fixed are given by \begin{widetext}
\begin{align}
    \frac{\delta_{\Omega_{\rm ax}}}{\sigma_{\Omega_{\rm ax}}} =&-\sigma_{\Omega_{\rm ax}} \left[
    \sum_{A,A' \in \{\text{TT,TE,EE}\}} 
    \sum_\ell f_{\text{sky}}
    \Delta C_\ell^A 
    \frac{\partial C_\ell^{A'}}{\partial \Omega_{\rm ax}}
    \left(\Xi_{\ell}^{-1}\right)_{AA'}
    \right],\label{eq:bias_ofix}\\
    \sigma_{\Omega_{\rm ax}}=&\sqrt{\frac{1}{F_{\Omega_{\rm ax}\Omega_{\rm ax}}}}.
\end{align}\end{widetext} 

Before computing an extensive set of numerical derivatives, it would be useful to establish the maximum bias on a single parameter, normalized to its Fisher-level error. Using Eq.~(\ref{eq:bias_ofix}) and the Cauchy-Schwarz inequality (see Appendix \ref{app:Zstat_deriv} for a derivation), it can be shown that \begin{widetext}
\begin{equation}
\left |\frac{\delta_{\lambda_{i}}}{\sigma_{\lambda_{i}}}\right| \leq Z\equiv 
\sqrt{\sum_{A,A' \in \{\text{TT,TE,EE}\}} \sum_{\ell=\ell_\text{min}}^{\ell_\text{max}}f_{\text{sky}}\Delta C_\ell^A\Delta C_\ell^{A'} \left(\Xi_{\ell}^{-1}\right)_{AA'}}
\label{eq:Zstat}.
\end{equation} \end{widetext} This $Z$ statistic \cite{hirata} has been used to estimate the impact of improvements to cosmic recombination history computations on CMB parameter estimation \cite{Hirata:2008ny,Hirata:2009qy,Grin:2009ik}. If $Z\ll 1$, we can safely conclude that the systematic errors in $C_{\ell}^{A}$ induced by use of the EFA make a negligible impact on the estimation of both $\Omega_{\rm ax}$ and standard cosmological parameters. Conversely, if $Z \gtrsim 1$, there could be large shifts in central values, and a bias estimate computed using Eq.~(\ref{eq:bias}) or (\ref{eq:bias_ofix}) is needed.\footnote{$Z$ is just the square root of the predicted change in the $\chi^{2}$ statistic induced by the use of the EFA. The interpretation of $Z$ as the maximum fractional bias in a single parameter (neglecting degeneracies) requires the added assumption of a Gaussian likelihood for the parameters.}

\subsection{Z-statistic for current and future CMB experiments}
\label{sec:zzsec}
We compute the Z statistic using Eqs.~(\ref{eq:data_covar}) and (\ref{eq:Zstat}), assuming noise properties for various CMB instruments (past, present, anticipated) given in Table \ref{tab:noises}. Of course even an ideal zero-noise experiment is affected by cosmic variance and is shown as the cosmic-variance limited (CVL) case. In this case, $Z$ quantifies the maximum impact of EFA-induced errors on CMB constraints to $\Omega_{\rm ax}$. 

\begin{table}
\centering
\caption{The parameters used to calculate the noise, as also used in Ref.~\cite{He:2015msa}. The FWHM column gives $\theta_{\text{FWHM}}$ in arcmin, the Noise column gives $\delta_{TT}$ in $\mu$K arcmin, and $\ell_\text{T,max}$ is the maximum harmonic of temperature fluctuations that was/will be measured. For Planck, we give $(\delta_{TT},\delta_{EE})$ in the Noise column. We assume $\delta_{EE}=2\delta_{TT}$ otherwise. For Planck and WMAP, the reciprocal of the noise we use is the reciprocal sum of the noises from each band. For CMB-S4, higher $l_{\rm max}$ values are used for polarization, as described in the main text.}
 \begin{tabular}{c |c c c c |c} 
 \hline
 Experiment & FWHM  & Noise & $f_\text{sky}$ & $\ell_\text{T,max}$ & Ref. \\ 
 \hline
WMAP V band & 21 & 434 & 0.65 & 2200 & Ref. \cite{wmapnoise}\\
WMAP W band & 13 & 409 & 0.65 & 2200 & Ref. \cite{wmapnoise}\\
Planck 143 GHz & 7.1 & (37, 78) & 0.65 & 2200 & Ref. \cite{Planck:2013cta}\\
Planck 217 GHz & 5.0 & (54, 119) & 0.65 & 2200 & Ref. \cite{Planck:2013cta}\\
ACTPol & 1.4 & 8.9 & 0.097 & 2200 & Ref. \cite{2010SPIE.7741E..1SN}\\
SPT-3G & 1.1 & 2.5 & 0.06 & 2200 & Ref. \cite{crawford}\\
CMB-S4 & 3.0 & 1.0 & 0.50 & 2200 & Ref. \cite{Abazajian:2013oma} \\
CVL & 0.0 & 0.0 & 1.0 & 2200 & \\
 \end{tabular}
 \label{tab:noises}
\end{table}

We create a 7~X~7 logarithmically spaced grid with $10^{-27}$ eV$\le m_{\rm ax} \le 10^{-24}$ eV and $10^{-2}\leq \Omega_{\rm ax}/\Omega_{\rm DM}\leq 1$, and evaluate the $Z$ statistic for the three different versions of the EFA. For temperature, we restrict ourselves to $\ell \le \ell_\text{T,max}=2200$, as secondary temperature anisotropies dominate the primordial CMB at smaller scales (although there are futuristic proposals to go well beyond this limit, e.g. Refs. \cite{Nguyen:2017zqu,Sehgal:2019ewc}). Since polarization foregrounds are expected to be less severe (than ones in the temperature) at small angular scales \cite{Abazajian:2016yjj}, we assume $\ell \le4000$ for polarization data and use the similar (polarization-only) expression
\begin{equation}\label{eq:Zgt2000}
    Z_{>\ell_\text{T,max}} = \sqrt{ \sum_{\ell=\ell_\text{2201}}^{\ell_\text{max}} f_{\text{sky}}\Delta C_\ell^\text{EE} \Delta C_l^\text{EE}(\Xi_\text{EE,EE})_\ell^{-1}}
\end{equation}
with $Z_\text{tot} = Z(\ell_\text{max}=\ell_\text{T,max})+Z_{>\ell_\text{T,max}}$. 

\begin{figure*}[ht]
    \centering
    \includegraphics[scale=0.4]{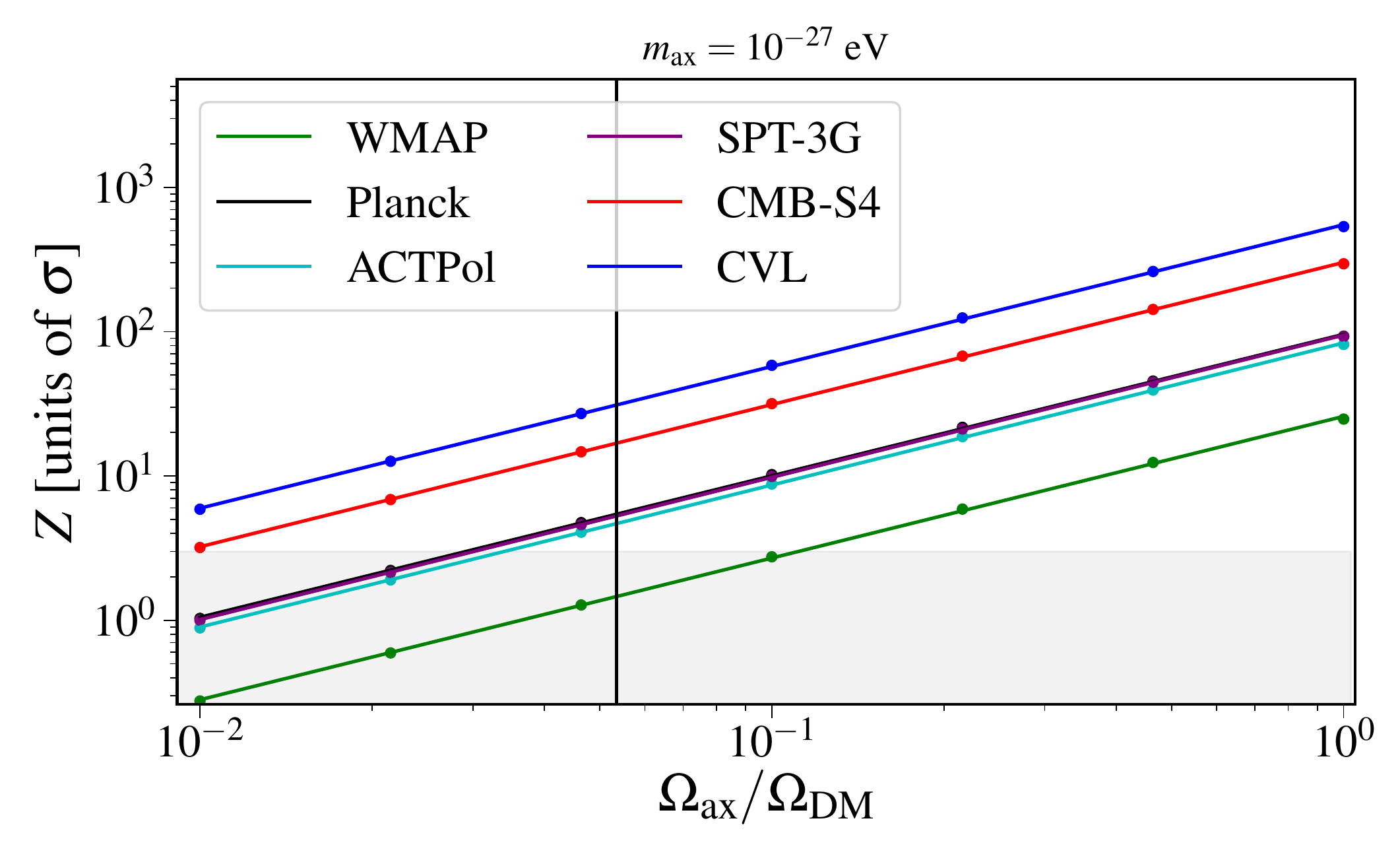}\includegraphics[scale=0.4]{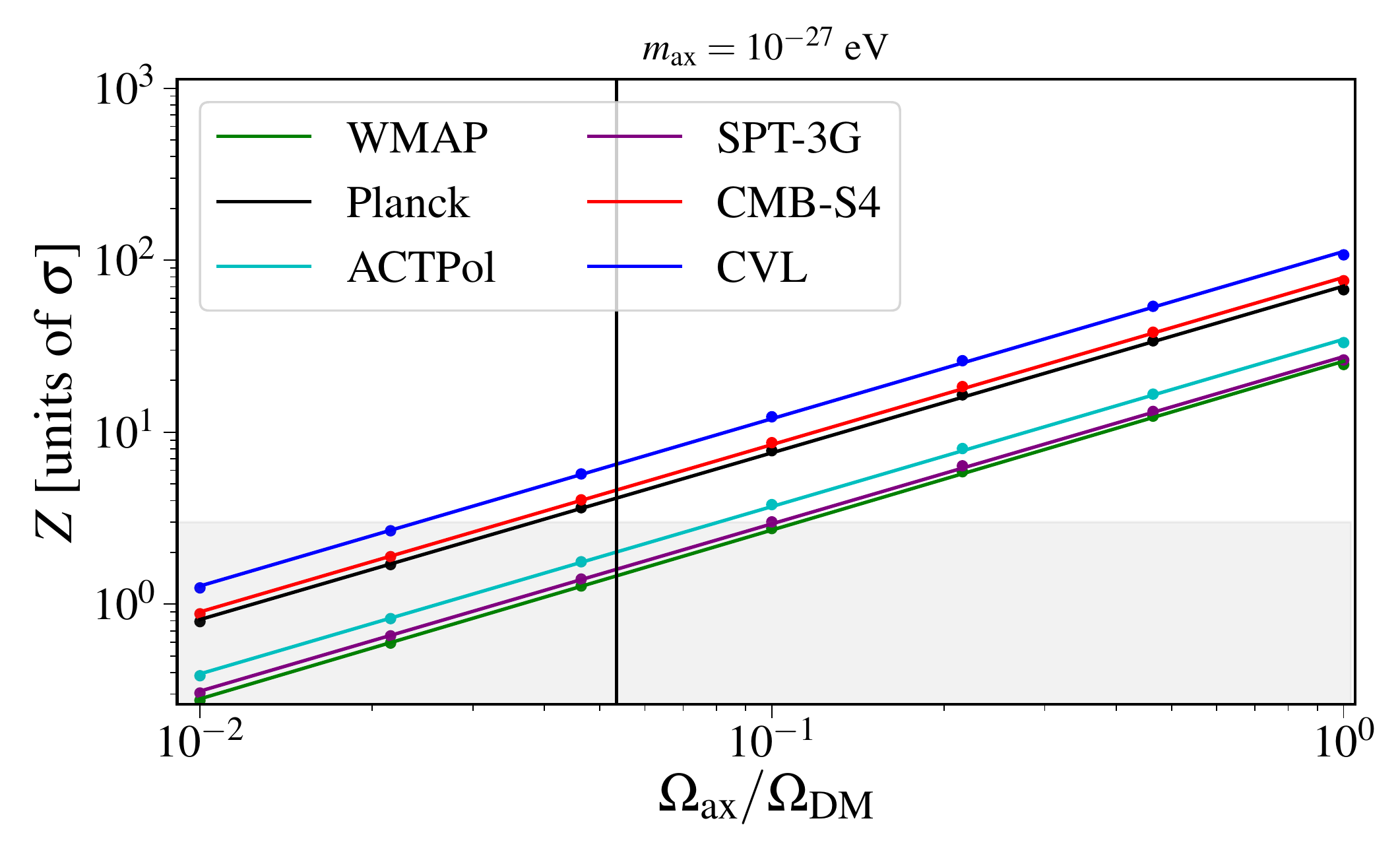}
    \caption{(Color online). The $Z$ statistic encoding deviations between the \bench and ``exact" treatments of ULA dynamics is shown; large $Z$ values indicate potentially large biases in cosmological parameters. The $Z$ is plotted at fixed ULA mass $m_{\rm ax}$ vs. the fraction of dark matter composed of axions, $\Omega_{\rm ax}/\Omega_{\rm DM}$. The black line indicates the 3$\sigma$ upper limit found in Ref.~\cite{Hlozek:2014lca}, and the gray shaded area is below $3\sigma$. In the left panel, we show results for temperature and polarization anisotropies. In the right panel, we show results for temperature anisotropies only.}
    \label{fig:ZsliceExtraDan}
\end{figure*}

\begin{figure}[h]
    \centering
    \includegraphics[width=\columnwidth]{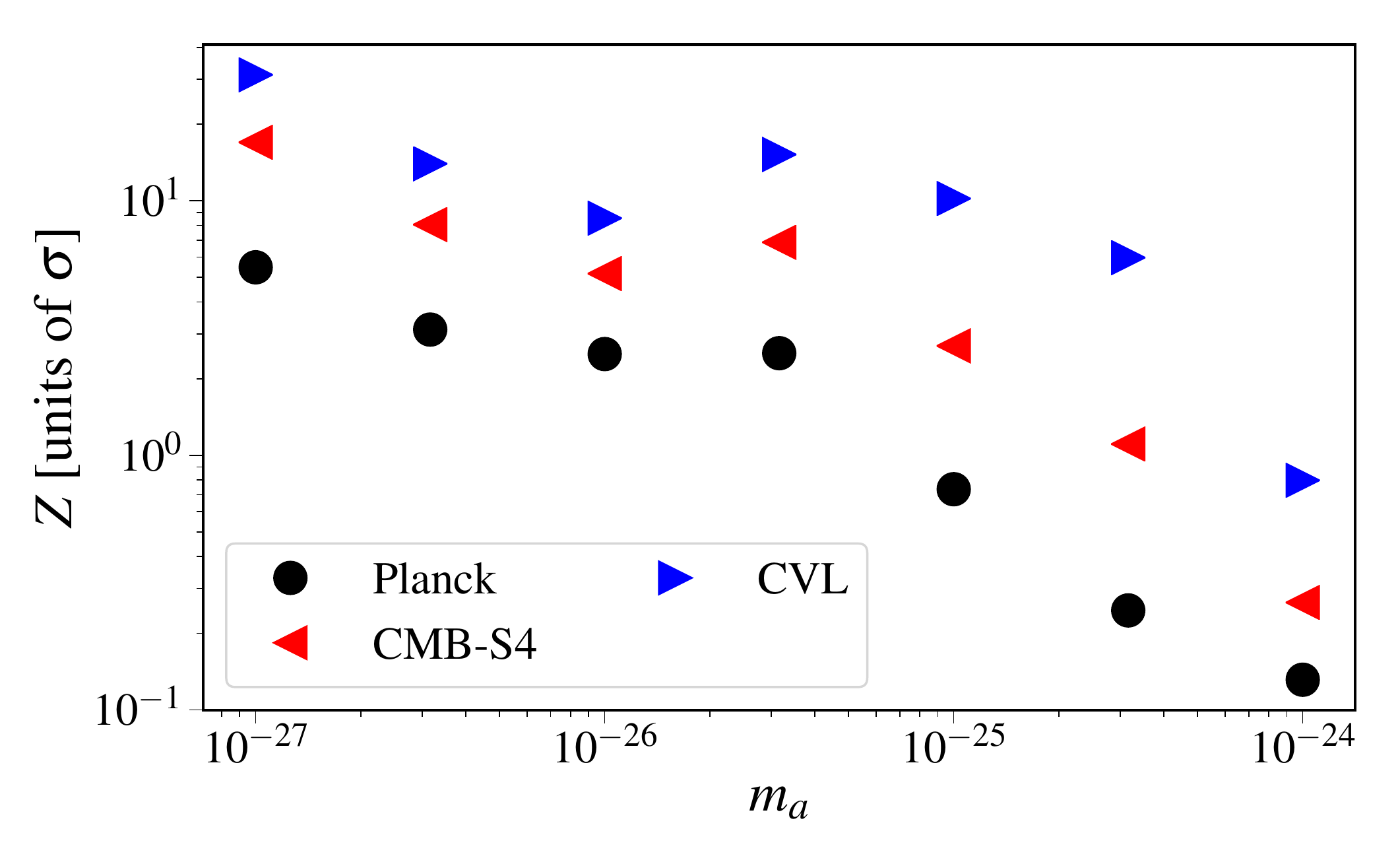}
        \caption{(Color online). The $Z$ statistic as a function of $m_{\rm ax}$ for $\Omega_{\rm ax}/\Omega_{\rm DM}$ values saturating the constraints in Ref.~\cite{Hlozek:2014lca}. Here temperature and polarization anisotropies are used.}
    \label{fig:Zmoney}
\end{figure}

As an example, in Fig.~\ref{fig:ZsliceExtraDan}, we show the $Z$ statistic as a function of $\Omega_{\rm ax}/\Omega_{\rm DM}$ for $m_{\rm ax}=10^{-27}~{\rm eV}$, where the constraints of Ref.~\cite{Hlozek:2014lca} are most stringent, and where the effects of the EFA are most severe. If the true ULA abundance saturates current constraints (from the CMB power spectra without lensing, as in Ref.~\cite{Hlozek:2014lca}), we see that (depending on the experiment), the $Z$ bound indicates potential biases as large as  $\sim 4\to 30\sigma$. The effect is significantly less pronounced if only temperature data are used, highlighting the importance of CMB polarization. If the actual ULA density is far \textit{lower} than present-day constraints, $Z$ could be much smaller, allowing the EFA-induced error to be neglected.

To succinctly capture the potential impact of the EFA (for the \bench implementation) at all masses, we plot $Z$ for $\Omega_{\rm ax}/\Omega_{\rm DM}$ values saturating the constraints of Ref.~\cite{Hlozek:2014lca} in Fig.~\ref{fig:Zmoney}, including temperature and polarization anisotropies. We see that for $m_{\rm ax}\lesssim 3.2\times 10^{-26}~{\rm eV}$, $\sim3\sigma$ and greater biases could occur for all three experimental scenarios considered there. For CMB-S4 or the CVL case, significant ($\gtrsim 1 \sigma$) $Z$ values occur for values as high as $m_{\rm ax}= 3.2\times 10^{-25}~{\rm eV}$. We also compute $Z$ for the \smith~and \urena implementations. These results are of the same order of magnitude, and are thus omitted for brevity. 

To check that the constraints of Ref.~\cite{Hlozek:2014lca} are robust and to assess the impact of the EFA on future CMB tests of ULA physics, we must thus evaluate the more complete Fisher-matrix based estimates [Eqs.~(\ref{eq:bias}) and (\ref{eq:bias_ofix})]. These are needed to properly include the detailed response of $C_{\ell}^{\rm A}$ to $\Omega_{\rm ax}$ variations in all implementations considered, and to properly assess the impact of parameter degeneracies.

\subsection{Bias}
\label{sec:bias}
Since the $Z$ statistic was $\mathcal{O}(1)$ for a significant part of parameter space, we directly compute the bias in the value of $\Omega_{\rm ax}$ induced by the use of the EFA. We compute the parameter bias for our full 7~X~7 grid of models using Eq.~(\ref{eq:bias}) and forecast the error on the best fit parameters via $\sigma_{\lambda_i}=\sqrt{(\mathcal{F}^{-1})_{ii}}$. As noted in Sec. \ref{sec:zzsec}, for $\ell>2200$ we expect temperature measurements to be dominated by secondary anisotropies, and restrict all Fisher sums to EE-only portions of the data covariance matrix.

We calculate the relevant numerical derivatives with respect to $\Lambda$CDM parameters by modifying the Planck values at the percent level as in Ref.~\cite{Eisenstein:1998hr} and using the two-point symmetric finite difference method. We also compute derivatives with respect to $\Omega_{\rm ax}$. These bias/error estimates are thus a first-order approximation, given the use of $\Omega_{\rm ax}=0$ for derivatives with respect to $\Lambda$CDM parameters. 

For the derivative with respect to $\Omega_{\rm ax} \ne 0$, we check convergence by comparing the two-point symmetric, left, and right finite-difference methods and verifying that the bias is converged at the $\sim~5\%-10\%$ level for $m_{\rm a}<10^{-24}~{\rm eV}$, where ULA sensitivity is very poor and the overall bias $\delta_{\Omega_{\rm ax}} \ll \sigma_{\Omega_{\rm ax}}$.  Below, we discuss how large the bias would be for $\Omega_{\rm ax}$ values saturating current constraints, but note that the bias will be less severe for $\Omega_{\rm ax}$ values well below the current upper limits. We confirm for the null (no ULA) hypothesis that \textit{Planck} parameter error levels are  reproduced. For the \bench implementation, we also confirm that if ULAs are included that the marginalized $\sigma_{\Omega_{\rm ax}}$ curve (as a function of $m_{\rm ax}$) is consistent (at order-of-magnitude level) with the constraints of Ref. \cite{Hlozek:2014lca}, which were obtained using the same EFA implementation.

\begin{figure}
    \centering
  \includegraphics[scale=0.4]{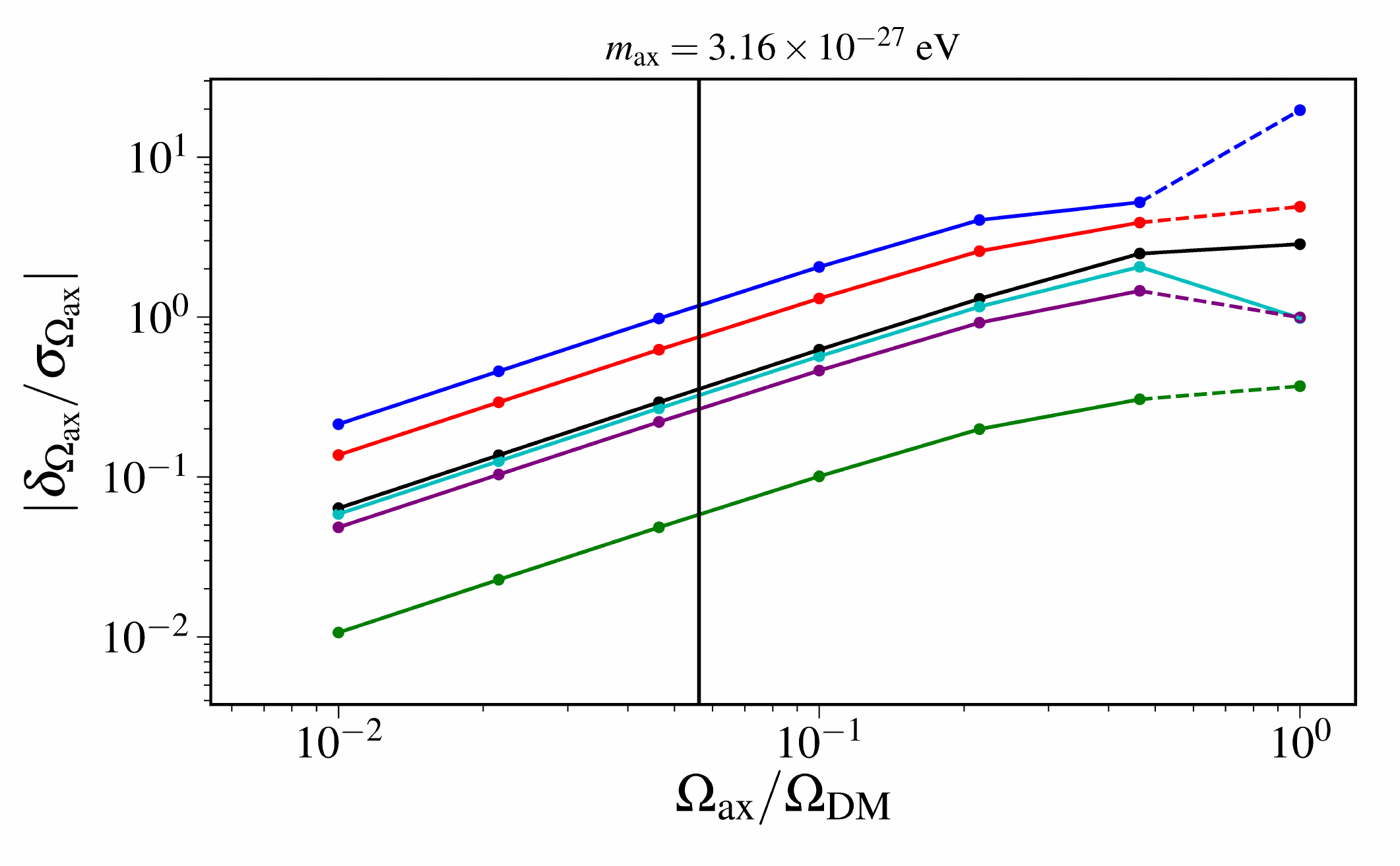}
    \caption{(Color online). The points show the dimensionless bias in $\Omega_{\rm ax}$ for the \bench implementation, calculated using Eq.~\eqref{eq:bias} (thus including parameter degeneracies), plotted as a function of the ULA dark matter fraction, $\Omega_{\rm ax}/\Omega_{\rm DM}$. The lines show a simple point-to-point linear interpolation on a log-log plot. The vertical black line shows the $3\sigma$ upper limit to $\Omega_{\rm ax}/\Omega_{\rm DM}$ from Ref.~\cite{Hlozek:2014lca} for the $m_{\rm ax}$ values shown. Color code as in Fig. \ref{fig:ZsliceExtraDan}. Dashed lines indicate a positive bias, while solid lines indicate a negative bias.}
    \label{fig:forextra}
\end{figure}

We begin by focusing our attention on the fiducial $\mathcal{N}=3$ case. As an example, in Fig. \ref{fig:forextra}, we show the bias for $m_{\rm ax}=3.16\times 10^{-27}~{\rm eV}$ using hypothetical temperature and polarization data, as a function of $\Omega_{\rm a}/\Omega_{\rm DM}$. We see that for this $m_{\rm ax}$ value,  the bias is negligible for $\Omega_{\rm ax}/\Omega_{\rm DM}$ values satisfying the constraints of Ref. \cite{Hlozek:2014lca}. As beam, noise, and $\Omega_{\rm ax}/\Omega_{\rm DM}$ are varied, different scales are emphasized in the calculation of bias, and so it is natural that there is some nonmonotonicity with respect to experimental ordering and $\Omega_{\rm ax}/\Omega_{\rm DM}$. Additionally, the bias can be positive or negative depending on the subtle interplay of some of these parameters.

\begin{figure}
    \includegraphics[scale=0.4]{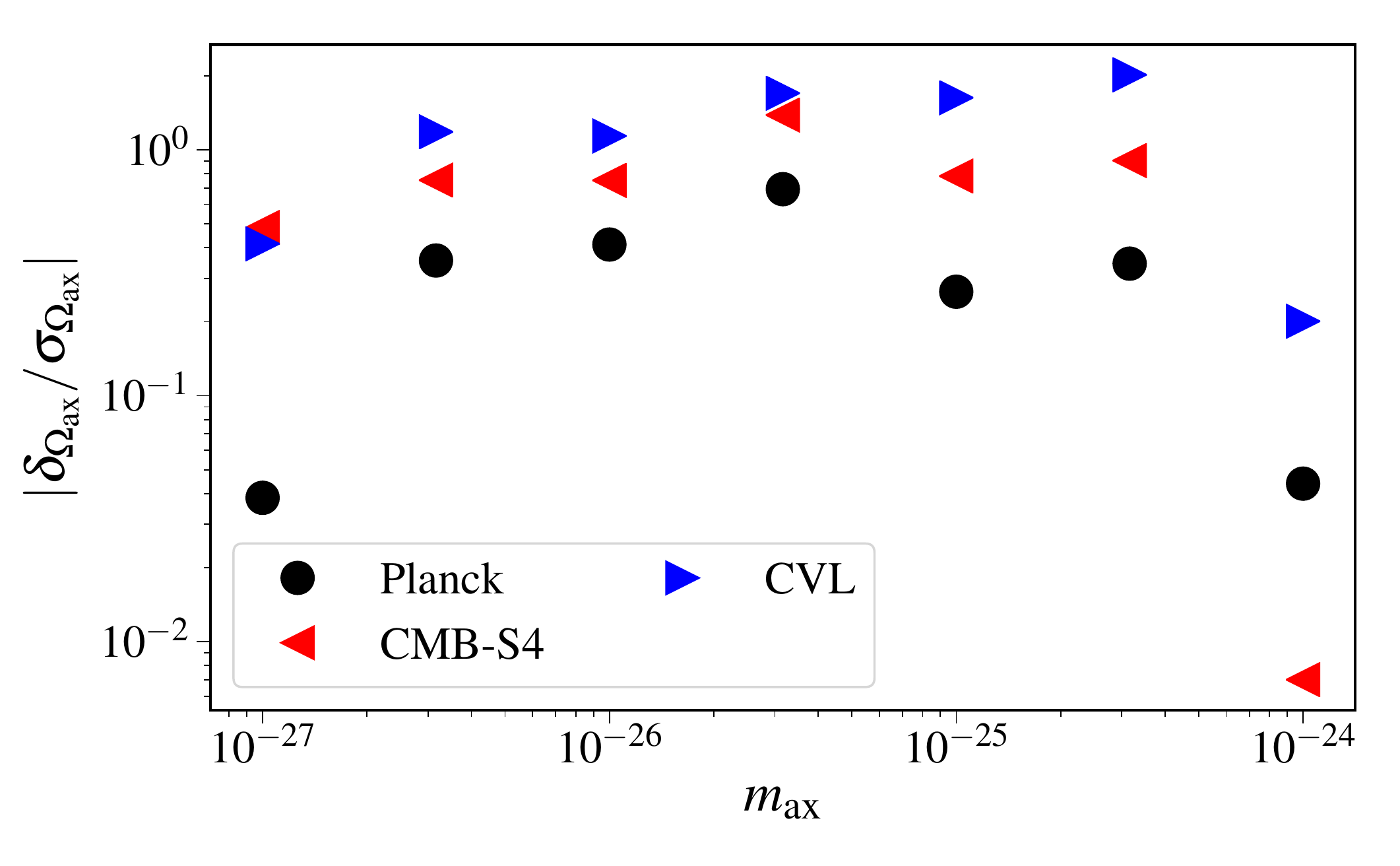}
    \caption{(Color Online). Dimensionless bias in $\Omega_{\rm ax}$ as a function of $m_{\rm ax}$ for the \bench implementation, calculated using Eq.~\eqref{eq:bias} (thus including parameter degeneracies), with $\Omega_{\rm ax}/\Omega_{\rm DM}$ set equal to the current (\textit{Planck}) $3\sigma$ upper limits.}
        \label{fig:full_money_plot}
\end{figure}

To better capture the information content of all these bias figures, we tabulate the dimensionless bias as a function of $m_{\rm ax}$, with $\Omega_{\rm ax}/\Omega_{\rm DM}$ fixed at the current $3\sigma$ constraint level (at that $m_{\rm ax}$ value). The results (for \textit{Planck}, CMB-S4, and the CVL case) are shown in Fig. \ref{fig:full_money_plot}. We see that small biases ($|\delta_{\Omega_{\rm ax}}|<\sigma_{\Omega_{\rm ax}}$) in ULA densities occur at \textit{Planck} noise levels. The CMB-only results of Ref. \cite{Hlozek:2014lca} are thus robust to the use of the EFA for $m_{\rm ax}\geq 10^{-27}~{\rm eV}$. For more futuristic noise levels (e.g.,~the CMB-S4 and CVL cases), the bias for the primary CMB remains relatively small ($\lesssim 2\sigma_{\Omega_{\rm ax}}$)  in the $m_{\rm ax}$ range shown. 

Our reference case for all this analysis is the ``exact" case, which in actuality still has a switch to the EFA at \textit{very} late times (when $\tilde{m}=10^{4}H$), long after recombination at all $m_{\rm ax}$ values where the primary CMB significantly constrains $\Omega_{\rm ax}$. Qualitatively, then, we expect that any transients in mode evolution introduced by the use of this switch will not alter our results. To test this expectation, we repeated all the bias analysis with the restriction that $\ell \geq \ell_{\rm late}$, where $\ell_{\rm late}$ is the angular size of the causal horizon when $\tilde{m}=10^{4}H$, and found that the conclusions implied by Fig. \ref{fig:full_money_plot} are unaffected by the late-time switch.

So far, our estimates of bias have been computed using Eq.~(\ref{eq:bias}), and these include parameter degeneracies (between ULAs and standard cosmological parameters). We may alternatively treat standard cosmological parameters as fixed and neglect parameter degeneracies, using Eq.~(\ref{eq:bias_ofix}) to compute the bias at maximum-likelihood parameter values. These results will be an upper bound to the absolute value of the bias when complementary datasets (e.g galaxy clustering or weak lensing, or CMB lensing) are used to break degeneracies. A complete treatment of this issue requires a combined Fisher analysis for CMB power spectra and other data, but we leave this for future work. To evaluate $\sigma_{\Omega_{\rm ax}}$ in Eq.~(\ref{eq:bias_ofix}), we use the relation $\sigma_{\lambda_{i}}=\sqrt{1/F_{ii}}$, valid for the variance of $\lambda_{i}$ with a multivariate Gaussian likelihood, if other parameters are held fixed.

\begin{figure}
\includegraphics[scale=0.4]{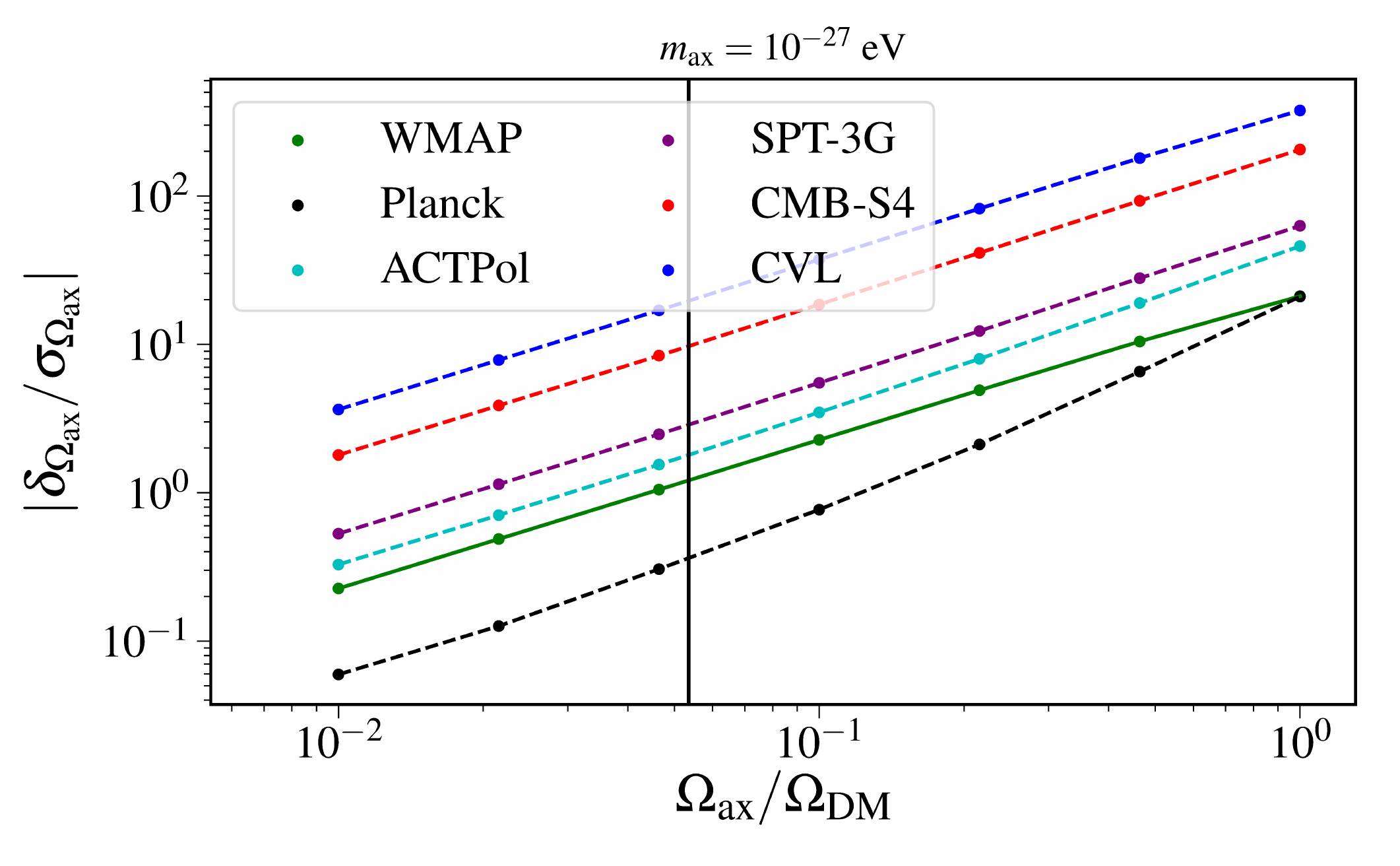}
    \caption{(Color online). The points show the dimensionless bias in $\Omega_{\rm ax}$ for the \bench implementation, calculated using Eq.~\eqref{eq:bias_ofix} (thus neglecting parameter degeneracies), plotted as a function of the ULA dark matter fraction, $\Omega_{\rm ax}/\Omega_{\rm DM}$. The lines show a simple point-to-point linear interpolation on a log-log plot. The vertical black line shows the $3\sigma$ upper limit to $\Omega_{\rm ax}/\Omega_{\rm DM}$ from Ref. \cite{Hlozek:2014lca}. Color code as in Fig. \ref{fig:ZsliceExtraDan}.  Dashed lines indicate a positive bias, while solid lines indicate a negative bias.}    \label{fig:forextra_fix}
\end{figure}

As an example, in Fig.~\ref{fig:forextra_fix}, we show the bias for $m_{\rm ax}=3.16\times 10^{-27}~{\rm eV}$ using hypothetical temperature + polarization data, as a function of $\Omega_{\rm a}/\Omega_{\rm DM}$. We see that $2\sigma$ or greater biases result at SPT-3G \& CMB-S4 noise levels, as well as in the CVL case. Biases are smaller than $2\sigma$ for the ACTPol, WMAP, and Planck cases.

One important aspect of the bias (see above) is that better experiments do not necessarily have larger magnitude bias, and that the bias can be positive or negative. We have verified that this occurs because $\Delta C_\ell$ and $\partial C_\ell/\partial \Omega_{\rm ax}$ can have different signs, and as a result, the terms in the bias sum can have either sign. In the case of Fig.~\ref{fig:forextra_fix}, on the scales that WMAP most accurately probes, the terms almost all have the same sign, but, when adding more accurate high-$\ell$ \textit{Planck} measurements, added terms with opposite signs reduce the bias amplitude. More precise (current and future) experiments reach even deeper into this high-$\ell$ regime, causing the amplitude of the fractional bias to retain its sign but increase in amplitude.

\begin{figure}
    \includegraphics[scale=0.4]{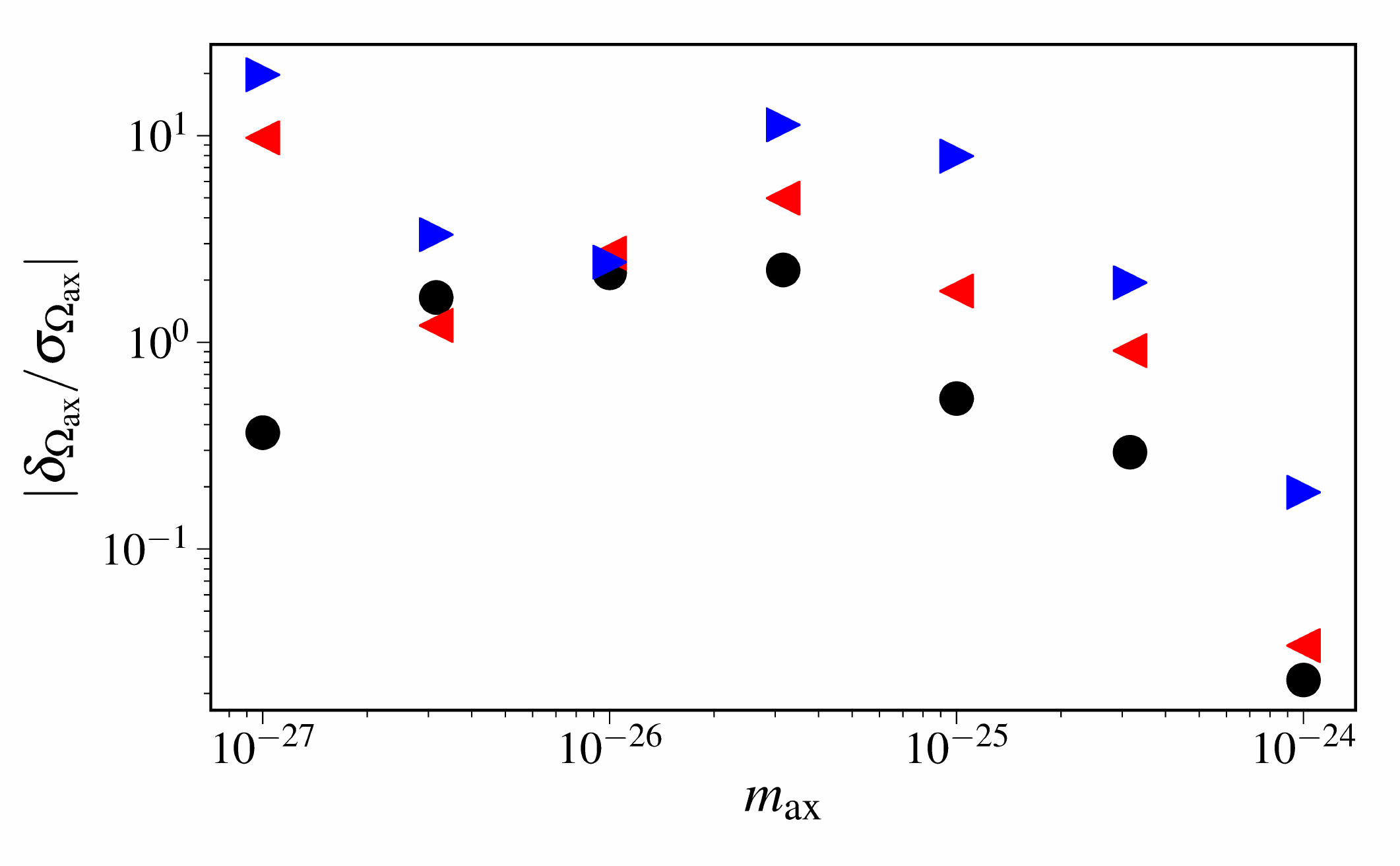}
    \caption{(Color online). Dimensionless bias in $\Omega_{\rm ax}$ as a function of $m_{\rm ax}$ for the \bench implementation, calculated using Eq.~\eqref{eq:bias_ofix} (thus neglecting parameter degeneracies), with $\Omega_{\rm ax}/\Omega_{\rm DM}$ set equal to the current (\textit{Planck}) $3\sigma$ upper limits.}
        \label{fig:full_money_plot_ofix}
\end{figure}

In Fig.~\ref{fig:full_money_plot_ofix}, we show the dimensionless bias (now neglecting degeneracies), as a function of $m_{\rm ax}$, with $\Omega_{\rm ax}/\Omega_{\rm DM}$ fixed exactly at the current $3\sigma$ constraint level. As in some of the cases above, the sharp dip in the CVL bias at $m_{\rm ax}=3.16\times 10^{-25}~{\rm eV}$ (from which the bias curve returns to a more standard ordering) is driven by physical sign changes in the summand of Eq.~(\ref{eq:bias_ofix}). We see that at some of the most constrained masses, ($3.16 \times 10^{-27}~{\rm eV}\leq m_{\rm ax}\leq 3.16\times 10^{-26}~{\rm eV}$), large biases ($|\delta_{\Omega_{\rm ax}}|>2\sigma_{\Omega_{\rm ax}}$) in ULA densities occur at \textit{Planck} noise levels. 

For more futuristic noise levels (as shown by the CMB-S4 and CVL cases), similar and even larger biases result if $m_{\rm ax}\leq 10^{-25}~{\rm eV}$. We note Fisher-matrix calculations that indicate biases greater than several $\sigma_{\Omega_{\rm ax}}$ should just be taken as an indication that the actual bias is severe and not a precise result, given the breakdown of the Gaussian likelihood approximation for large deviations from central values. 

We see that biases are large when parameters are fixed, but not when they are marginalized over; this implies that the systematic errors in $C_{\ell}$ values induced by the EFA are significantly degenerate with shifts in other cosmological parameters. It is natural to wonder then, if future more precise measurements of standard cosmological parameters will push
bias toward the larger values shown in Fig.~\ref{fig:full_money_plot_ofix}. 

We explored this issue further using forecasts for matter power spectrum measurements by the Large Synoptic Survey Telescope (LSST) \cite{Zhan:2017uwu}, Dark Energy Spectroscopic Instrument (DESI) \cite{Casas:2017eob}, Euclid satellite \cite{Casas:2017eob,Sartoris:2015aga}, and Square Kilometer Area (SKA) \cite{Bacon:2018dui}, as well as more direct distance ladder-based/gravitational-wave inferred measurements of the Hubble constant $H_{0}$ \cite{Beaton:2019jwm}. We imposed priors to cosmological parameters via a diagonal modification to the Fisher matrix, $F_{ij}\to F_{ij}+\delta_{ij}/\sigma_{i}^{2}$, where $\sigma_{i}$ is the forecasted error to a $\Lambda$CDM parameter from one of these efforts (priors were applied to $A_{s}$, $n_{s}$, $H_{0}$, and $\Omega_{c}$ as appropriate given Refs. \cite{Zhan:2017uwu,Casas:2017eob,Sartoris:2015aga,Bacon:2018dui,Beaton:2019jwm}). 

We then used Eq.~(\ref{eq:prebias}) to compute the bias to $\Omega_{a}$ resulting from the use of the EFA in the presence of priors, and found in all cases that the results reproduced those of Fig. \ref{fig:full_money_plot} at the $\sim~10\%-20\%$ level, and were not comparable to the large bias in Fig. \ref{fig:full_money_plot_ofix}. In other words, the  primary CMB Fisher matrix exhibits sufficient parameter degeneracy for the impact of the EFA on ULA parameters to be absorbed by variations in other parameters, even in the presence of sensible priors from upcoming experiments. Future work to extend this analysis should include CMB lensing and the full off-diagonal Fisher matrices from galaxy surveys (and other efforts), including the impact of ULAs on galaxy and cosmic shear clustering power spectra.

\begin{figure}
    \includegraphics[scale=0.4]{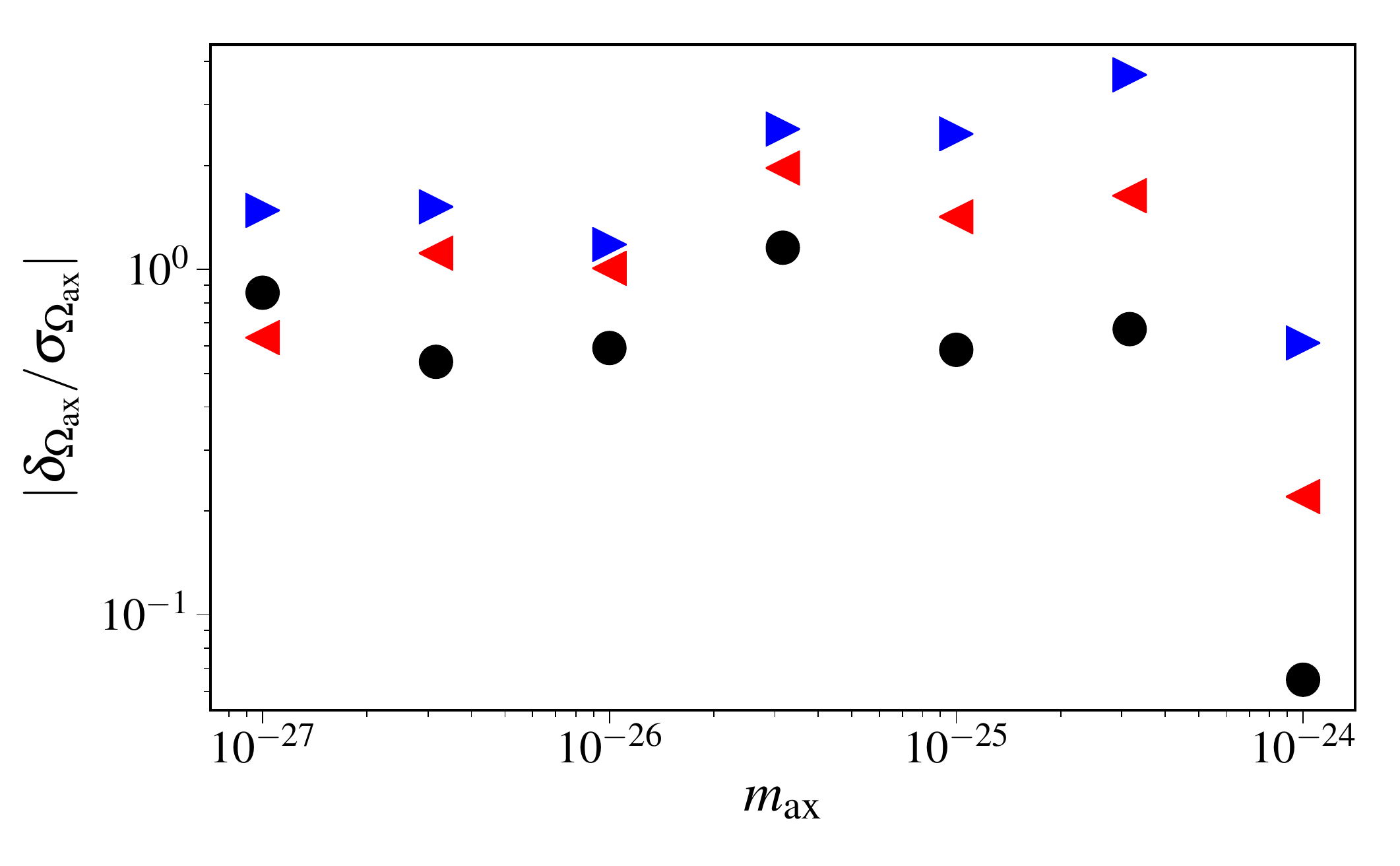}
    \caption{(Color Online). Dimensionless bias in $\Omega_{\rm ax}$ as a function of $m_{\rm ax}$ for the \smith~implementation, calculated using Eq.~\eqref{eq:bias} (thus including parameter degeneracies), with $\Omega_{\rm ax}/\Omega_{\rm DM}$ set equal to the current (\textit{Planck}) $3\sigma$ upper limits. This figure is similar to Fig. \ref{fig:full_money_plot}, but uses the \smith~EFA implementation.}
        \label{fig:full_money_plot_smith}
\end{figure}

\begin{figure}
    \includegraphics[scale=0.4]{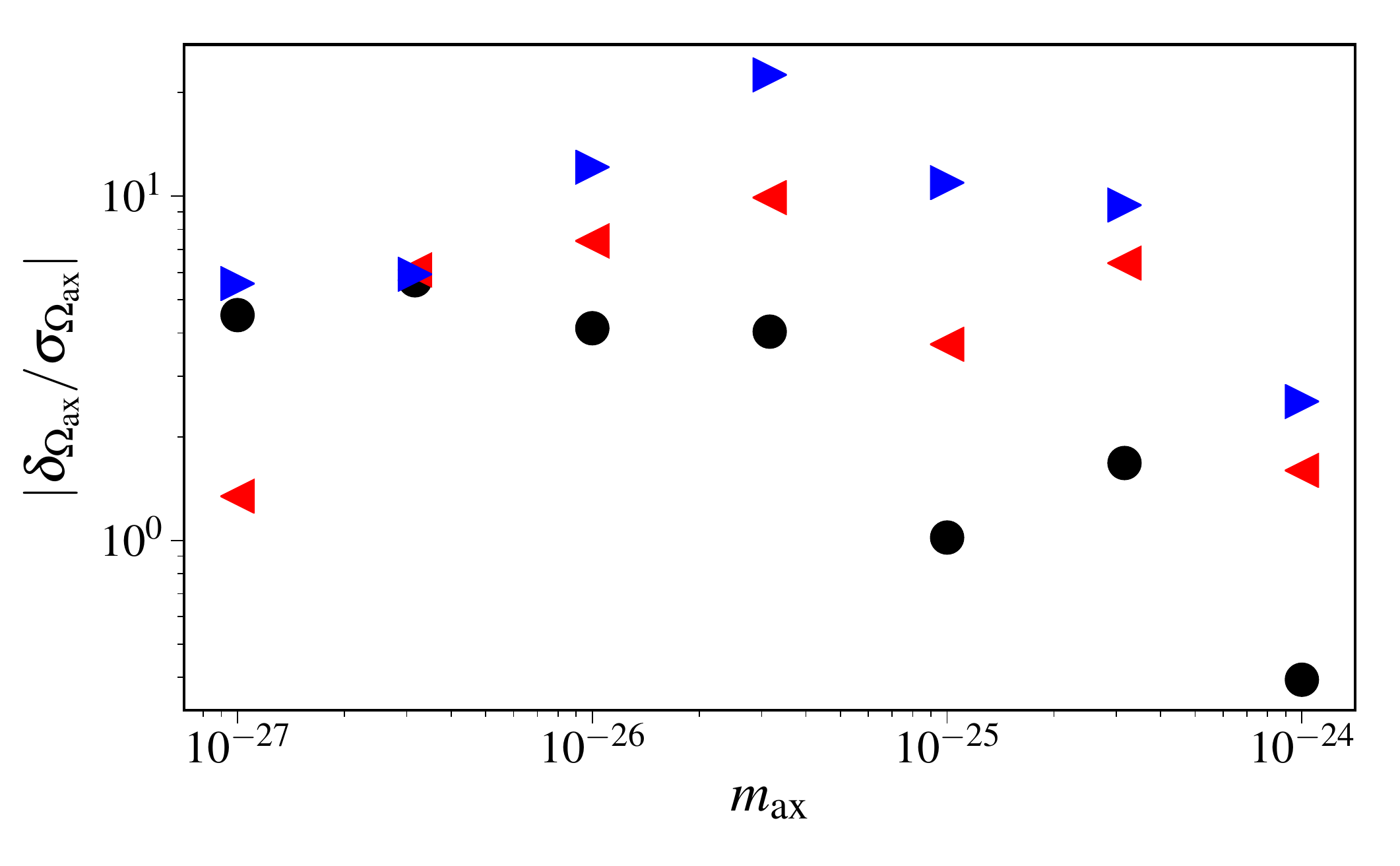}
    \caption{(Color online). Dimensionless bias in $\Omega_{\rm ax}$ as a function of $m_{\rm ax}$ for the \smith~implementation, calculated using Eq.~\eqref{eq:bias_ofix} (thus neglecting parameter degeneracies), with $\Omega_{\rm ax}/\Omega_{\rm DM}$ set equal to the current (\textit{Planck}) $3\sigma$ upper limits. This figure is similar to Fig. \ref{fig:full_money_plot_ofix}, but uses the \smith~EFA implementation.}
        \label{fig:full_money_plot_ofix_smith}
\end{figure}

\begin{figure}
    \includegraphics[scale=0.4]{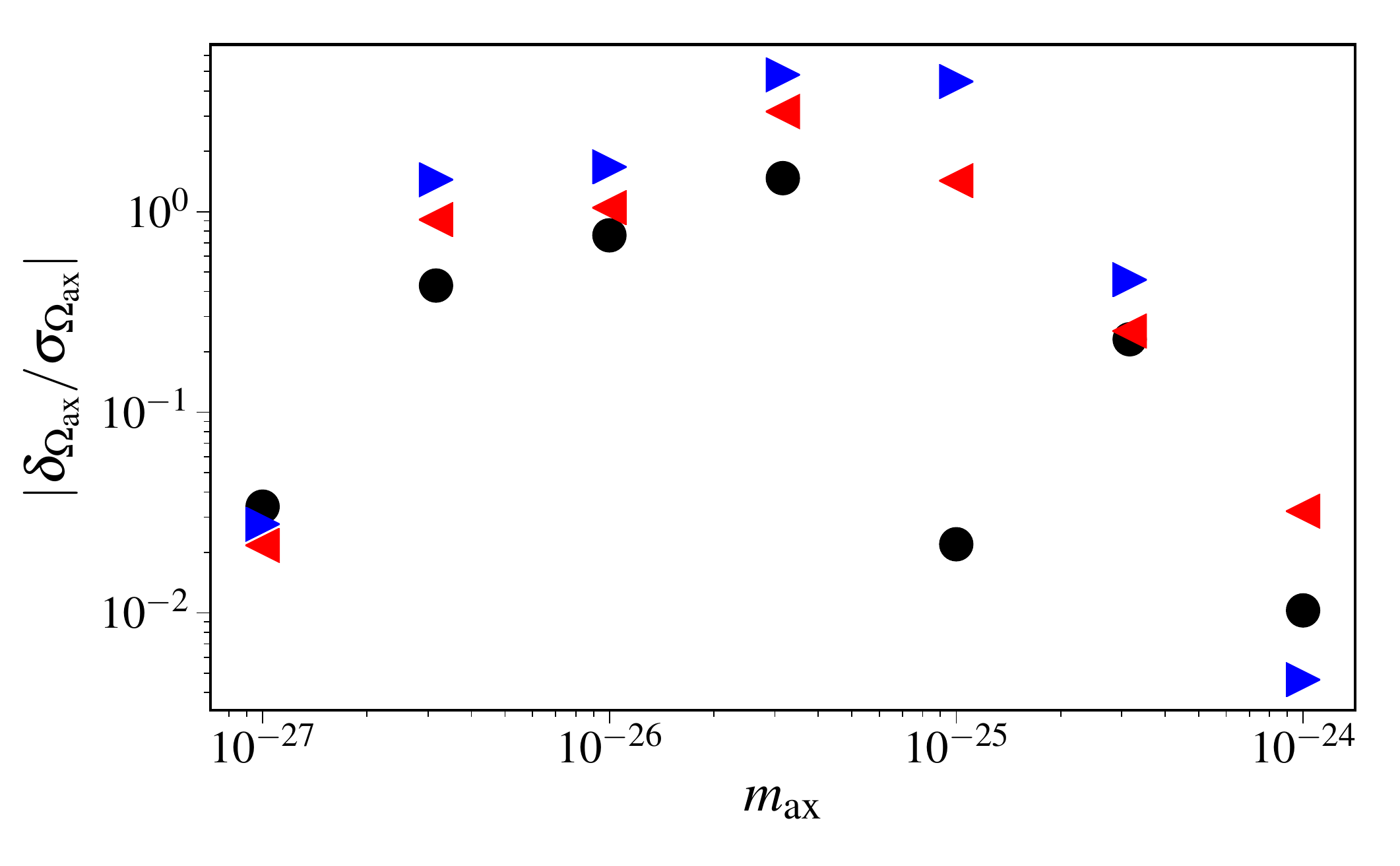}
    \caption{(Color Online). Dimensionless bias in $\Omega_{\rm ax}$ as a function of $m_{\rm ax}$ for the \urena implementation, calculated using Eq.~\eqref{eq:bias} (thus including parameter degeneracies), with $\Omega_{\rm ax}/\Omega_{\rm DM}$ set equal to the current (\textit{Planck}) $3\sigma$ upper limits. This figure is similar to Fig. \ref{fig:full_money_plot}, but uses the \urena~EFA implementation.}
        \label{fig:full_money_plot_urena}
\end{figure}

\begin{figure}
    \includegraphics[scale=0.4]{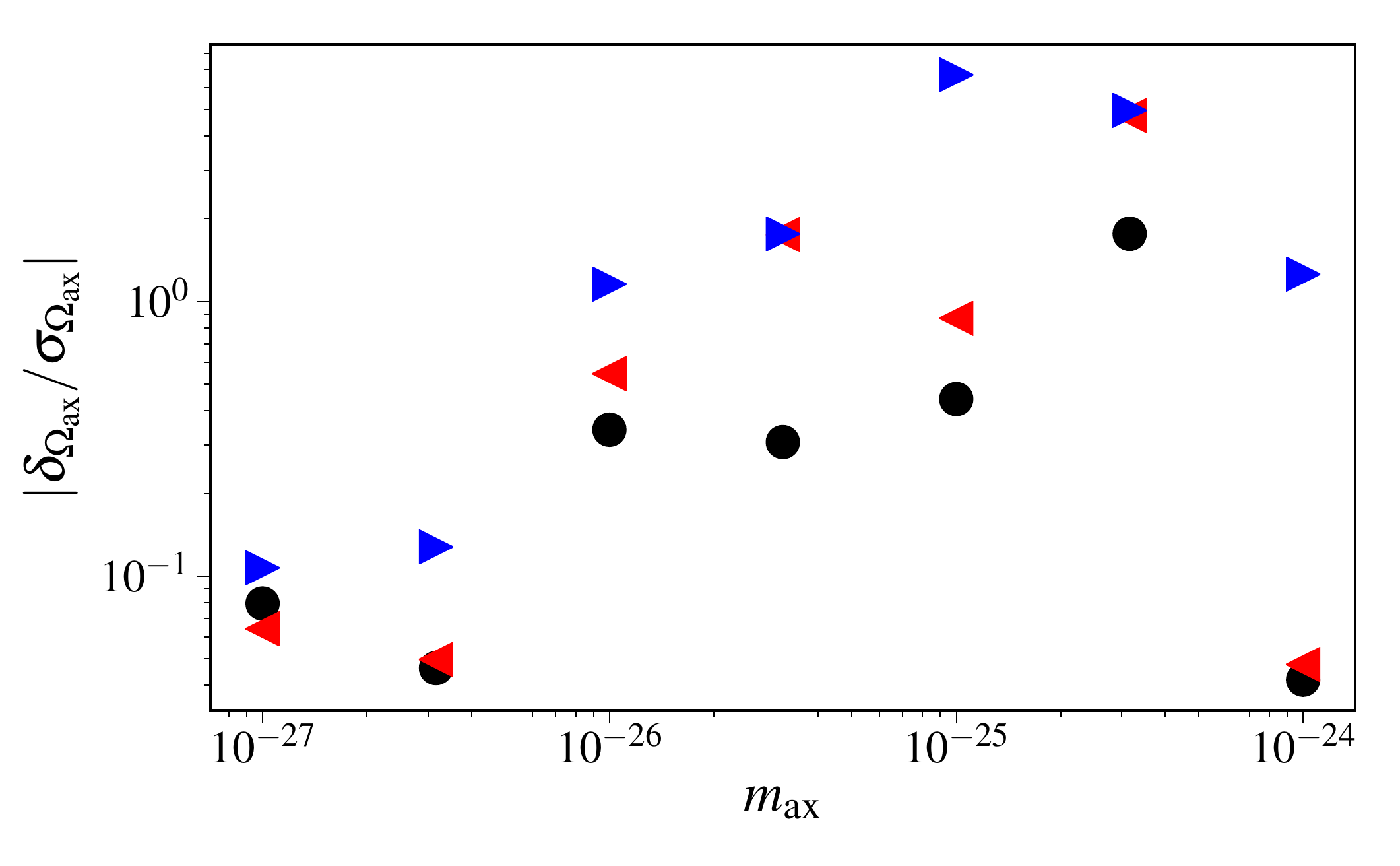}
    \caption{(Color online). Dimensionless bias in $\Omega_{\rm ax}$ as a function of $m_{\rm ax}$ for the \urena implementation, calculated using Eq.~\eqref{eq:bias_ofix} (thus neglecting parameter degeneracies), with $\Omega_{\rm ax}/\Omega_{\rm DM}$ set equal to the current (\textit{Planck}) $3\sigma$ upper limits. This figure is similar to Fig. \ref{fig:full_money_plot_ofix}, but uses the \urena~EFA implementation.}
        \label{fig:full_money_plot_ofix_urena}
\end{figure}

We also evaluated bias with no priors for the \smith~and~\urena implementations, and show the results in Figs. \ref{fig:full_money_plot_smith}-\ref{fig:full_money_plot_ofix_urena}. At \textit{Planck} noise levels for the~\smith~implementation, $|\delta_{\Omega_{\rm ax}}|\lesssim \sigma_{\Omega_{\rm ax}}$ for all $m_{\rm ax}$ values if we marginalize over cosmological parameters (see Fig. \ref{fig:full_money_plot_smith}). For the CVL case here, $2\sigma_{\Omega_{\rm ax}}\lesssim |\delta_{\Omega_{\rm ax}}|\lesssim 4\sigma_{\Omega_{\rm ax}}$. If cosmological parameters are fixed, the bias in the CVL case is large: $3\sigma_{\Omega_{\rm ax}}\lesssim |\delta_{\Omega_{\rm ax}}|\lesssim 20\sigma_{\Omega_{\rm ax}}$ (see Fig. \ref{fig:full_money_plot_ofix_smith}).

For the \urena implementation at \textit{Planck} noise levels including cosmological parameter marginalization, $|\delta_{\Omega_{\rm ax}}|\lesssim \sigma_{\Omega_{\rm ax}}$, while $|\delta_{\Omega_{\rm ax}}|\lesssim 2\sigma_{\Omega_{\rm ax}}$ if cosmological parameters are assumed fixed, at the same noise level (see Fig. \ref{fig:full_money_plot_urena}). For the CVL case with cosmological parameter marginalization $\sigma_{\Omega_{\rm ax}}\lesssim |\delta_{\Omega_{\rm ax}}|\lesssim 4\sigma_{\Omega_{\rm ax}}$ if $m_{\rm ax}\lesssim 10^{-25}~{\rm eV}$ and $|\delta_{\Omega_{\rm ax}}|\ll \sigma_{\Omega_{\rm ax}}$ otherwise. For the \urena implementation with cosmological parameters held fixed in the CVL case, $\sigma_{\Omega_{\rm ax}}\lesssim |\delta_{\Omega_{\rm ax}}|\lesssim 7\sigma_{\Omega_{\rm ax}}$ if $m_{\rm ax}\gtrsim 10^{-26}~{\rm eV}$ and $|\delta_{\Omega_{\rm ax}}|\ll \sigma_{\Omega_{\rm ax}}$ otherwise.

We note that the ~\urena implementation exhibits an unusual property in the nonmarginalized bias computation [see Eq.~(\ref{eq:bias_ofix}) and Fig. \ref{fig:full_money_plot_ofix_urena}]. Unlike the \bench and \smith~implementations (see Figs. \ref{fig:full_money_plot_ofix} and \ref{fig:full_money_plot_ofix_smith}), the low $m_{\rm ax}$ single-parameter bias computed in the \urena implementation is far smaller than the bound implied by the order of magnitude of $Z$ (see Fig. \ref{fig:Zmoney}), i.e.  $|\delta_{\Omega_{\rm ax}}/\sigma_{\Omega_{\rm ax}}|\ll Z$.  The dimensionless bias in this case [Eq.~(\ref{eq:bias_ofix}] is essentially a dot product between the vectors $dC_{\ell}/d\Omega_{a}$ and $\Delta C_{\ell}$, under a metric set by the data covariance. 

As discussed in Appendix \ref{app:Zstat_deriv}, the Cauchy-Schwarz inequality thus imposes the bound $|\delta_{\Omega_{\rm ax}}/\sigma_{\Omega_{\rm ax}}|\leq Z$, but does not require the bias to saturate this bound. Examining the summand of this dot product, we note that the \urena~implementation yields $\Delta C_{l}$ values that oscillate about zero with alternating sign, compared with the \bench \&~\smith~implementations, which typically show a negative semidefinite offset ($\Delta C_{l}\leq 0$) at $\ell$ values that contribute significantly to the sum. This sign structure appears to be responsible for the single-parameter $|\delta_{\Omega_{\rm ax}}/\sigma_{\Omega_{\rm ax}}|\ll Z$ behavior of the \urena implementation. The \urena implementation also exhibits a different ordering of single and multiparameter (degenerate) bias levels at some masses than the other $2$ implementations, likely due to the different covariance level of $\Omega_{\rm DM}h^{2}$ with $\Omega_{\rm ax}$ in the Fisher matrix in the \urena implementation.

\section{Conclusions}
\label{sec:disc-conc}
In this work, we have quantitatively compared predictions for cosmological observables in ultra-light axion (ULA) models obtained from the effective fluid approximation (EFA) with the results of a full Klein-Gordon (KG) treatment of the dynamics. Along the way, we found that an alternative treatment of ULA perturbations \cite{Urena-Lopez:2015gur} is, in fact, equivalent to the EFA, when written in terms of appropriate variables; in future work, it would be valuable to undertake a more systematic comparison of our methods. 

We forecasted the resulting bias to ULA parameters from primary (unlensed) CMB anisotropy power spectrum measurements, using well-established Fisher-matrix techniques to estimate how the numerical discrepancies between EFA and KG predictions will offset the centroids of the parameter likelihood, assuming that a ULA detection is hiding just underneath the sensitivity level of completed experimental analyses.

For the \bench version of the EFA, we find that if the full field dynamics are neglected, primary CMB anisotropy constraints to ULAs from \textit{Planck} are robust, while constraints to/measurements of $\Omega_{\rm ax}$ from future CMB experiments (e.g.~CMB-S4 \cite{Abazajian:2019eic}) will be moderately biased at the $\delta/\sigma \lesssim 2$ level, if marginalization over the usual cosmological parameters is assumed. If an idealistic external dataset can break all the relevant degeneracies, the bias could be as high as $\delta/\sigma \sim \mathcal{O}(10)$. We find that no immediately planned effort will approach this bias level. 

These conclusions could be altered when the full covariance matrix of future large-scale structure efforts (including degeneracies among standard cosmological parameters and their covariance with ULA parameters) is used, and we will address this issue in future work. For the \smith~and~\urena implementations in the marginalized case, we find comparable (but slightly worse, by a factor of $\sim 2$) bias levels when cosmological parameters are marginalized over. 

It is interesting to note that the fractional differences between the different choices of $\mathcal{N}$ - which generate residuals well in excess of the $3/\ell$ threshold - lead to relatively small biases at least, in part, because the residuals have support over a large range of $\ell$. On the other hand, it stands to reason that if the residuals were more localized in $\ell$ we might have found a much larger bias (as recently pointed out in a different context in Ref. \cite{Smith:2019ihp}).

Measurements of CMB weak lensing are a key scientific driver for upcoming experiments like CMB-S4 \cite{Abazajian:2019eic} or more futuristic ideas like CMB-HD \cite{Nguyen:2017zqu}. These efforts have the promise of sensitivity to ULAs in the window $m_{\rm ax}\sim 10^{-22}~{\rm eV}$, where they might comprise all the DM. In future work, we will thus extend our work to include measurements of the lensing potential power spectrum and higher $m_{\rm ax}$ values. The EFA generally works best at late times (where the lensing kernel peaks \cite{Hanson:2009kr}), but it is also possible that the large information content of the full CMB $4$-point correlation function (which drives constraints to the lensing potential power spectrum $C_{L}^{\phi \phi}$ \cite{Hanson:2009kr}) induces larger biases.

The tools used in this work to compute observables for the ``exact" computational benchmark only solve the Klein-Gordon equation until $\tilde{m}_{\rm ax}=10^{4}H_{0}$, which is more than sufficient for the time and length scales probed by primary CMB anisotropies (as discussed in Sec. \ref{sec:bias}). Future work will require us to extend the ``exact" case to later times, in order to conduct definitive comparisons with the EFA in the case of CMB lensing, and other observables sensitive to the matter power spectrum (e.g.~the galaxy correlation and shear power spectra to be probed with exquisite accuracy by LSST \cite{Abell:2009aa} and other comparable efforts).

In this work, we have assumed purely adiabatic initial conditions. If the relevant $U(1)$ global symmetry is broken before the end of the inflationary era, the ULA field (as a nearly massless scalar spectator) will carry quantum fluctuations with $\phi_{1}\sim H_{\rm I}/(2\pi)$, where $H_{\rm I}$ is the Hubble expansion rate during inflation. These fluctuations will in turn source primordial DM isocurvature (entropy) fluctuations (See Refs. \cite{Hlozek:2017zzf} and references therein). These are observationally known to be subdominant \cite{Ade:2015lrj} (while still allowing large $\Omega_{\rm ax}$ values), but could be detected by ongoing/upcoming CMB experimental efforts \cite{Abazajian:2019eic}.

The CMB will thus provide interesting constraints to $H_{I}$ and $\Omega_{a}$, complementing the impact of CMB polarization-based tests of the inflationary tensor-to-scalar ratio $r$. Entropy fluctuations are more directly sensitive to dark-sector fluctuations than the adiabatic CMB power spectrum, and it is thus important to fully evaluate the EFA-induced bias in ULA isocurvature scenarios.

\begin{acknowledgments}
T.~L.~S. and D.~G. acknowledge support in part by NASA grant 80NSSC18K0728. D.~G. and J.~C. acknowledge support from the Provost's office at Haverford College. J.~C. thanks Haverford College for hospitality during the completion of this work. T.~L.~S. acknowledges support from the Provost's office at Swarthmore College.  This work was supported in part by the National Science Foundation under Grant No. NSF PHY-1125915 at the Kavli Institute for Theoretical Physics (KITP) at UC Santa Barbara. We thank A.~Aoki, R.~Hlo\v{z}ek, W.~Hu, L.~Hui, M.~Kamionkowski, D.~J.~E.~Marsh, C.~Owen, V.~Poulin, and M.~Raveri for useful conversations. D.~G. thanks KITP for its hospitality during the completion of this work.  
\end{acknowledgments}

\appendix
\section{WKB}\label{app:WKB}
\begin{figure*}[h]
\centering
    \includegraphics[scale=0.4]{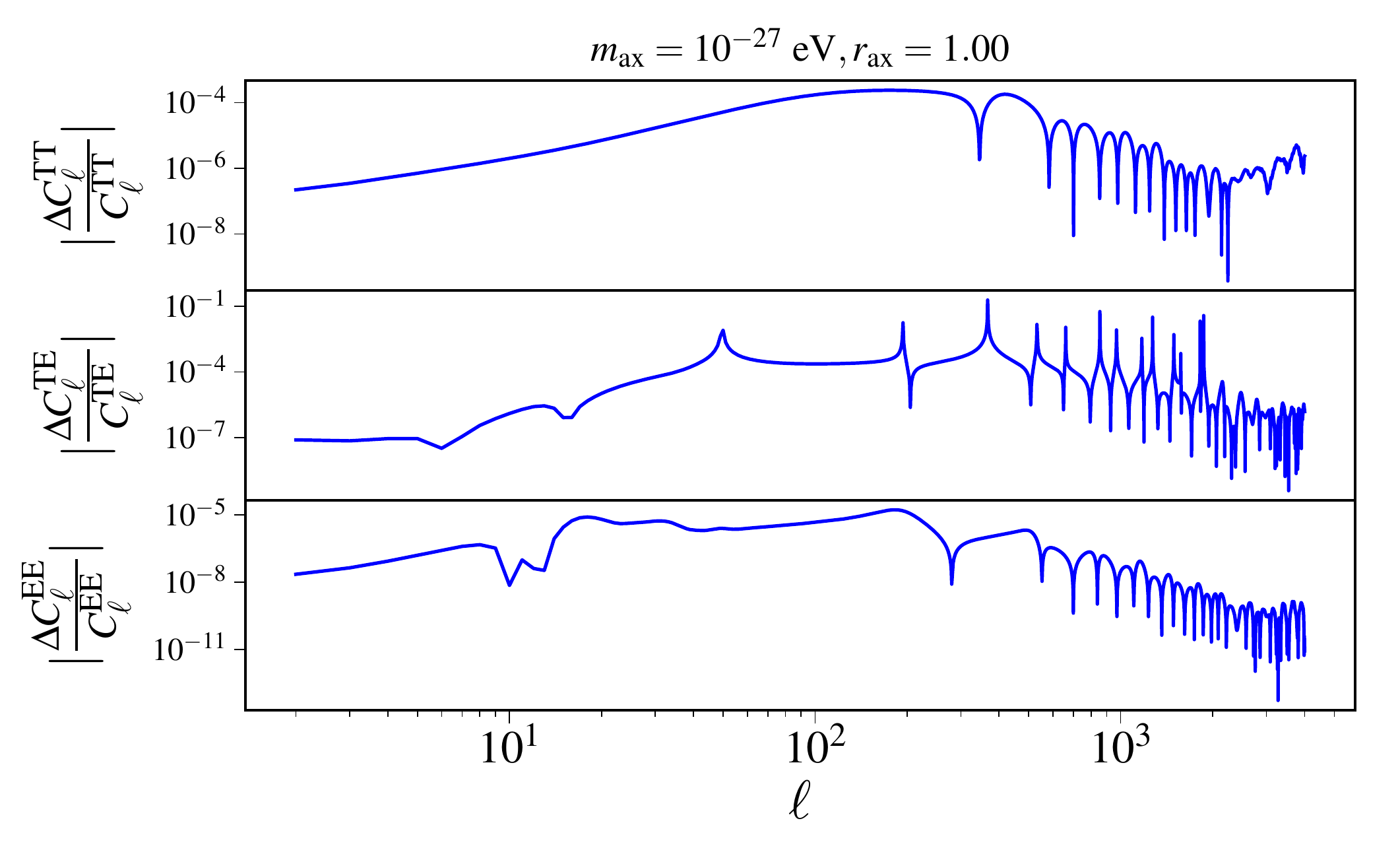}    \includegraphics[scale=0.4]{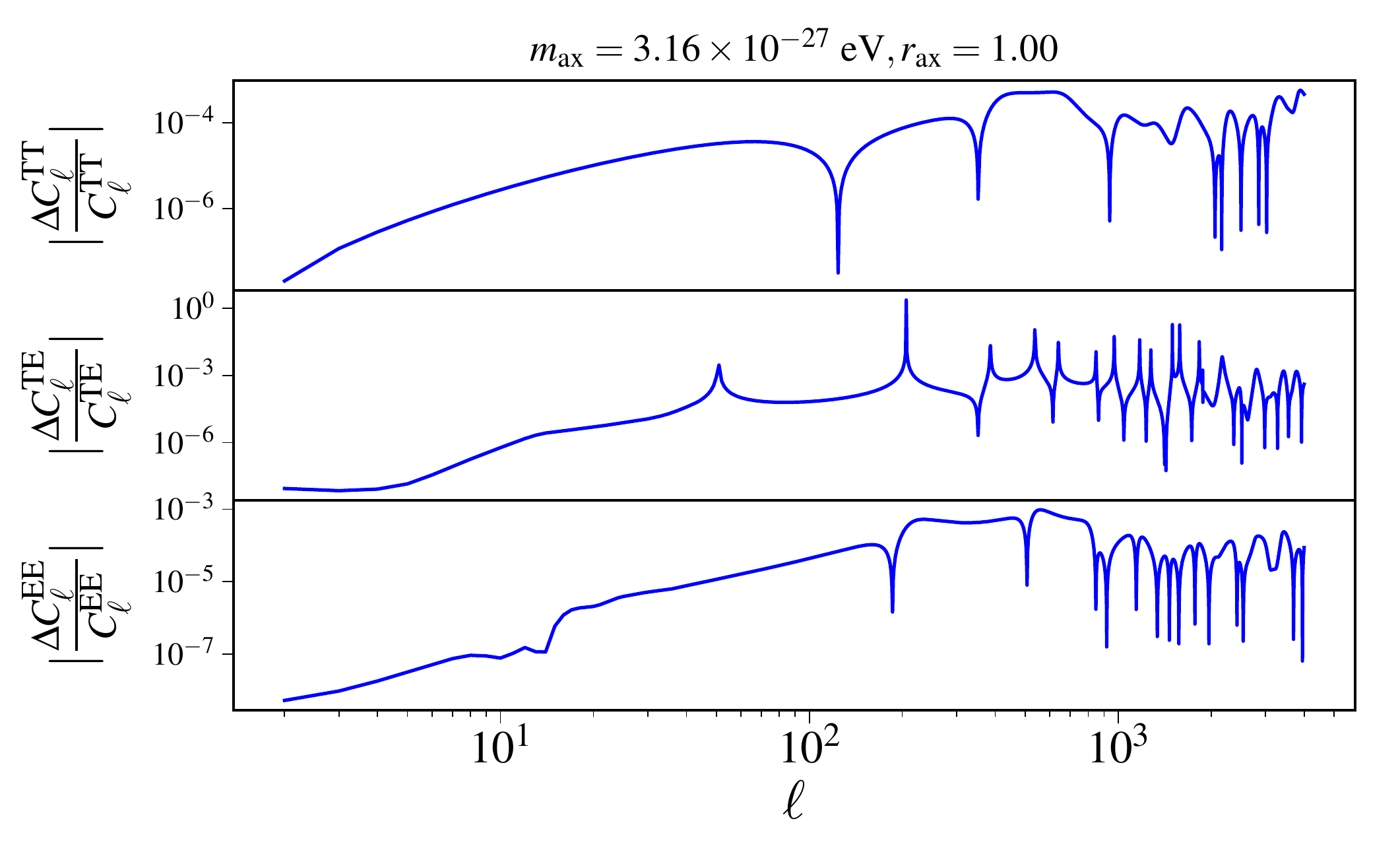}
        \includegraphics[scale=0.4]{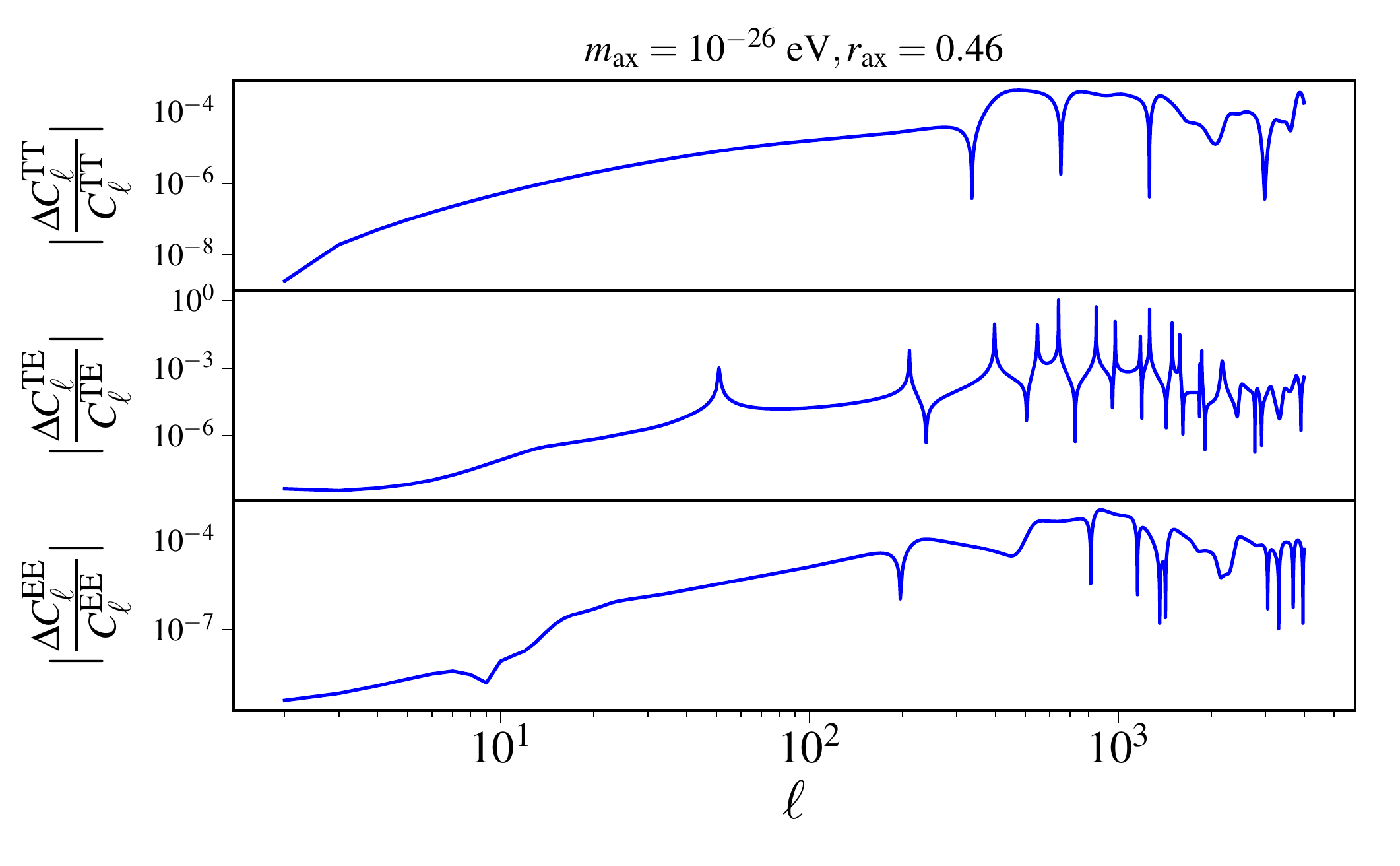}    \includegraphics[scale=0.4]{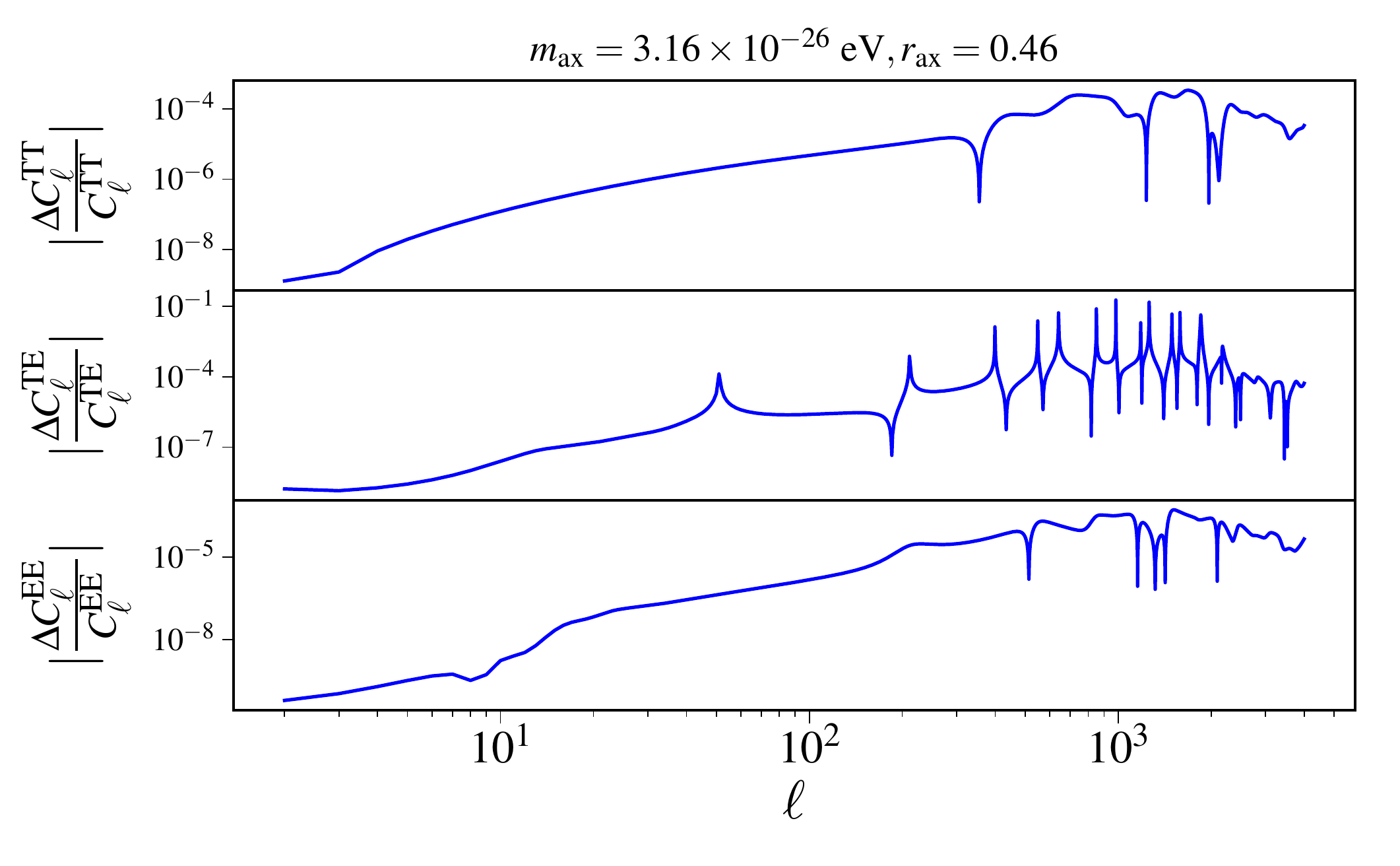}
          \includegraphics[scale=0.4]{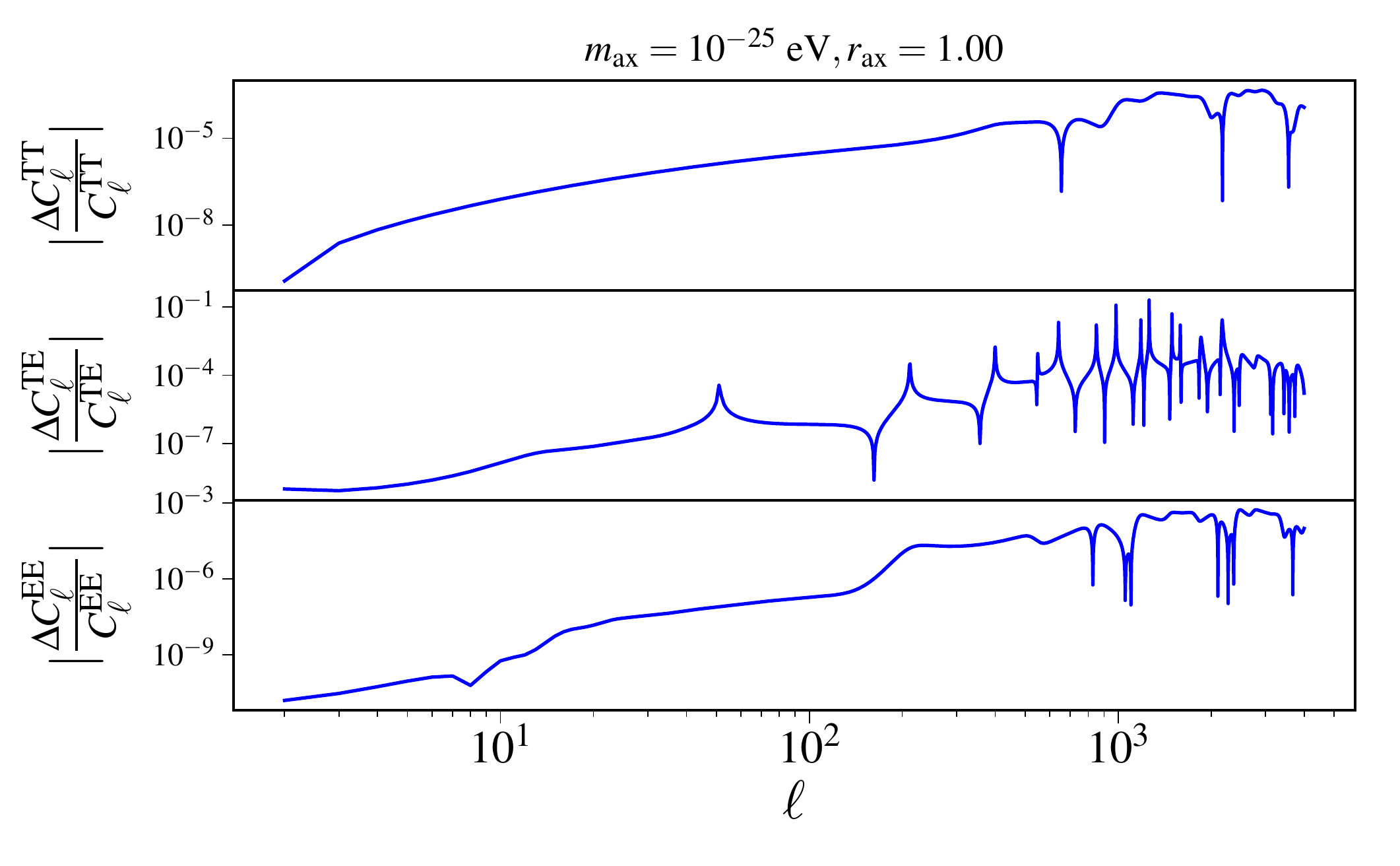}
          \includegraphics[scale=0.4]{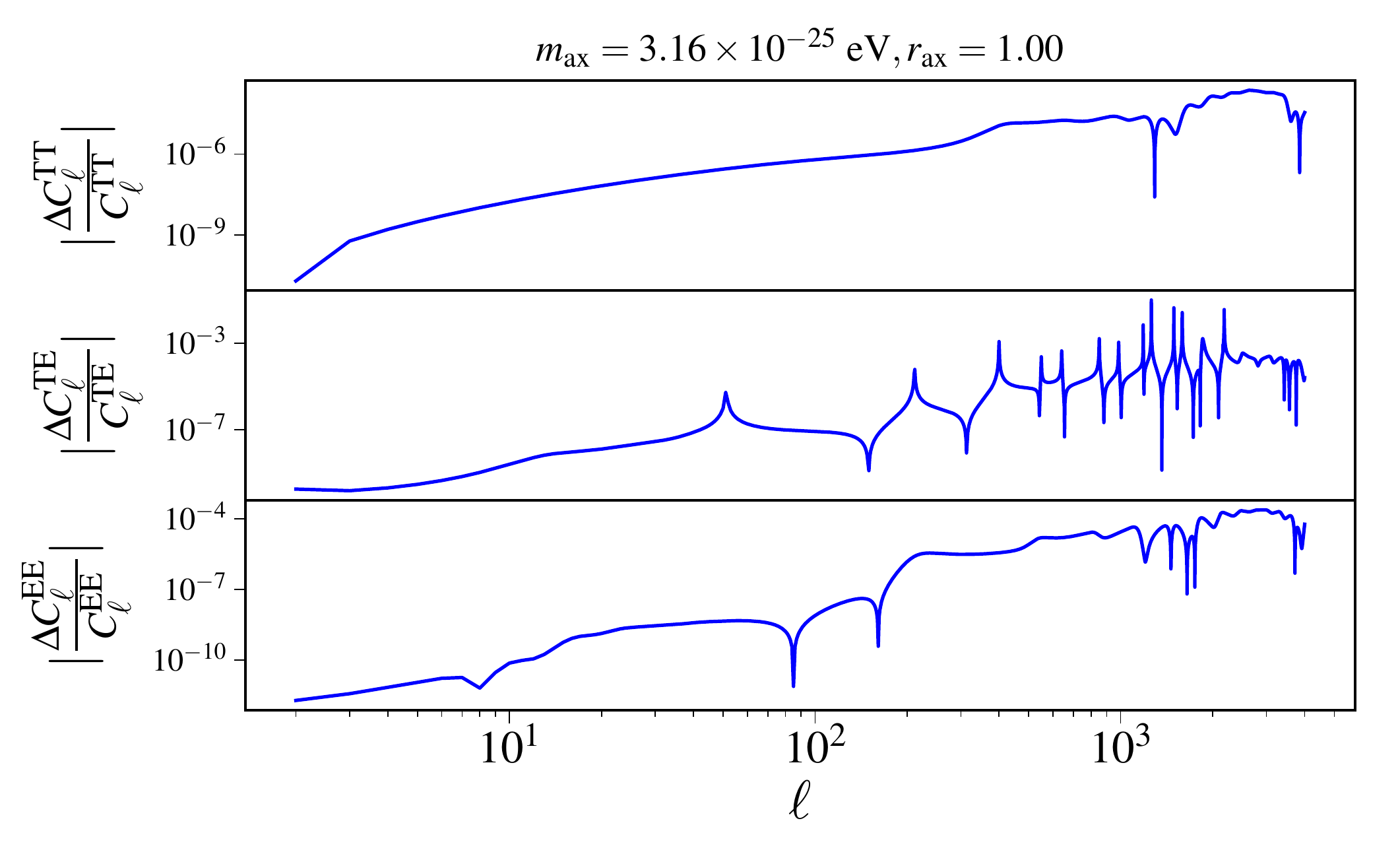}
     \caption{Plots of CMB power spectra comparing different implementations of the \urena implementation, as explained in Appendix \ref{app:URnum}.     \label{fig:ulvplot}
}
\end{figure*}
Recall that for the ULA background, we want to solve
\begin{equation}
 \phi_0'' + 2\frac{ a'}{a}  \phi_0' + a^2 \tilde{m}^2 \phi_0 = 0.
\end{equation}
Substituting $\tilde \phi = a \phi$, we find a differential equation solvable with the WKB method (see, for example, Ref.~\cite{bender:2013advanced}). It has the solution,
\begin{equation}\label{uapp}
\begin{aligned}
\phi_0 = \frac{1}{a Q^{1/4}}\Bigg(& C_1 \exp\left(i \int_{\eta_*}^\eta \sqrt{Q}d \eta'+...\right) \\
+ &C_2 \exp\left(-i \int_{\eta_*}^\eta\sqrt{Q} d\eta' + ...\right)\Bigg),
\end{aligned}
\end{equation}
with $Q=a^2 \tilde{m}^2-a''/a$. Self-consistently, if $H/\tilde{m} \ll 1$, then we assume $\rho_{\rm ax} \sim 1/a^3$ which implies $a''/a^3 \lesssim H^2$, so we can assert $Q \approx a^2 \tilde{m}^2$. Then, the usual \textit{ansatz} for the ULA field is recovered $\phi_0 = A \cos (\tilde{m} t+\theta)/a^{3/2}$ for some constant $A$ and $\theta$. It then immediately follows that $\rho_{\rm ax} = A^2 \tilde{m}^2 / (2a^3)$ to leading order in $H/\tilde{m}$.

\section{Alternative cycle averaging procedures}
\label{app:URnum}
We note that our implementation of the \urena implementation [which applies a $2\tilde{m}t=100$ switch with equations of motion (EOMs) given by Eqs.~(\ref{eq:fluidax})-(\ref{eq:euler}), a $\langle w_{\rm ax}\rangle =0$ approximation, and the cycle-averaged sound speed, Eq.~(\ref{eq:cycavsound})] has additional gauge terms coming from the transformation from the fluid rest frame to the synchronous gauge. These terms are not present in Eqs.~(\ref{eq:fluidaxul})-(\ref{eq:eulerul}), which represent the formalism of Ref. \cite{Urena-Lopez:2015gur} in terms of fluid variables. 

There are also differences between our expression for the cycle-averaged sound speed [Eq.~(\ref{eq:cycavsound})], and the simpler expression $c_{s}^{2}=k^{2}/(4\tilde{m}^{2}a^{2})$ obtained in Ref.~\cite{Urena-Lopez:2015gur}.

As a result of the scaling of various terms with $k$, the first set of differences can only be large for $k\ll H$ (at superhorizon scales), while the difference between the two different cycle-averaged sound speeds only grows large when $k\gg \tilde{m}a$, deep in the Jeans-suppressed regime of perturbation evolution. To be sure that these implementation differences do not affect our conclusions, we wrote a modified code that exactly implemented Eqs.~(\ref{eq:fluidaxul})-(\ref{eq:eulerul}).

The resulting differences in observables are shown in Fig.~\ref{fig:ulvplot}. Fractional differences in power spectra between the different versions of the \urena implementation are several orders of magnitude smaller than differences between it and other implementations at scales contributing to the sums in expressions for $Z$ and $\delta_{\Omega_{\rm ax}}$. Our conclusions about the relative accuracies of different EFA implementations are thus robust. 

\section{$Z$ statistic derivation}\label{app:Zstat_deriv}
We define the following inner product (which clearly satisfies the usual axioms),
\begin{equation}
\langle \vec a, \vec b \rangle =
\sum_{X,Y \in \{\text{TT,TE,EE}\}} \sum_{\ell=\ell_\text{min}}^{\ell_\text{max}} f_{\text{sky}}a_\ell^X b_\ell^Y(\Xi^{-1}_{XY})_\ell
\end{equation}
where the components of $\vec a$ are $a_l^X$ for $\ell =\ell_\text{min},\ell_\text{min}+1,...,\ell_\text{max}$ and $X \in \{\text{TT},\text{TE},\text{EE}\}$ in any arbitrary order. The Cauchy-Schwarz inequality holds for $\langle \vec a, \vec b \rangle$, as it is an inner-product. If we consider a one parameter model (so that $\sigma_{\lambda} = 1/\sqrt{F}$ for $F$ the Fisher matrix), we see
\begin{equation}
\begin{aligned}
    \left|\frac{\delta_{\lambda}}{\sigma_{\lambda}^2}\right|  &= |\langle \vec \Delta C_\ell , \frac{\partial \vec C_\ell }{ \partial \lambda }\rangle| \\
    &\le \sqrt{\left|\left\langle \frac{\partial \vec C_\ell }{ \partial \lambda},\frac{\partial \vec C_\ell }{ \partial \lambda}\right\rangle \right| |\langle \vec \Delta C_\ell,\vec \Delta C_\ell\rangle |} = \left|\frac{Z}{\sigma_{\lambda}}\right|
\end{aligned}
\end{equation}
where we used Eqs.~(\ref{eq:bias}) and (\ref{eq:Zstat}). Since $Z,\sigma_\lambda \ge 0$, we see $Z \ge |\delta_\lambda/\sigma_\lambda|$. If there are multiple parameters being varied for the calculation of $\sigma_{\lambda_i}=\sqrt{(F^{-1})_{ii}}$, the explicit bound may not hold, but the $Z$ statistic can still give an estimate of the dimensionless bias.

\newpage\onecolumngrid\newpage
\section{Additional CMB power spectra}
\begin{figure}[ht]
    \begin{center}
    \includegraphics[scale=0.42]{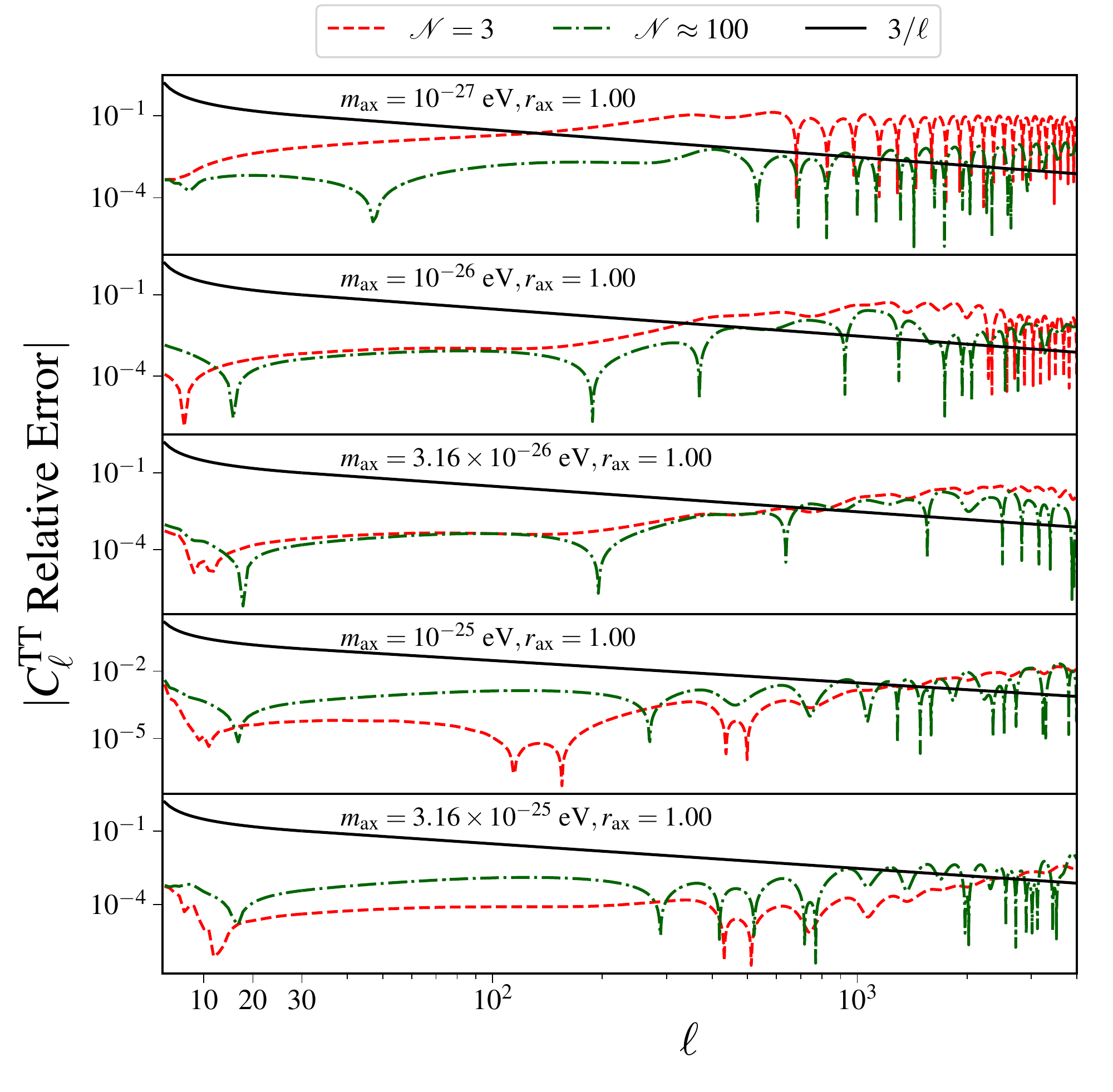}
    \includegraphics[scale=0.396]{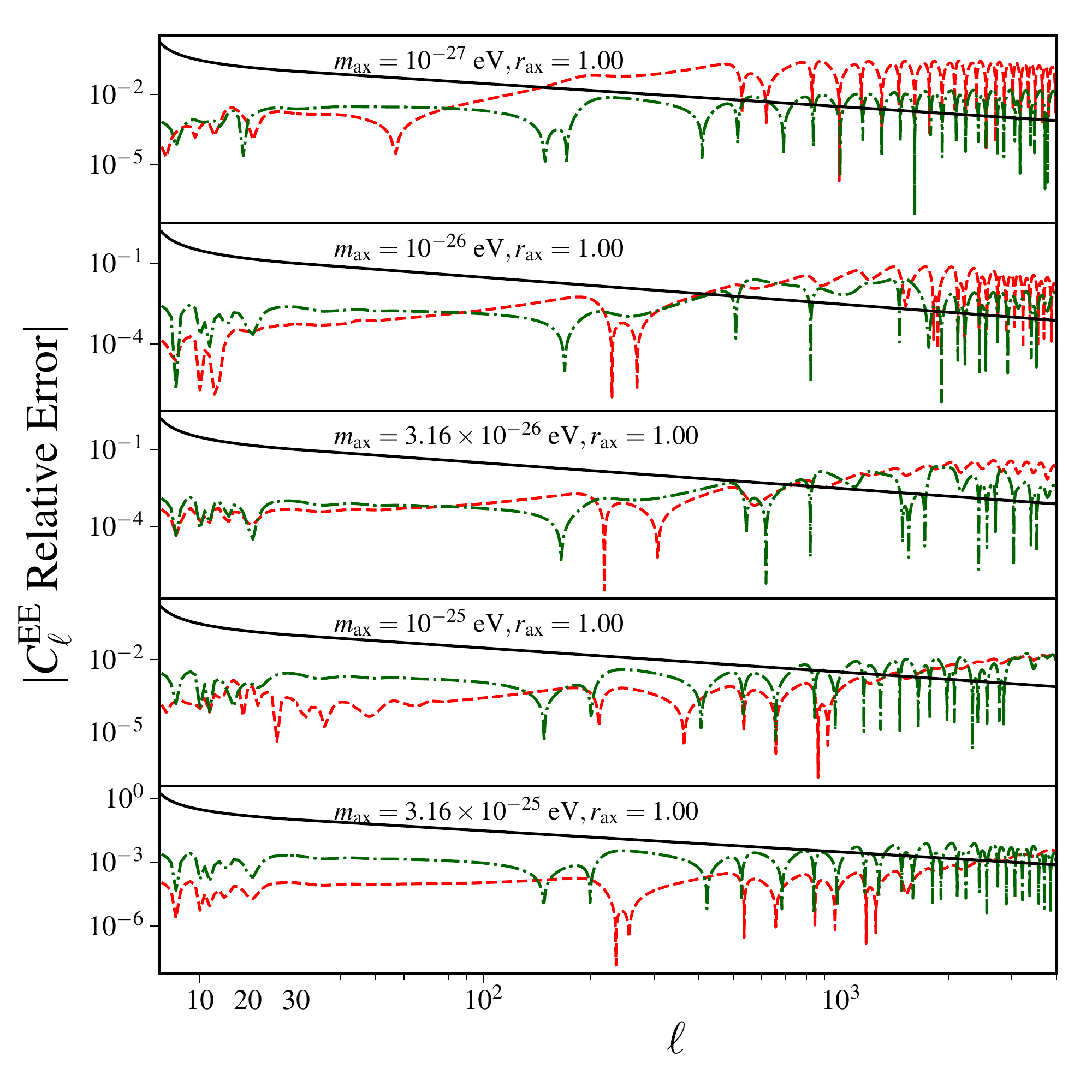}\\\end{center}
    \includegraphics[scale=0.42]{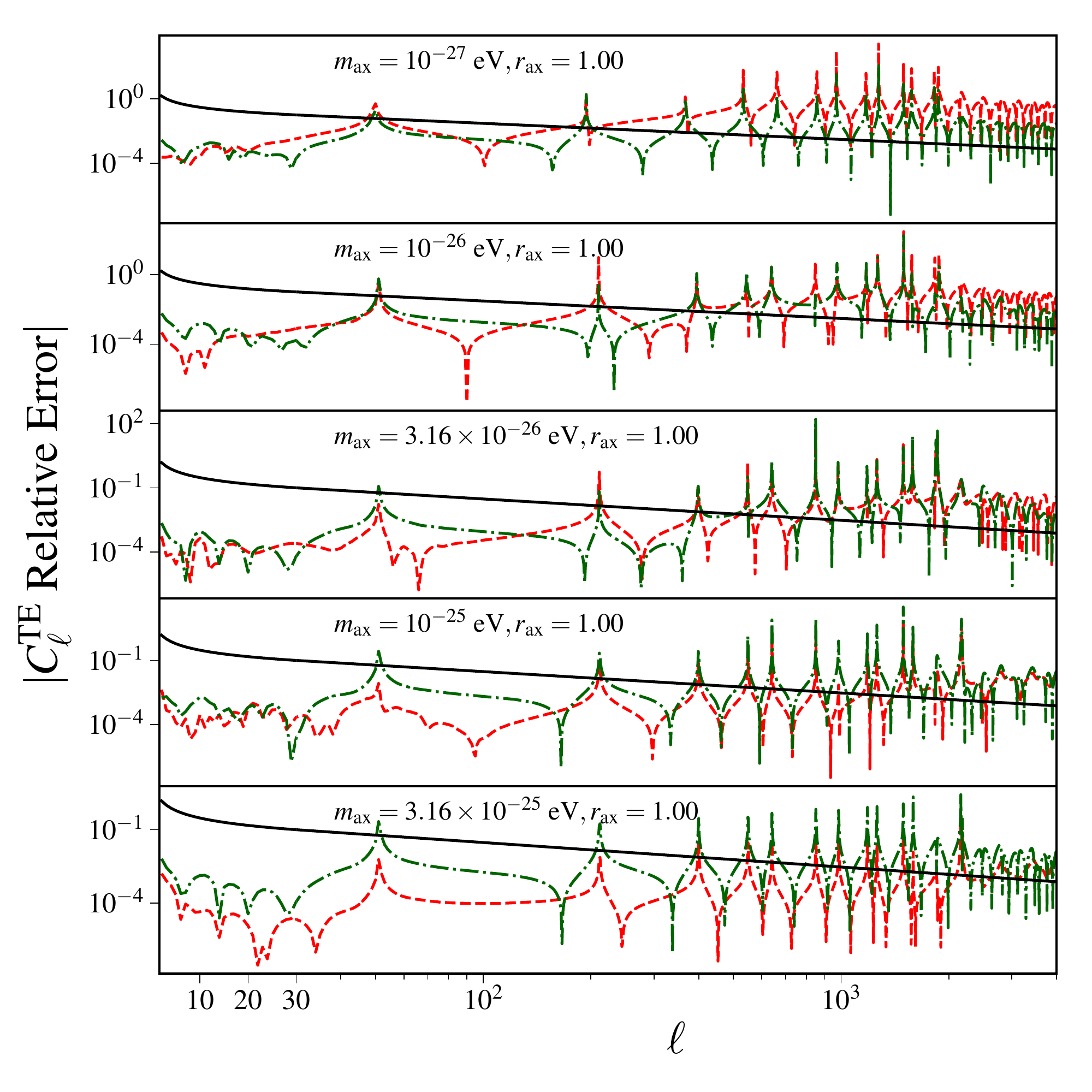}
    \caption{(Color online). Relative error of CMB anisotropy power spectra computed with different EFA implementations ($\mathcal{N}=3$ and $\mathcal{N}=100$), computed in comparison to the exact solution. The constant $\mathcal{N}$ defines the moment at which the exact equations are switched to the EFA, using the criterion $m_{\rm ax} /\hbar> \mathcal{N}H$ (in units where $c=1$). The black curve ($3/\ell$) is a rough precision threshold beyond which parameter biases may be significant \cite{Seljak:2003th}. If this curve is exceeded at many $\ell$ values by the actual EFA relative errors, an explicit computation of bias is needed to assess the full implications of these errors for cosmological parameter inference and ULA constraints.} \label{fig:CMBlow_extra0}
\end{figure}
\label{sec:exfigs_cmb}

\newpage
\twocolumngrid
\end{document}